\documentclass[11pt,a4paper]{article}
\usepackage[utf8]{inputenc}
\usepackage{jheppub}
\usepackage{amsthm,amsbsy,amsfonts,mathrsfs,enumerate,float,wrapfig,amsmath}
\usepackage[utf8]{inputenc}
\usepackage{subfigure}
\usepackage{amsmath}
\usepackage{tikz}
\usetikzlibrary{arrows,decorations.pathmorphing,backgrounds,positioning,fit,shapes,chains,shapes.gates.logic.US,trees,tikzmark,decorations.text,calc,matrix,chains,positioning,decorations.pathreplacing,arrows,mindmap,decorations.pathreplacing,decorations.markings}

\usepackage{nicefrac}
\usepackage[enableskew,vcentermath]{youngtab}
\usepackage{epstopdf}
\usepackage{url}
\usepackage{float}
\usepackage{slashed,framed,xcolor,tikz}
\usepackage{amsmath}
\makeatletter
\newcommand\xleftrightarrow[2][]{%
	\ext@arrow 9999{\longleftrightarrowfill@}{#1}{#2}}
\newcommand\longleftrightarrowfill@{%
	\arrowfill@\leftarrow\relbar\rightarrow}
\makeatother

\usepackage{float}

\def\a{{\alpha}}

\def\b{{\beta}}

\def\r{{\gamma}}

\def\0{{\emptyset}}
\def\u{{\mu}}
\def\v{{\nu}}
\def\h{{\eta}}
\def\la{{\lambda}}

\def\p{{\rho}}
\def\tZ{{\tilde{Z}}}

\def\sqqt {{ \sqrt{\frac{q}{t}}}}
\def\sqtq {{ \sqrt{\frac{t}{q}}}}

\def\inf{{\infty}}

\def\half{\frac{1}{2}}
\def \r){\right)}

\def\half{\text{half}}

\def\>{{  \succcurlyeq} }
\newcommand{\bsmall}{\begin{small}   }
	\newcommand{\esmall}{ \end{small}   }

\def\qbrane{{q\text{-brane}}}
\def\qbarbrane{{\bar{q}\text{-brane}}}
\def\tbrane{{t\text{-brane}}}
\def\tbarbrane{{\bar{t}\text{-brane}}}

\def\PE{\mathrm{PE}}

\title{
	Refined open topological strings revisited
}

\author[a]{Shi Cheng} 
\author[a,b]{and Piotr Su{\l}kowski}

\affiliation[a]{Faculty of Physics, University of Warsaw, ul. Pasteura 5, 02-093 Warsaw, Poland}
\affiliation[b]{Walter Burke Institute for Theoretical Physics, California Institute of Technology, Pasadena, CA 91125, USA}

\emailAdd{scheng@fuw.edu.pl} \emailAdd{psulkows@fuw.edu.pl}

\abstract{In this work we verify consistency of refined topological string theory from several perspectives. First, we advance the method of computing refined open amplitudes by means of geometric transitions. Based on such computations we show that refined open BPS invariants are non-negative integers for a large class of toric Calabi-Yau threefolds: an infinite class of strip geometries, closed topological vertex geometry, and some threefolds with compact four-cycles. Furthermore, for  an infinite class of toric geometries without compact four-cycles we show that refined open string amplitudes take form of quiver generating series. This generalizes the relation to quivers found earlier in the unrefined case, implies that refined open BPS states are made of a finite number of elementary BPS states, and asserts that all refined open BPS invariants associated to a given brane are non-negative integers in consequence of their relation to (integer and non-negative) motivic Donaldson-Thomas invariants. Non-negativity of motivic Donaldson-Thomas invariants of a symmetric quiver is therefore crucial in the context of refined open topological strings. Furthermore, reinterpreting these results in terms of webs of five-branes, we analyze Hanany-Witten transitions in novel configurations involving lagrangian branes.

\rule{0pt}{0pt}
\\
\rule{0pt}{0pt}
\\
\\
\\
\\
\\
\\
\\
\\
\\
\\
\\
\\
\\
\\
\\
\\
CALT-2021-013
}

\begin{document}
\maketitle

\newpage

\section{Introduction}

Refined topological string theory is an intriguing and mysterious creature. It is expected to arise as a deformation (refinement) of an ordinary (unrefined) topological string theory. However, the status of the unrefined and refined theory is very different. 

Unrefined topological string theory has a worldsheet definition in genus expansion in the topological string coupling $g_s$, and a mathematical formulation as the Gromov-Witten theory \cite{Hori:2003ic}. Unrefined amplitudes are related to the low energy effective action for superstring theory \cite{Bershadsky:1993cx,Antoniadis:1993ze}; encode integral closed (Gopakumar-Vafa) and open (Ooguri-Vafa) BPS invariants \cite{Gopakumar:1998ii,OoguriV}; for some class of manifolds they are related to Chern-Simons theory \cite{Gopakumar:1998ki} and -- in consequence of this relation -- for toric Calabi-Yau threefolds can be computed in the formalism of the topological vertex \cite{AKMV}; for some particular Calabi-Yau manifolds, topological string amplitudes are captured by Nekrasov partition functions \cite{Nekrasov}, and can be also reinterpreted in terms of webs of five-branes and 5-dimensional supersymmetric theories \cite{Leung:1997tw,Hanany:1996ie,Benini:2009gi}. Relations between these systems arise from an underlying physical picture based on embedding them in M-theory. Equality of various quantities computed from these different perspectives assures consistency of this whole picture.

Refined topological string theory is a generalization of the unrefined theory, in which the dependence on the string coupling $g_s$ is replaced by the dependence on two parameters $\epsilon_1$ and $\epsilon_2$. One of the main motivations for the existence of such a theory is its expected relation to Nekrasov partition functions, which naturally depend on $\epsilon_1$ and $\epsilon_2$ that encode the $\Omega$-background \cite{Nekrasov}. However, there is a crucial difference in comparison to unrefined theory: there is no general worldsheet definition of refined topological strings, and no rigorous corresponding mathematical formulation in terms of Gromov-Witten theory. Nonetheless, in various dual perspectives mentioned in the previous paragraph one can naturally introduce a dependence on both $\epsilon_1$ and $\epsilon_2$, and then predict the form of certain amplitudes in such a hypothetical theory from computations in these dual pictures. In fact, these dual pictures are often regarded as providing a definition of refined topological string theory, at least for some specific types of Calabi-Yau manifolds. For example, such definitions, based on the relation to low energy effective action for superstring theory, have been proposed in \cite{Antoniadis:2010iq,Nakayama:2011be,Antoniadis:2013bja}. Other formulations of refined topological string amplitudes are proposed in terms of refined topological vertex \cite{Iqbal:2007ii,Awata:2005fa} (for toric manifolds), through the relation to refined Chern-Simons theory \cite{Aganagic:2012hs}, etc. 

Once one computes candidates for refined topological string amplitudes, one strong test of their consistency is provided by integrality of associated BPS numbers, which generalizes the original unrefined Gopakumar-Vafa and Ooguri-Vafa integrality conjectures. More precisely, while unrefined closed BPS invariants are integer numbers, for toric manifolds they turn out to be combinations (with extra signs) of refined BPS numbers, which are non-negative integers and count certain particle states. It was verified in various examples, e.g. in \cite{Iqbal:2007ii}, that refined closed BPS invariants are indeed non-negative integers. Other consistency tests involve reproducing properties of 5-dimensional supersymmetric theories from refined topological vertex calculations \cite{Hayashi:2013qwa}. All these are strong consistency tests of refined topological string theory, at least in the \emph{closed} sector.

\bigskip

A general aim of this paper is to verify consistency of refined \emph{open} topological strings from several interrelated perspectives. First, we verify integrality of refined open topological string amplitudes for a wide class of toric manifolds. Such tests have been conducted only in a few simple, special cases in \cite{DGH,Kozcaz:2018ndf}. In a related, albeit a bit different context, integrality of refined BPS states associated to  knots, which can be engineered by branes in the resolved conifold, was analyzed in \cite{Kameyama:2017ryw}. However such tests have not been conducted for more complicated toric manifolds with toric Aganagic-Vafa branes. The first aim of this work is to fill this gap. Analogously to refined closed BPS invariants, refined open BPS invariants for toric manifolds should also count certain particle states, and thus we expect that they should be non-negative integers. We confirm this claim in various examples.

Recall that complexity of topological string amplitudes for toric manifolds depends on whether they contain compact four-cycles or not. There are two types of toric manifolds without compact four-cycles: an infinite series of so-called strip geometries (also referred to as generalized conifolds) \cite{Iqbal:2004ne,Panfil:2018faz} and one other interesting manifold referred to as the closed topological vertex (also referred to as $T_2$-geometry)  \cite{Bryan:2003yd,Sulkowski:2006jp}. In this work we analyze strip geometries and the closed topological vertex geometry, as well as various examples of toric manifolds with compact four-cycles. The fact that refined open BPS invariants are non-negative integers in all these cases provides a non-trivial consistency check of refined topological strings. 

The second perspective and a parallel aim of this work is to advance the method of computing refined amplitudes based on the geometric transition. Computations of refined amplitudes using the refined topological vertex \cite{Iqbal:2007ii,Awata:2005fa} in some cases are subtle and may lead to inconsistent results, which e.g. violate integrality of associated BPS numbers. In some situations such problems may be fixed by appropriate modification of refined holonomies \cite{Kozcaz:2018ndf}. It is however important to develop other calculational schemes. In this work we develop the method based on the geometric transition \cite{DGH}, which enables to determine refined amplitudes with branes from the closed amplitudes for more complicated Calabi-Yau manifolds, by appropriate specialization of certain K\"{a}hler parameters. In particular, this approach leads to the same results as arise from appropriate modifications of refined holonomies (whenever both methods can be used). In the gauge theory interpretation, fixing K\"{a}hler parameters in the geometric transition method has an interpretation of Higgsing. While this method was proposed before, in this work we implement it in much more complicated systems and verify underlying integrality of resulting amplitudes, which itself is a strong consistency check of refined topological string theory.

The third important aim of this work concerns threefolds without compact four-cycles: we show that for such manifolds open refined topological string amplitudes take form of generating series for symmetric quivers, and we identify corresponding quivers. Such a relation to quivers was originally found in the context of knots-quivers correspondence \cite{Kucharski:2017poe,Kucharski:2017ogk}, its links with topological strings were further elucidated in \cite{Ekholm:2018eee,Ekholm:2019lmb}, and (still in the unrefined case) it was generalized to Aganagic-Vafa branes in strip geometries \cite{Panfil:2018faz,Kimura:2020qns}; related results are also discussed in \cite{Bousseau:2020fus,Bousseau:2020ryp}. Our results can be regarded as generalization of \cite{Panfil:2018faz,Kimura:2020qns} to the refined case. In the present work we show that quivers corresponding to strip geometries in the refined case are the same as in the unrefined case, while dependence on two parameters arises in some universal way in the quiver generating series; this also implies some particular structure of refined BPS invariants. The existence of corresponding quivers implies that open BPS invariants are expressed in terms of motivic Donaldson-Thomas invariants. Moreover, in case of refined amplitudes, it is particularly important that for symmetric quivers motivic Donaldson-Thomas invariants are non-negative integers \cite{Kontsevich:2010px,efimov2012} -- this implies that all refined open BPS invariants associated to a given lagrangian brane are also non-negative integers, as expected. For open unrefined amplitudes and in the knots-quivers correspondence positivity does not play such a role, because unrefined BPS invariants arise as combinations of refined ones with signs, and in general can be negative -- however, for refined amplitudes positivity is crucial. Moreover, using the formulae from \cite{Panfil:2018sis}, we can also write down explicit expressions for all classical refined open BPS numbers for any strip geometry.

One other aspect of our work is reinterpretation of refined amplitudes for toric geometries in terms of webs of five-branes, or 5-dimensional supersymmetric theories with defects \cite{Leung:1997tw,Gaiotto:2014ina}. From this viewpoint one can consider Hanany-Witten transitions, which involve extra D7-branes and processes of creation and annihilation of branes \cite{Hanany:1996ie,Benini:2009gi,Hayashi:2013qwa,Cheng:2018aa}. Topological string partition functions for two systems related (from the five-branes perspective) by a Hanany-Witten transition should be the same or simply related. We verify this statement in the refined setting, for various systems that involve the closed topological vertex geometry ($T_2$-geometry). Recall that $T_2$-geometry is a particular example of $T_N$-geometries that engineer non-lagrangian theories, therefore their analysis is particularly important. Our analysis also involves additional lagrangian branes that engineer defects in 5-dimensional supersymmetric theories; it appears that apart from analysis of one such example in the unrefined setting in \cite{Kim:2020npz}, such processes have not been analyzed before. Agreement of partition functions for such systems with predictions arising from Hanany-Witten transitions provides yet another non-trivial consistency check of refined topological string theory. 

To conduct various calculations in this work we use an improved version of a Mathematica notebook \emph{schurcancellation.nb}, which we also make available for others \cite{schurcancellation}.

The plan of this work is as follows. In section \ref{sec-refined} we introduce various aspects of refined topological string theory: the formalism of refined topological vertex, geometric transitions, refined holonomies, refined BPS states. We also summarize the relation to quivers. In section \ref{sec-strip} we determine refined open amplitudes for toric strip geometries, show that associated open BPS numbers are non-negative integers, and identify corresponding quivers. In section \ref{sec-T2} we analyze the closed topological vertex geometry, as well as Hanany-Witten transitions in various systems involving this threefold and in presence of extra branes. In section \ref{sec-4} we determine refined amplitudes for several threefolds with compact four-cycles and also show that associated open BPS numbers are non-negative integers. In the appendix we collect various identities useful in earlier calculations.


\section{Aspects of refined topological string theory}   \label{sec-refined}

Refined open topological string amplitudes considered in this work are computed using the geometric transition method, which was proposed in \cite{DGH} and which we further develop. To obtain an open partition function for a brane in a toric Calabi-Yau threefold we take the following steps. First, we engineer an appropriate more complicated toric threefold and compute its closed partition function using the formalism of refined topological vertex. If necessary, we simplify the result using \emph{schurcancellation.nb} notebook \cite{schurcancellation}. Next, upon the geometric transition, we fix certain K\"{a}hler parameters to appropriate special values $Q^*$; this operation effectively removes some local $\mathbb{P}^1$'s, while one particular local $\mathbb{P}^1$ is replaced by a brane of our interest. In this way obtain an open partition function for a desired threefold with a brane. Finally, we determine refined open BPS invariants associated to such a partition function.

In this section we review various aspects of refined topological string theory necessary to conduct calculations mentioned above; some of these results were known before, while some are new. We briefly summarize the formalism of refined topological vertex, geometric transitions, refined holonomies, and refined BPS states. We also summarize the relation to quivers, which automatically asserts that all open refined BPS numbers associated to a given brane (in a toric threefold without compact four-cycles) are non-negative integers. The formalism presented in this section is used in various calculations in subsequent sections.


\subsection{Refined topological vertex}    \label{sec:TVT2}

A-model topological string amplitudes for toric Calabi-Yau manifolds can be conveniently computed in the formalism of topological vertex \cite{AKMV}. Refined version of the topological vertex, relevant for computation of refined amplitudes, was introduced in \cite{Iqbal:2007ii,Awata:2008ed}. For appropriate choices of toric manifolds, topological string computations agree with Nekrasov partition functions for corresponding supersymmetric gauge theories \cite{Nekrasov,Nekrasov:2003rj}, which among others provides an important guiding principle for constructing the refined topological vertex. Refined or unrefined topological vertex computations have been recently used for calculation of partition functions of more involved 5d $\mathcal{N}=1$ SCFTs e.g. in \cite{Kim:2015jba,Hayashi:2016abm,Taki:2007dh,Hayashi:2018bkd,Cheng:2018aa,Hayashi:2013qwa,Hayashi:2015xla,Kim:2017jqn,Bao:2013pwa}.  Moreover, in \cite{Hayashi:2013qwa,Hayashi:2015xla,Cheng:2018aa} Higgsing techniques have been introduced, which enable computation of refined topological string amplitudes for some class of non-toric manifolds. 

In this section we briefly summarize the formalism of refined topological vertex, mainly following the notation in \cite{Cheng:2018aa,Bao:2013pwa}. To facilitate various computations in this formalism we also take advantage of the Mathematica notebook \emph{schurcancellation.nb} \cite{schurcancellation}, which we used to obtain various results presented in what follows. Among others, this notebook automatically implements Cauchy identities and cancels various Schur functions in refined vertex calculations.

Recall that the structure of a toric Calabi-Yau manifold can be encoded in a toric diagram, which consists of trivalent vertices connected by edges (also called internal legs). In addition, some edges (external legs) extend from vertices to infinity. In the refined setting we also need to choose a preferred direction, which is denoted by $||$ in a toric diagram, while to legs in other directions we assign parameters $q$ and $t$. The assignment of $q$ and $t$ does not play a role in computation of closed string amplitudes -- two possible choices of such an assignment yield the same refined closed string partition functions. However, it turns out that the assignment of $q$ and $t$ plays an important role in computation of refined open amplitudes. We find that definitions of various types of branes (which we introduce in what follows) and refined open BPS degeneracies that they encode depend on the above assignment. For this reason the assignment of $q$ and $t$ plays a prominent role in this work -- so to start with we carefully explain how it should be made. 

\begin{figure}[]
	\centering
	\includegraphics[width=4in]{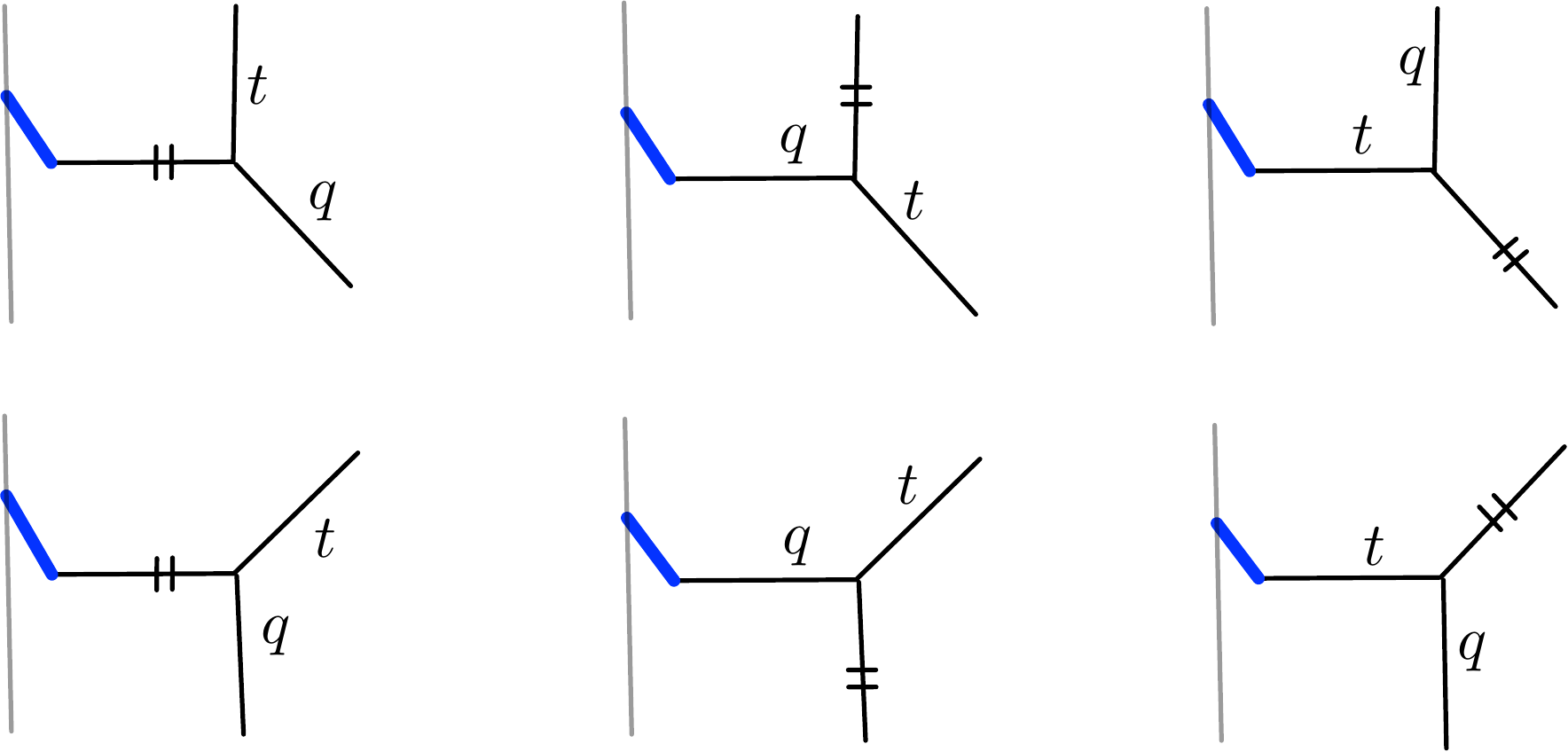}
	\caption{Choices of preferred direction together with a standard assignment of $q$ and $t$ at the vertex to which a topological brane (in blue) is attached. These are the assignments that we consider in this paper, together with appropriate definitions of brane types and refined open BPS states. The assignment of $q$ and $t$ in the rest of a toric diagram (for simplicity not shown in this figure) follows uniquely from the above choices. The geometric transition that produces the brane, discussed in section \ref{ssec-geomtran}, is made through the direction of the vertical line (in grey). We refer to opposite assignments of $q$ and $t$ as alternative assignments; for these opposite choices, definitions of brane types and refined open BPS states must be appropriately adjusted.}	\label{fig:convenassign}
\end{figure}

First, note that it is sufficient to choose a preferred direction and to assign $q$ and $t$ at three legs of one particular vertex in a toric diagram -- such a choice uniquely fixes assignments along all other legs in the diagram. Let us therefore focus on the vertex to which a topological brane of our interest (whose partition function we are going to determine) is attached. For such a vertex, having chosen the preferred direction first, we consider all possible assignments of $q$ and $t$. We now treat this vertex as a part of the resolved conifold, and compute the open partition function for this configuration, as well as the open partition function after the flop transition accompanied by a change of preferred direction. It is natural to expect that open partition functions before and after this transition are the same. It  turns out that this is the case for 6 particular choices of preferred direction and assignments of $q$ and $t$, shown in fig. \ref{fig:convenassign}. In the rest of the paper we consider these choices and regard them as standard assignments. Opposite assignments of $q$ and $t$ can also be made, and we refer to them as alternative assignments; however partition functions for a brane in the conifold are then not invariant under the flop.

\begin{figure}[]
	\centering
	\includegraphics[width=5in]{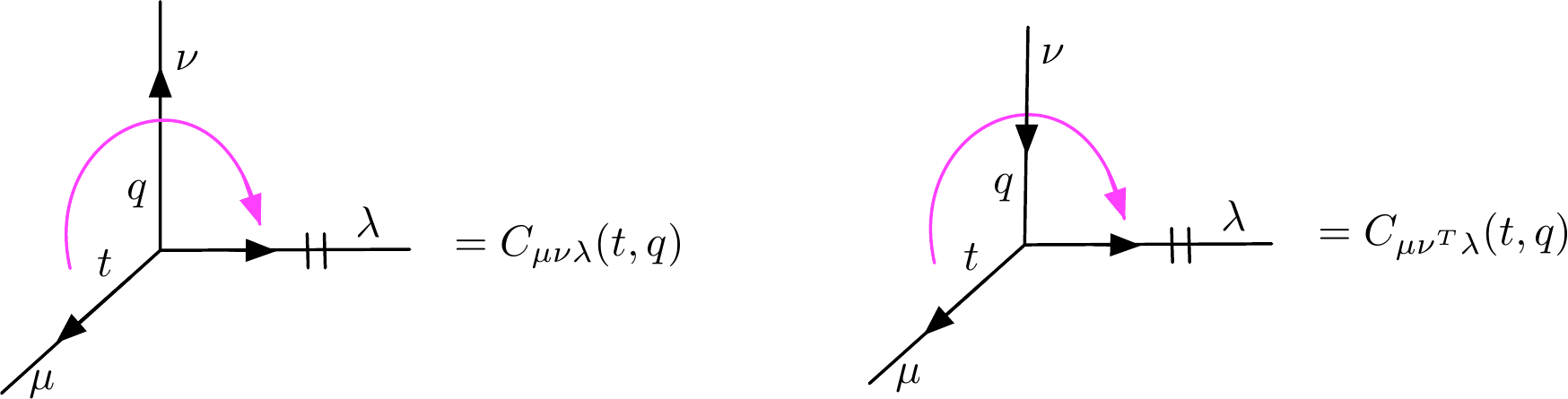}
	\caption{ Assignment of the vertex factor.  
		The direction of arrows on edges can be chosen arbitrarily, and the associated Young diagrams get transposed when the arrow is flopped. Parameters $q=e^{-\epsilon_2}$ and $t=e^{\epsilon_1}$ parametrize the $\Omega$-deformation.} 
	\label{fig:Cuvr}
\end{figure}

\begin{figure}[]
	\centering
	\includegraphics[width=5.5in]{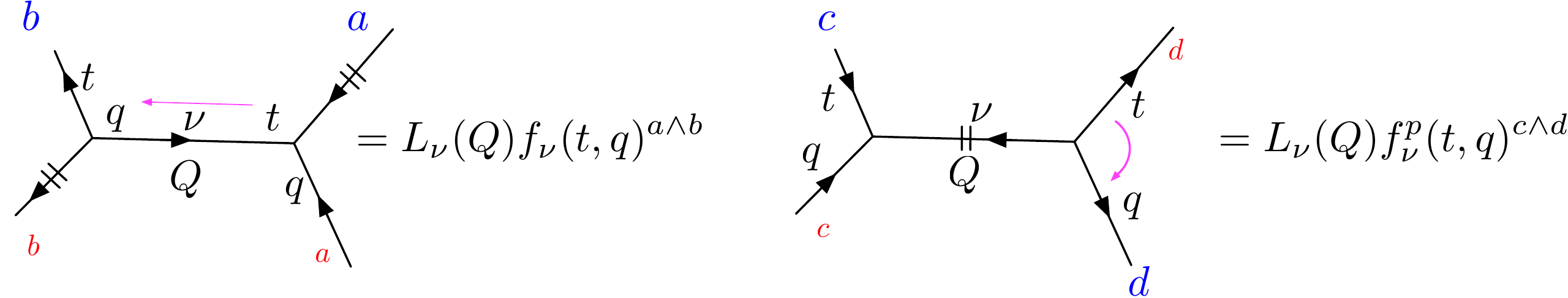}
	\caption{Factors assigned to an edge along a non-preferred direction (left) and preferred direction (right). 
		The directions of external legs are specified by vectors $a,b,c,d$, whose cross products are denoted by a wedge $\wedge$. Consistency conditions impose that the value $a \wedge b$ is the same for both blue and red $a$ and $b$, and similarly $c\wedge d$ is the same for both blue and red $c$ and $d$.}
	\label{fig:framing}
\end{figure}

Having made the above choices, we assign to all edges of a toric diagram their directions (represented by arrows, see e.g. fig. \ref{fig:Cuvr} and \ref{fig:framing}), and to edges around each vertex we assign Young diagrams $( \mu, \nu,\lambda,\ldots)$ (for outgoing edges) or their transpose $( \mu^T, \nu^T,\lambda^T,\ldots)$ (for incoming edges), as well as framing numbers, which encode boundary conditions for open topological strings. Furthermore, to internal edges we assign K\"{a}hler parameters $Q_\bullet$, where $\bullet$ in the subscript stands for an appropriate label of a given leg. Then, to each vertex, with edges labeled by Young diagrams $(\mu,\nu,\lambda)$, we assign a refined topological vertex amplitude defined as 
\begin{align}
C_{\la \u \v } (t,q ) = q^{ \frac{||\u||^2+ ||\v||^2} {2 } } t^{ -\frac {||\u^T||^2} {2}}  \tZ_{\v} (t,q) \sum_{\h}  \left(\frac{q}{t} \right)^{ \frac{|\h| + |\la| -|\u|}{2}  } s_{\la^T/ \h } (t^{-\p} q^{-\v} ) s_{\u /\h}(q^{-\p} t^{-\v^T})\,,
\end{align}
where $s_{\lambda/\eta}$ are skew Schur functions, $t^{\rho}=(t^{-1/2}, t^{-3/2}, t^{-5/2},\ldots)$, and
\begin{align}
\tZ_{\v}(t,q)=\prod\limits_{(i,j)\in \v   } \left(     1-q^{  \v_i-j} t^{\v_j^T -i+1}       \r)^{-1}.
\end{align}
This amplitude is a function of $t=e^{\epsilon_1}$ and $q=e^{-\epsilon_2}$, where $\epsilon_1$ and $\epsilon_2$ parametrize the $\Omega$-background. Similarly, to various edges we assign appropriate factors 
\begin{align}
{f^{\bullet}_{\v}(t, q)}^{\text{framing number}} L_{\v}(Q), \qquad L_\v(Q)= (-Q)^{|\v|},
\end{align}
where $f^{\bullet}(t,q)$ denotes either $f^p(t, q)$ for the edges along the preferred direction, or $f(t, q)$ for other edges of non-preferred directions, such that
\begin{align}
&f^p_\v(t,q)= (-1)^{|\v|}t^{\frac{||\v^T||^2}{2} } q^{-\frac{||\v||^2}{2} },\qquad f_\v(t,q)=\left(\frac{q}{t}\r)^{-\frac{|\v|}{2}  } f^p_\v(t,q).
\end{align}
The assignment of vertex factors and edge factors is illustrated in fig. \ref{fig:Cuvr} and \ref{fig:framing}; in particular, pink arrows in these figures denote the ordering of diagrams $\mu,\nu,\lambda$ at a given vertex and the ordering of arguments $q$ and $t$ in the function $f^{\bullet}(\cdot,\cdot)$. For more details see e.g. \cite{Bao:2013pwa}. 

Finally, the topological string partition function schematically takes form
\begin{align}\label{vertexformu}
Z^{\rm top} = \sum_{\lambda_i}\,\prod\, ({\rm Edge ~factor}) \cdot \prod \,({\rm Vertex~ factor})\,.
\end{align}
After summing over Young diagrams along non-preferred directions and many contractions of Schur functions through Cauchy identities, the above expression generically reduces to
\begin{eqnarray}\label{Z^M}
Z^{\text{top}}(Q_{\bullet}, t,q )&=&Z^{M}\cdot Z^{\text{sum}}\,,
\end{eqnarray}
where $Z^{M}$ is a product 
\begin{align}   \label{Z^M-factor}
Z^{M}=\frac{\prod M(Q_\bullet, t,q)}{\prod M(Q_\bullet, t,q)}
\end{align}
of refined MacMahon functions with arguments being appropriate K\"ahler parameters $Q_{\bullet}$
\begin{align}
M(Q, t,q)=& \prod\limits_{i, j=1}^\inf  (1-Q~q^i t^{j-1})  , 
\end{align}
while $Z^{\rm sum}$ is the sum over Young diagrams along preferred directions, which has the following structure
\begin{align}\label{Z^sum}
& Z^{\text{sum}}=\sum_{\u_\bullet,~\v_\bullet } Q_{\bullet}^{|\u_{\bullet}|}  \prod\limits_{\u_\bullet} ||\tZ_{\u_{\bullet}}(t,q)||^2  \dfrac{  \prod N_{\v_{\bullet}}^{\rm half,-
	}{(Q_{\bullet},t^{-1},q^{-1}) }N_{\u_{\bullet} \v_{\bullet}}(Q_{\bullet},t^{-1},q^{-1})   }{  \prod N_{\u_{\bullet} \v_{\bullet}}(Q_{\bullet},t^{-1},q^{-1})   } \,,
\end{align}
where 
\begin{align}
||\tZ_{\u}(t,q)||^2=& \tZ_{\u^T}(t,q) \tZ_{\u}(q,t)\,,
\end{align}
$N_{\u \v}(Q; t, q)$ is called the Nekrasov factor
\begin{align}
N_{\u \v}(Q; t, q) =&   \prod\limits_{i, j=1}^\inf \frac{1- Q~ q^{\v_i -j}~ t^{\u_j^T-i+1}	}{	1- Q~ q^{-j} ~t^{-i+1}	}\,,  
\end{align}
and half-Nekrasov factors are defined by 
\begin{align}
N_{\v}^{\rm half,-}(Q; t,q)= N_{\v \0}(Q\sqrt{ \frac{q}{t}}, t,q) ,\qquad
	N_{\v}^{\rm half,+}(Q; t,q)= N_{\0 \v}(Q\sqrt{ \frac{q}{t}}, t,q)  . 
\end{align}
Note that $Z^M$ is an overall factor in $Z^{\text{top}}$ and it can be obtained by setting Young diagrams along preferred directions to $\emptyset$, i.e. $Z^M=Z^{\rm top}|_{\mu_i=\emptyset}$. 

There are a few other issues that need to be taken into account in computations in the formalism of refined topological vertex. First, in the computation of open amplitudes using topological vertex, in certain configurations holonomies need to be additionally modified, as already discussed e.g. in \cite{Kozcaz:2018ndf}. Such modifications are automatically taken into account in the approach based on geometric transition, as we discuss in more detail in section \ref{ssec-holonomies}. Furthermore, if there are parallel external lines in a toric diagram, then some closed string contributions breaking closed Gopakumar-Vafa formula should be removed by hand, since these extra strings decouple from the local geometry; more details and examples of such extra closed states can be found in \cite{Kim:2012gu,Hayashi:2013qwa,Kim:2015jba,Cheng:2018aa}. 

\subsection{Refined geometric transitions and topological branes}   \label{ssec-geomtran}

Geometric transition is a duality between A-model open and closed topological strings, which is accompanied by a transformation of an underlying Calabi-Yau manifold \cite{Gopakumar:1998ki}. As proposed in \cite{DGH,Taki:2010bj}, geometric transitions can be used to introduce surface defects engineered by lagrangian branes in topological A-model. In this context, and from the viewpoint of 5-dimensional $\mathcal{N}=1$ gauge theories, they play the role of Higgsing fundamental or bifundamental hypermultiplets, which tunes the theories to the roots of Higgs branches. They can be also interpreted in terms of 3-dimensional $\mathcal{N}=2$ theories, with contributions from underlying 5-dimensional $\mathcal{N}=1$ theories interpreted as background flavors. In this section we summarize these ideas and provide some new details, which we take advantage of in computations in subsequent sections.

One needs to be careful when identifying branes introduced upon the geometric transition. In the $\Omega$-deformed background the rotations of complex coordinates $z_1$ and $z_2$ are parametrized respectively by $q=e^{-\epsilon_2}$ and $t=e^{\epsilon_1}$
\begin{align}
(z_1,z_2) \rightarrow(q z_1, t^{-1} z_2),
\end{align}
and correspondingly there are various types of lagrangian branes. The brane along $z_1$ is called a $q$-brane, the brane along $z_2$ is called a $\bar{t}$-brane (anti-$t$-brane); there exist also their partner $\bar{q}$-brane and $t$-brane. In this section we show how to identify all these branes through analysis of geometric transitions and Nekrasov factors that arise in topological string partition functions, and discuss relations between them.

The crucial feature of the geometric transition that we take advantage of is that closed string partition functions with certain K\"{a}hler parameters tuned to specific values $Q^*$
can be identified as open string partition functions, which represent the presence of a lagrangian brane with open modulus $z$ identified with one K\"{a}hler parameter $Q$ before the transition (dependence on other K\"{a}hler parameters is not shown)
\begin{align}
Z^{\rm closed}(Q, Q^*,t,q)=Z^{\rm open}(z,t,q).   \label{ZclosedZopen}
\end{align}

\begin{figure}[]
	\centering
	\includegraphics[width=5in]{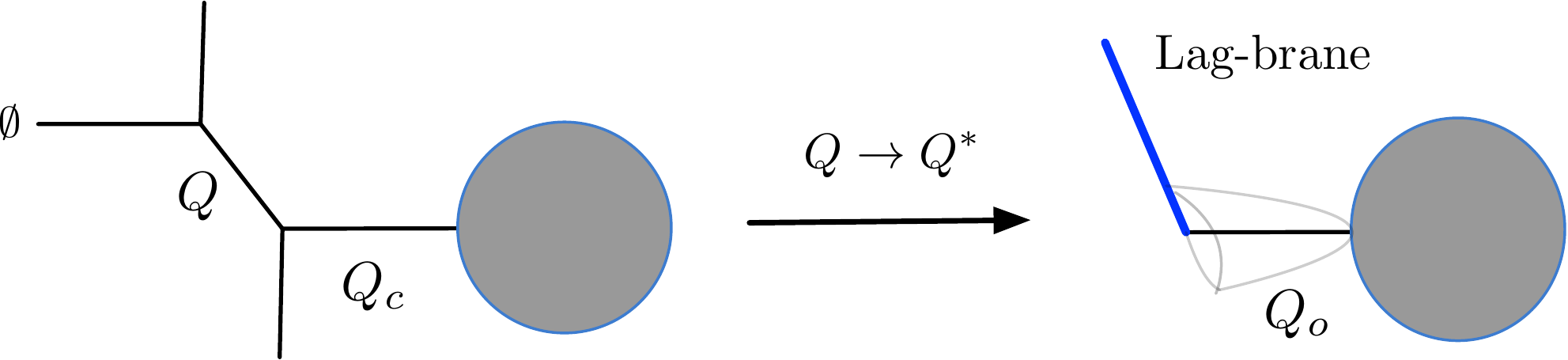}
	\caption{Upon the geometric transition, tuning the K\"ahler parameter $Q$ of a local conifold to a specific value $Q^*$, results in a simpler geometry with a lagrangian brane with open  K\"{a}hler parameter $Q_c=Q_o$. A grey circle represents the remaining part of the toric diagram.}
	\label{fig:geotr}
\end{figure}

For example, in fig. \ref{fig:geotr} one particular closed string K\"{a}hler parameter $Q_c$ is identified as an open K\"{a}hler parameter $Q_c\equiv Q_o$ of a lagrangian brane after the geometric transition (or, equivalently, as the FI parameter of vortex particles from the viewpoint of 3-dimensional $\mathcal{N}=2$ gauge theory on a surface defect \cite{DGH}), while  another K\"{a}hler parameter is tuned to some specific value $Q^*$. 
In general, such specific values $Q^*$ can be determined through constraints involving the half-Nekrasov factors $N_\v^{\half,\pm}(Q^*,t^{-1},q^{-1})$ that appear in the computation of partition functions \cite{DGH,Cheng:2018aa}. 
If no brane is created during the geometric transition (in other words closed strings become massless, and the corresponding 5-dimensional $\mathcal{N}=1$ gauge theory is tuned to the root of the Higgs branch), the values $Q^*=(q/t)^{\pm 1/2}$ are fixed through the constraints 
\begin{equation}\label{constraint1}
N_{\v}^{\rm half, +} \left(\sqqt;~ t^{-1},q^{-1} \r) =
\begin{cases}
1   & \v=\0\\
0   & \v\neq\0
\end{cases}
\qquad	N_{\v^T}^{\rm half, -} \left(\sqtq;~ t^{-1},q^{-1}  \r) =
\begin{cases}
1  & \v=\0\\
0  & \v\neq\0
\end{cases}
\end{equation} 
If a single brane is created during the transition, the constraints on half-Nekrasov factors take form
\begin{align}
\begin{split}
N_\v^{\half, +}\left( q\sqqt;~ \frac{1}{t},\frac{1}{q}  \r) \neq  0 \qquad &\text{only if}~~ \v=\{n\} \label{Higgsing1}\\
N_\v^{\half, +}\left(\frac{1}{ t} \sqqt;~\frac{1}{t},\frac{1}{q}  \r) \neq  0 \qquad &\text{only if}~~ \v=\{1,1,...,1\}   \\
N_\v^{\half, -}\left( t\sqtq;~\frac{1}{t},\frac{1}{q}  \r) \neq  0 \qquad &\text{only if}~~ \v=\{1,1,...,1\}    \\
N_\v^{\half, -}\left(\frac{1}{q}\sqtq;~\frac{1}{t},\frac{1}{q}  \r) \neq  0 \qquad &\text{only if}~~ \v=\{n\} 
\end{split}
\end{align}
where $\v=\{n\}$ and $\v=\{1,1,...,1\}\equiv1^n$ denote respectively antisymmetric  and symmetric representations. These conditions fix four possible values of $Q^*$ (given as the first argument of $N_\v^{\half, \pm}$),
which we identify with four types of topological branes mentioned above
\begin{align}
\begin{split}
\label{defqqtt}
& q\text{-brane}: ~~Q^*=q\sqqt  \,,
\qquad t\text{-brane}: ~~Q^*=t\sqtq\,,\\
& \bar{ q}\text{-brane}: 
~~Q^*=\frac{1}{q}\sqtq\,,
\qquad  \bar{t}\text{-brane}: ~~Q^*=\frac{1}{t}\sqqt\,.
\end{split}
\end{align}
This identification is consistent with and completes the identification of lagrangian branes in refined Chern-Simon theories \cite{Aganagic:2012hs}.  
In fact, for these special values of $Q^*$, the above half-Nekrasov factors can be expressed in terms of $q$-Pochhammers $(x; q)_n=\prod_{i=0}^{n-1} (1-x q^i)$
\begin{align}
\begin{split}
&N_{\v=\{n\}}^{\half, +}\left( q\sqqt;~\frac{1}{t},\frac{1}{q}  \r) = (q;t)_n   \label{higgsing1}\\
&N_{\v=\{1,1,1,...,1\}}^{\half, +}\left(\frac{1}{ t} \sqqt;~\frac{1}{t},\frac{1}{q}  \r) =\left(  \frac{1}{t}; \frac{1}{q } \r)_n        \\
&N_{\v=\{1,1,1,...,1\}}^{\half, -}\left( t\sqtq;~\frac{1}{t},\frac{1}{q}  \r) = (t;q)_n \\
&N_{\v=\{n\}}^{\half, -}\left(\frac{1}{q}\sqtq;~\frac{1}{t},\frac{1}{q}  \r) = \left( \frac{1}{q};\frac{1}{t} \r)_n. 
\end{split}
\end{align}	
For future reference, we also note the following relations for $\v=\{n\}$
\begin{align}
\begin{split}
& |\v|=n,\quad ||\v||^2=n,\quad ||\v^T||^2=n^2, \quad \tZ_\v (t,q)=\frac{1}{(t;t)_n  },\quad \tZ_\v (q,t)=\frac{1}{(q;q)_n  } ,\\
& N_\v^{\half,+}(Q,t^{-1},q^{-1})= \Big( Q\sqtq; t \Big)_n ,\quad  N_\v^{\half,-}(Q,t^{-1},q^{-1})= \Big( Q\sqqt; \frac{1}{t} \Big)_n ,
\label{simply1}
\end{split}
\end{align}
and analogous ones for $\v=\{ 1,1,...,1\}\equiv 1^n$
\begin{align}
\begin{split}
& |\v|=n,\quad ||\v||^2=n^2,\quad ||\v^T||^2=n, \quad \tZ_\v (t,q)=\frac{1}{(t;q)_n  },\quad \tZ_\v (q,t)=\frac{1}{(q;t)_n  } ,\\
& N_\v^{\half,+}(Q,t^{-1},q^{-1})= \Big( Q\sqtq; \frac{1}{q} \Big)_n, \quad  N_\v^{\half,-}(Q,t^{-1},q^{-1})= \Big( Q\sqqt; q \Big)_n .
\label{simply2}
\end{split}
\end{align}
Some other useful identities involving Nekrasov factors are listed in appendix \ref{notation}.

\begin{figure}[h!]
	\centering
	\includegraphics[width=3.5in]{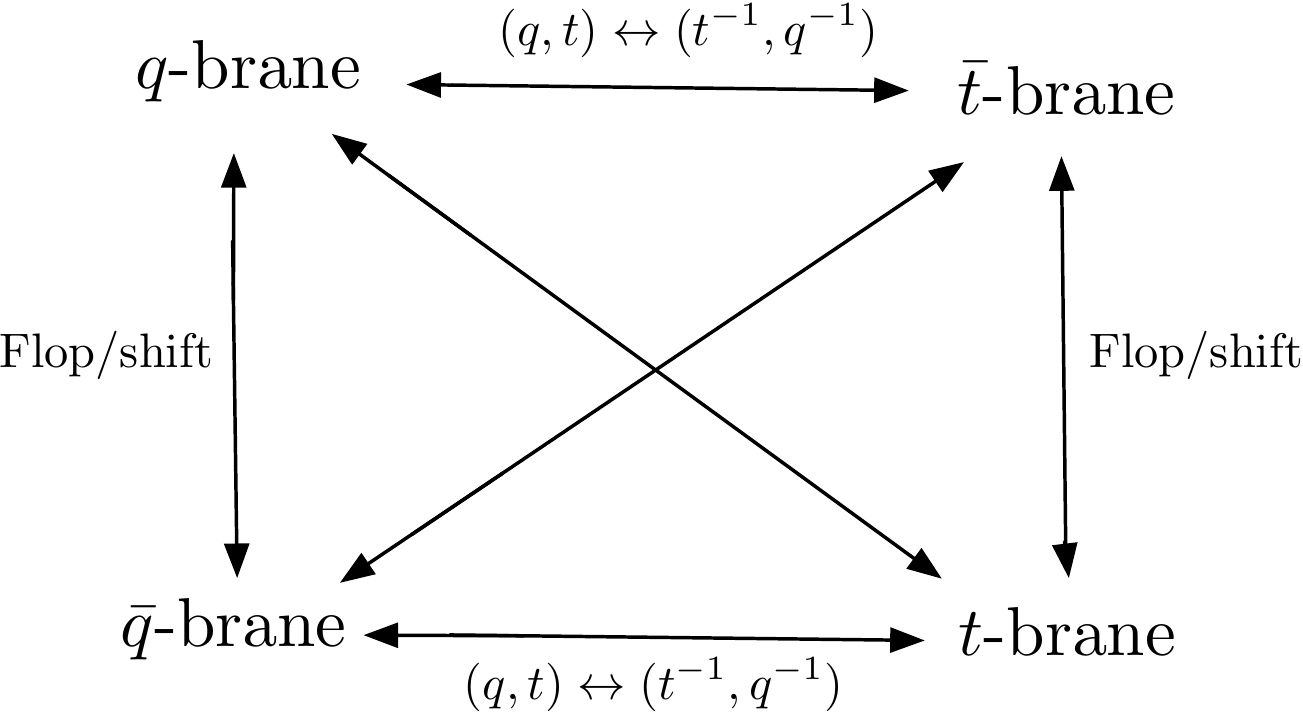} 
	\caption{Relations between various types of topological branes.}
	\label{branes-types}
\end{figure}

Having identified the above branes, we now show that they are related by two types of operations. First, there is an exchange symmetry $(q\to t^{-1}, t\to q^{-1})$ that relates $t$-brane to $\bar{q}$-brane and $q$-brane to $\bar{t}$-brane. Second,
$q$-brane and $t$-brane are related to their anti-branes by flop transitions. These relations are summarized in fig. \ref{branes-types}.

The exchange symmetry is a consequence of the following relation between half-Nekrasov factors
\begin{align}
&N_{\v}^{\rm half,+}  \left(Q; t^{-1},q^{-1}  \r) =N_{\v^{T}}^{\rm half,-}\left(Q; q^{-1},t^{-1} \r),
\end{align}
which implies equality of open string partition functions
\begin{align}
\begin{split}
Z_{q\text{-brane}} (z, Q;t,q) =  Z_{\bar{t}\text{-brane}} \left( z, Q; q^{-1}, t^{-1}  \r),  \label{qbarteql} \\
Z_{t\text{-brane}} (z,Q;t,q) =  Z_{\bar{q}\text{-brane}} \left(z, Q; q^{-1}, t^{-1}  \r),
\end{split}
\end{align}
where $z$ is the open parameter and $Q$ are closed K\"{a}hler parameters. We verified above relations in numerous examples. This exchange symmetry was found independently in \cite{Aganagic:2012hs} by the analysis of the partition function of refined Chern-Simon theory on $S^3$.

On the other hand, a flop transition leads to the following relations between partition functions for branes and anti-branes that involve a shift of the open K\"{a}hler parameter $z$
\begin{align}
\begin{split}
Z_{q\text{-brane}} \Big(z ~\frac{1}{q} \sqtq,Q;   t, q\Big) =  Z_{\bar{q}\text{-brane}} \left(z, Q; t,q \r),\label{qtoqbar} \\
Z_{t\text{-brane}} \Big(z~\frac{1}{t} \sqqt,Q;  t,q \Big) =  Z_{\bar{t}\text{-brane}} \left( z, Q; t,q\r). 
\end{split}
\end{align}

We can also combine the exchange symmetry and the flop transition, which yields the following relation between brane partition functions
\begin{align}
\begin{split}
Z_{q\text{-brane}} \Big(z ~t\sqtq,Q;   \frac{1}{q}, \frac{1}{t}\Big) =  Z_{t\text{-brane}} \left(z, Q; t,q \r), \\
Z_{t\text{-brane}} \Big(z ~q\sqqt, Q;  \frac{1}{q},\frac{1}{t} \Big)=Z_{q\text{-brane}} (z ,Q; t,q).
\end{split}
\end{align}
Note that upon taking the unrefined limit $q=t$ we are left with two types of branes, $q$-brane and $t$-brane, which are related by $q\rightarrow 1/q$, and thus can be identified with a brane and its anti-brane.


\begin{figure}[]
	\centering
	\includegraphics[width=6in]{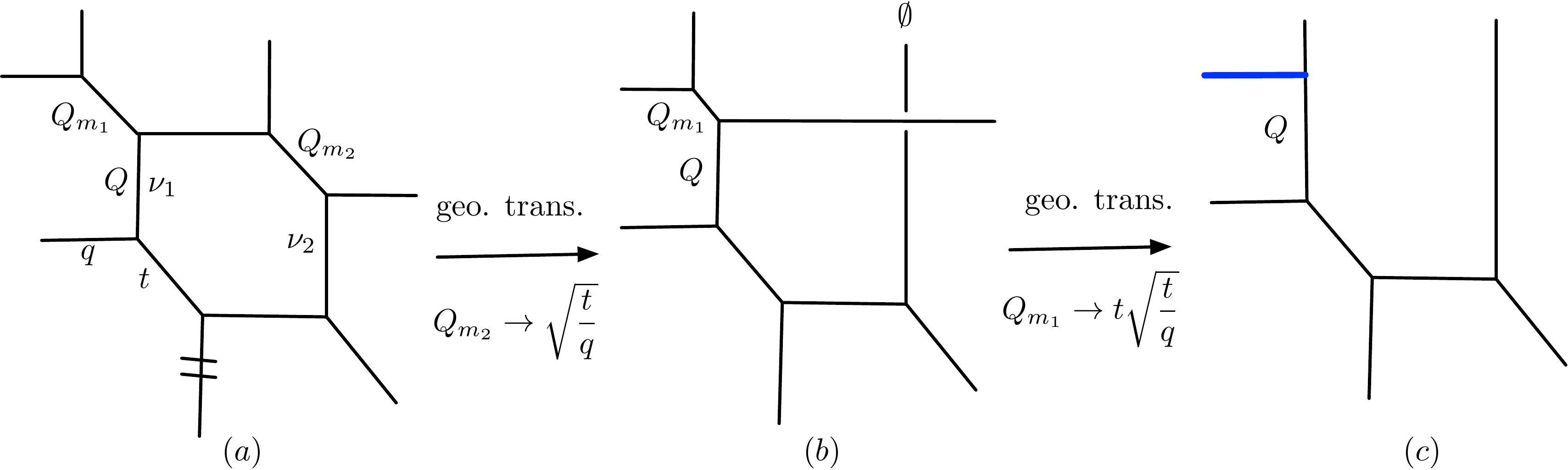}
	\caption{Geometric transitions can be used to engineer insertions of topological branes upon a specific choice of K\"{a}hler parameters $Q^*$ that appear in Nekrasov factors. In this example, a toric Calabi-Yau manifold in ($a$) that engineers $SU(2)$ with three massive fundamental hypermultiples, after two geometric transitions and appropriate choice of K\"{a}hler parameters is turned into a double-$\mathbb{P}^1$ strip geometry with an insertion of a lagrangian brane ($c$).	
}   \label{fig:geotrans}
\end{figure}

Let us illustrate the above considerations in a more involved example, shown in fig.  \ref{fig:geotrans}. 
In this case, there are terms $N_{\v_1}^{\half,-}(Q_{m_1};t^{-1},q^{-1}  )$ and $ N_{\v_2}^{\half,-}(Q_{m_2};t^{-1},q^{-1}  )$  in the closed partition function. If we want to introduce a lagrangian brane at the position of the local resolved conifold with K\"{a}hler parameter $Q_{m_1}$, the constraints \eqref{constraint1} and \eqref{Higgsing1} imply that we have to set $Q_{m_1}= t \sqtq$ or $\frac1q\sqtq$ and $Q_{m_2}=\sqtq$. This also identifies the resulting brane as a $t$-brane or a $\bar{q}$-brane.


Finally, we stress that the above identification of $q$-branes or $t$-branes is appropriate in the assignment of $q$ and $t$ to toric legs presented in fig. \ref{fig:convenassign}. If one changes the assignment $(q,t) \leftrightarrow (t,q)$ then the Nekrasov factors in partition functions are exchanged as follows
\begin{align}
N_{\u \v} \left(Q, t^{-1},q^{-1} \r) &\leftrightarrow  N_{\v^T \u^T} \left(Q, t^{-1},q^{-1} \r),\\
N^{\half,+}_{\v} \left(Q, t^{-1},q^{-1} \r) &\leftrightarrow N^{\half,-}_{\v^T}  \left(Q, t^{-1},q^{-1} \r) ,\\
N_{\v}^{\half,+}( Q^*, t^{-1}, q^{-1})
&\leftrightarrow 
N_{\v^T}^{\half,-}( Q^*, t^{-1}, q^{-1}),
\end{align}
which exchanges the definition of $t$-brane and $q$-brane, as well as $\tbarbrane$ and $\qbarbrane$. 
%

For concreteness and future reference, let us classify all possible choices of preferred directions in a local resolved conifold geometry and resulting identification of topological branes. First, for a horizontal preferred direction we get 
\begin{align}
&\includegraphics[width=6in]{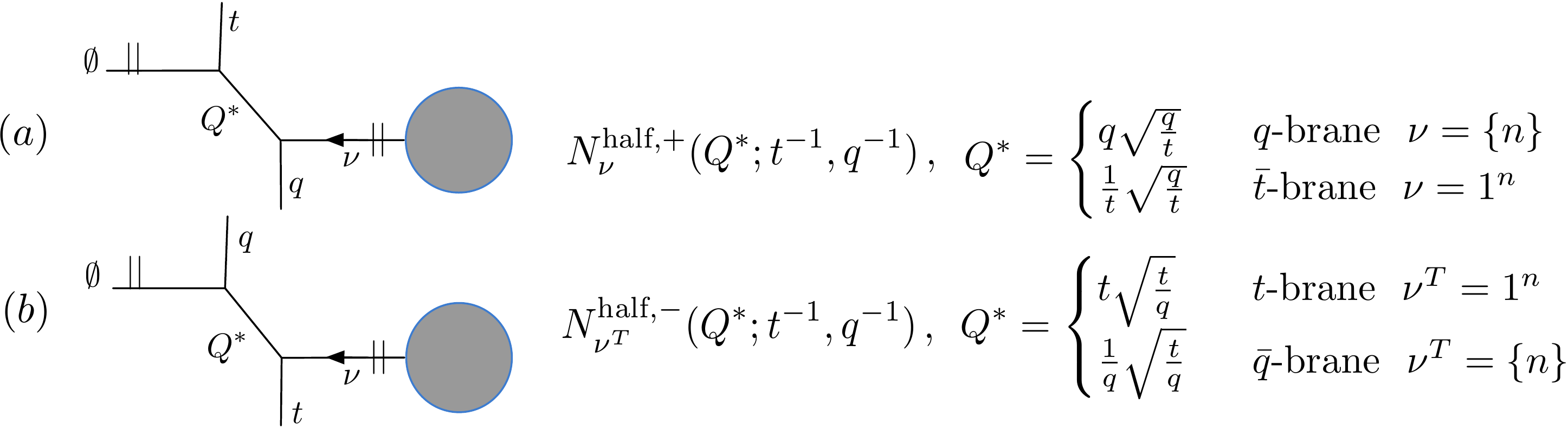}
 \nonumber  \\
&\includegraphics[width=6in]{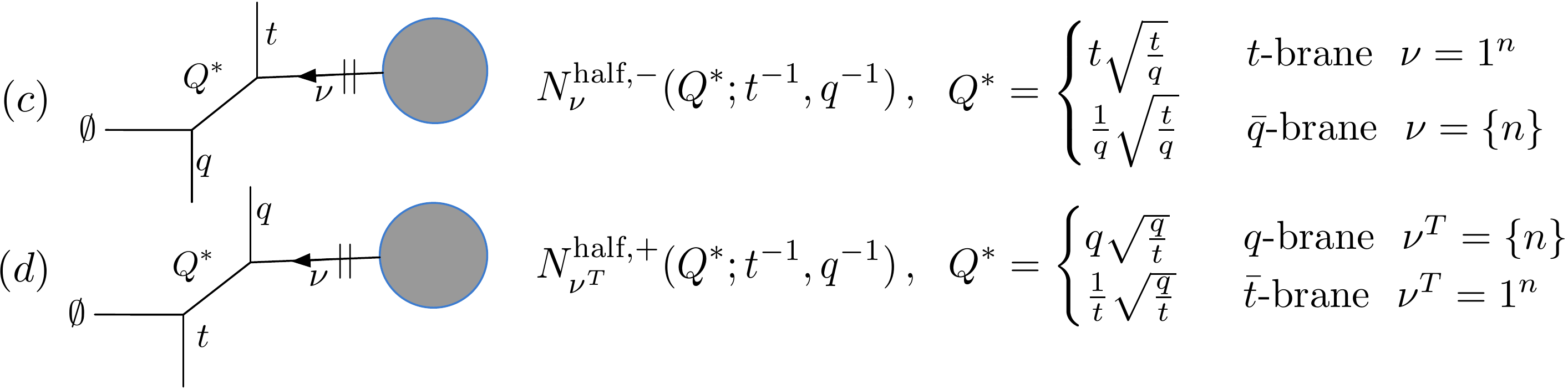}
\label{ttqq-brane}
\end{align}
Note that $(a)$ and $(c)$, as well as $(b)$ and $(d)$, are related by flops. 
For vertical preferred direction we find
\begin{align}
\includegraphics[width=6in]{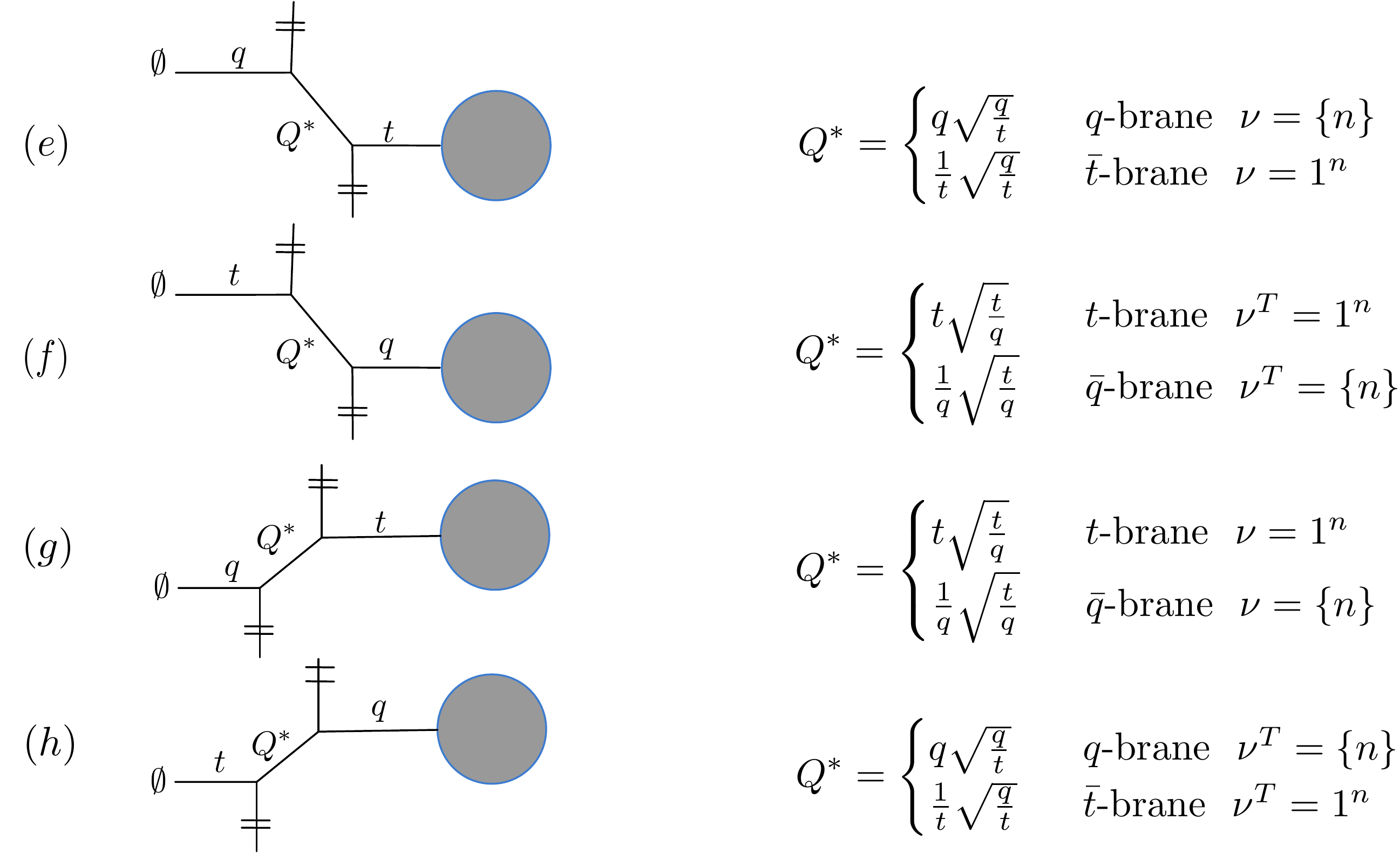}\label{qqtthorizontal}
\end{align}
Finally, for the third choice of preferred direction, we get
\begin{align}
\includegraphics[width=6in]{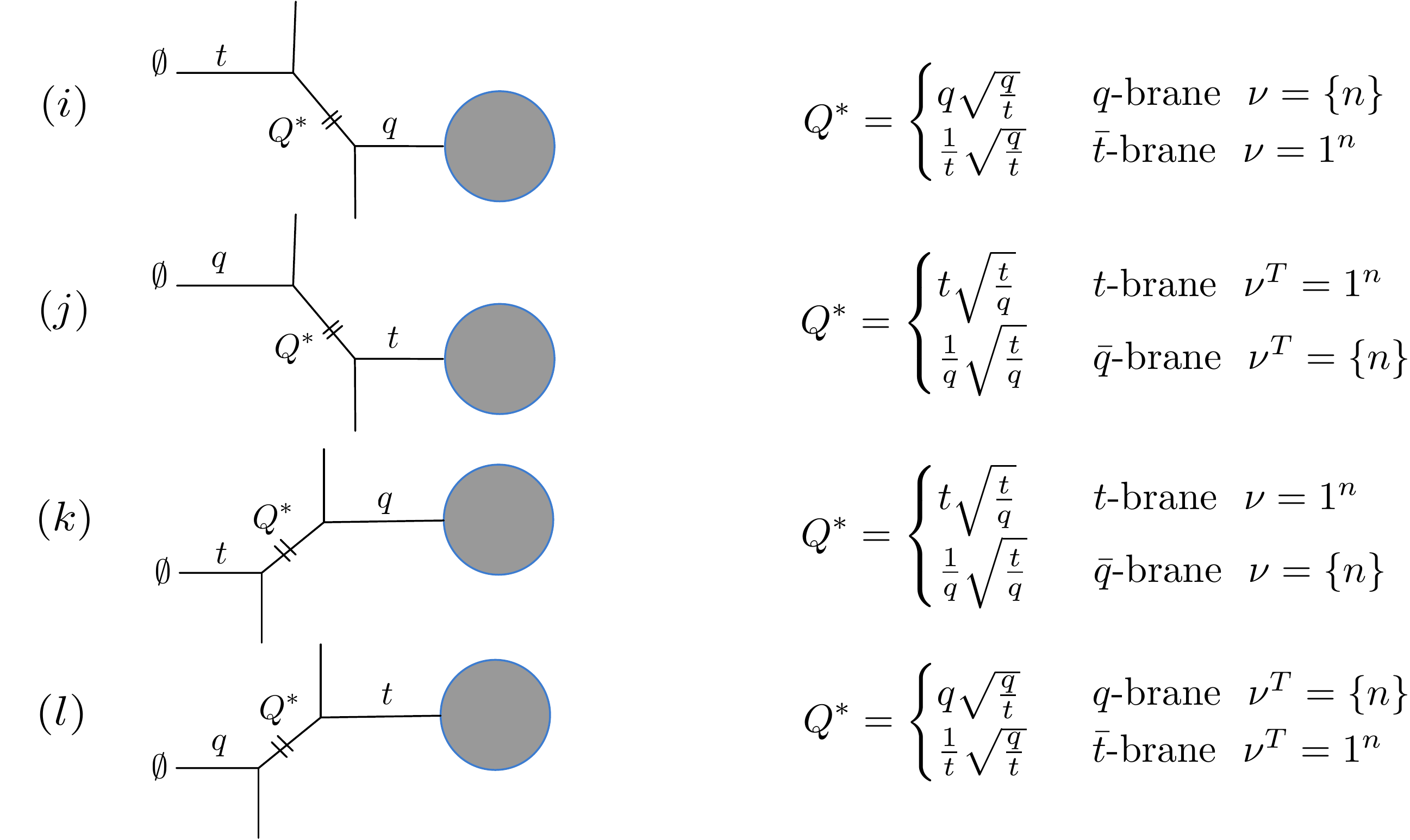}\label{qqttdiagl}
\end{align}
Note that there is the standard assignment of $q$ and $t$ to toric legs (as in fig. \ref{fig:convenassign}) in diagrams $(a)$ and $(c)$ in (\ref{ttqq-brane}), $(e)$ and $(g)$ in (\ref{qqtthorizontal}), and $(i)$ and $(k)$ in (\ref{qqttdiagl}). On the other hand, the assignment is alternative in diagrams $(b)$ and $(d)$ in (\ref{ttqq-brane}), $(f)$ and $(h)$ in (\ref{qqtthorizontal}), and $(j)$ and $(l)$ in (\ref{qqttdiagl}).


\subsection{Refined holonomies}    \label{ssec-holonomies} 

There is an important subtlety of the refined vertex formalism, concerning the form of refined holonomies associated to topological branes. It turns out that in certain situations they need to be additionally modified, as discussed e.g. in \cite{Kozcaz:2018ndf}. We find that in our approach based on geometric transitions such modifications arise automatically, in agreement with \cite{Kozcaz:2018ndf}. As a check that such modifications are correct, we verify that they are consistent with the relations in fig. \ref{branes-types}, and they lead to integer degeneracies of refined open BPS invariants that are of our main interest in this work. 

\begin{figure}[htb]
	\centering
	\includegraphics[width=1.7in]{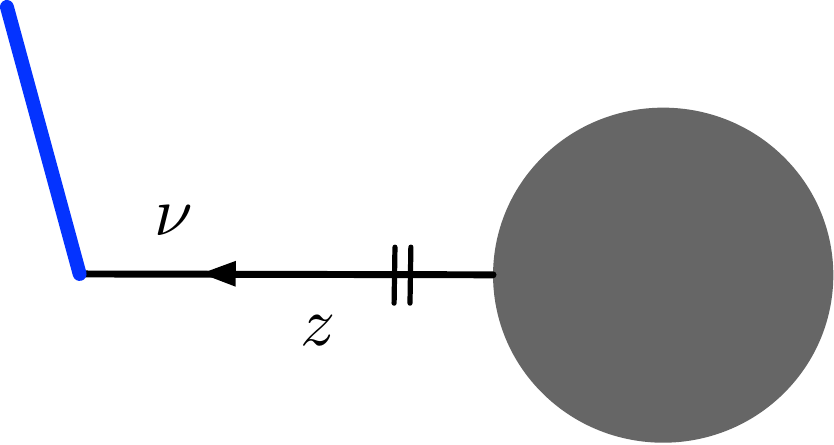}
	\caption{A blue edge represents a lagrangian brane with the open K\"{a}hler parameter $z$, labeled by a Young diagram $\v$. The preferred direction is horizontal. }
	\label{fig:laginsert}
\end{figure}

In unrefined topological vertex formalism, a holonomy $\text{Tr}_{\nu} V$ associated to a single lagrangian brane (see fig. \ref{fig:laginsert}) is identified with the Schur function $s_{\v^T}(z)$ \cite{AKMV} (a transposition of $\v$ in our convention is a consequence of the arrow on the horizontal edge pointing leftwards in fig. \ref{fig:laginsert}). 
To find a modification of this factor in the refined setting, and from the viewpoint of geometric transitions, we follow section \ref{ssec-geomtran} and identify a partition function of a brane in fig. \ref{fig:laginsert} with the conifold partition function with a single edge labeled by $\v$ shown in diagram ($d$) in (\ref{ttqq-brane}) with appropriately adjusted K\"{a}hler parameter $Q^*$ (ignoring grey circles in both these diagrams); this implies the following identification of $\text{Tr}_{\v}V$
\begin{align}\label{3.52case}
\text{Tr}_{\v} V\rightarrow (-z)^{\v} q^{  \frac{||\v^T||^2}{2}   } Z_{\v^T}(t,q) ~N_{\v^T}^{\text{half},+}  \left(Q^*,t^{-1},q^{-1} \right),
\end{align}
in agreement with \cite{DGH,Kozcaz:2018ndf}. 
%
For antisymmetric Young diagram $\v^T=\{1,1,...,1\}\equiv 1^n$, which represents a $\tbarbrane$ corresponding to $Q^*=\frac{1}{t} \sqqt$, the result \eqref{3.52case}
reduces to
\begin{align}
\text{Tr}_{\v} V\rightarrow  s_{\v^T}\Big(  \frac{\sqrt{q}}{t}  z\Big) = \Big(  \frac{\sqrt{q}}{t}  z\Big)^n.
\end{align}
For a symmetric Young diagram  $\v^T=\{n\}$, which represents a $\qbrane$ at $Q^*=q\sqqt$, \eqref{3.52case} reduces to
\begin{align}\label{t-operator}
& \text{Tr}_{\v} V \rightarrow\left(- \sqrt{q} z  \right)^n \frac{(q,t)_n}{(t,t)_n  }  \equiv S_{\v} \Big(\frac{q}{\sqrt{t}} z;q,t  \Big),
\end{align}
where we define $S_\v(z;t,q)=\big(-z \sqqt \big)^n \frac{(t,q)_n}{(q,q)_n  }$, which satisfies
\begin{align}
S_\v\Big(z;\frac{1}{t},\frac{1}{q} \Big)=S_\v(z; t,q).
\end{align}
It is not hard to find analogous refinements of holonomies for other types of branes. 



\subsection{Refined open BPS invariants}  \label{openGVcon}

An important property of topological string amplitudes is their underlying integrality. 
M-theory interpretation implies that the structure of (unrefined) closed and open topological string partition functions is captured by integer closed and open BPS invariants, also referred to as Gopakumar-Vafa and Ooguri-Vafa invariants \cite{Gopakumar:1998ii,OoguriV}. M-theory interpretation implies that analogous integrality properties should hold for refined topological string partition functions. In particular, for Calabi-Yau manifold with closed K\"ahler parameters $Q$, refined partition functions for closed strings are expected to take form
\begin{align}    \label{Z-closed}
Z^{\text{closed}} =\PE\bigg[  \sum\limits_{\b\in H_2(X,\mathbb{Z})}  \sum\limits_{j_L,j_R}      
(-1)^{ 2 j_L +2 j_R }   N_\b^{  ( j_L,j_R)}
f(j_L,j_R) Q^\b
\bigg],
\end{align}
where 
\begin{align}    \label{f-closed}
f(j_L,j_R)=\frac{\chi_{j_L}(tq) ~\chi_{j_R}(t/q)}
{ (t^{1/2}-t^{-1/2})  (q^{1/2}-q^{-1/2})},\qquad \chi_j(x)=x^{-j}+x^{-j+1}+\cdots+x^{j},
\end{align}
and the plethystic exponent is defined by 
\begin{align}
\PE\big[ f( a )\big]=\text{exp}\Bigg[  \sum\limits_{n=1}^{\inf}
\frac{f(a^n )}{ n}   \label{plethistic}
\Bigg],
\end{align}
so that $\PE[f(a)]\cdot\PE[f(b)]=\PE[f(a)+f(b)]$. The above formula encodes refined closed BPS invariants $N_\b^{  ( j_L,j_R)}$ that count particles of charge $\beta$ and $SU(2)_L\times SU(2)_R$ spin $(j_L,j_R)$, therefore they are non-negative integers. Note that the BPS invariants that arise in the unrefined limit take form $N^{j_L}_\b = \sum_{j_R } (-1)^{2j_R} (2j_R+1) N_\b^{  ( j_L,j_R)}$, so they are still integer, however not necessarily positive \cite{Iqbal:2007ii}.

The partition function (\ref{Z-closed}) can be also easily rewritten as a product of refined MacMahon functions with appropriate arguments. Refined MacMahon function can be written as
\begin{align}
\begin{split}
		M(Q, t,q) &= \prod\limits_{i, j=1}^\inf  (1-Q~q^i t^{j-1}) =\text{exp} \left( -\sum\limits_{n=1}^{\inf}  \frac{ Q^n \left( \frac{q}{t} \r)^{  \frac{n}{2}}  }{ n(q^ {  \frac{n}{2}}-q^{-  \frac{n}{2}}  ) (t^{  \frac{n}{2}} -t^{  -\frac{n}{2}}   )  }  \r)  = \\
	& =\PE \Big[-\frac{ Q \sqqt }{ \left( q^{1/2} -q^{-1/2} \r)  \left(  t^{1/2}  -t^{-1/2} \r)   }   \Big],   \label{M-Qtq}
\end{split}
\end{align}
and satisfies the relations
\begin{align}\label{flopidentity}
	M( Q,q,t ) = M(Q^{-1},t,q)=M\left(Q,t^{-1},q^{-1}\r)= M\big(Q \frac{t}{q},t, q\big).
\end{align}

There is an important qualitative difference between closed partition functions for Calabi-Yau threefolds with compact four-cycles and without compact four-cycles (such as strip geometries and closed topological vertex). In the former case, the partition function (\ref{Z-closed}) encodes an infinite number of refined closed BPS invariants $N_\b^{  ( j_L,j_R)}$; equivalently, it can be written as a product of infinite number of refined MacMahon functions. In the latter case, (\ref{Z-closed}) encodes a finite number of closed BPS invariants. Refined closed BPS invariants $N_\b^{  ( j_L,j_R)}$, in both of these cases, were shown to be non-negative integers in numerous examples e.g. in \cite{Iqbal:2007ii}. 

A consistency of M-theory interpretation should lead to analogous integrality properties of open BPS invariants (Ooguri-Vafa invariants) encoded in refined open amplitudes, which however have not been analyzed in the literature beyond the simplest cases. One important goal of this paper is to reveal that these invariants, which in M-theory interpretation also count appropriate particles, are non-negative integers for a wide class of Calabi-Yau manifolds. For basic Aganagic-Vafa branes it is natural to postulate a general form of expected integral expansion of refined open topological string amplitudes, see e.g. \cite{DGH}. 
In our notation, for a $t$-brane, such an expansion takes form
\begin{align}\label{openGV}
\begin{split}
Z_{\tbrane}(z,Q,t,q)&=
\prod\limits_{(d,\b) \in H_2(X,L,\mathbb{Z})}
\prod\limits_{s,r \in \mathbb{Z}/2  }  \prod\limits_{ n=0}^{\inf}
\left(  1-q^{-s+n+\frac{1}{2}  } t^{r+\frac{1}{2}} ~
z^dQ^{\b}             \r)^{ (-1)^{2 s} N_{(d,\b)}^{(s,r)}} = \\
& = \exp\Big(  \sum\limits_{(d,\b)\in H_2(X,L, \mathbb{Z})} \sum\limits_{s,r \in \mathbb{Z}/2}  \sum\limits_{n=1}^{\inf}
\frac{
	(-1)^{2 s}  N_{(d,\b)}^{(s,r)}
	q^{-n s} t^{n \left(r+\frac{1}{2} \r)}
}{n  \left(q^{n/2} -q^{-n/2} \r)}  z^{nd}Q^{n \b} \Big) = \\
& = \prod_{(d,\b)\in H_2(X,L, \mathbb{Z})} \prod_{s,r \in \mathbb{Z}/2}  \PE\Bigg[    \frac{ (-1)^{2s} N_{(d,\b)}^{(s,r)} q^{-s}  t^{r+\frac{1}{2}}    } { q^{
		\frac{1}{2}}-q^{-\frac{1}{2}}  }   z^dQ^{\b}\Bigg] \equiv \\
& \equiv \prod\limits_{(d,\b) \in H_2(X,L,\mathbb{Z})} \prod\limits_{s,r \in \mathbb{Z}/2  }  \PE\Big[z^dQ^\b, N_{(d,\b)}^{(s,r)},s,r \Big]_{\tbrane}
\end{split}
\end{align} 
where $Q$ are closed  K\"ahler parameters of Calabi-Yau manifold $X$ and $z$ is an open parameter associated to a lagrangian brane, while $N_{(d,\b)}^{(s,r)}$ are refined open BPS invariants (refined Ooguri-Vafa invariants). We conjecture that $N_{(d,\b)}^{(s,r)}$ are non-negative integers. In the third line we rewrite this partition function as a product of plethystic exponents (\ref{plethistic}) with the function $f(\cdot)$ in the argument given by
\begin{align}
f(z^dQ^\b,q,t)= \frac{ (-1)^{2s} N_{(d,\b)}^{(s,r)} q^{-s}  t^{r+\frac{1}{2}}    } { q^{
		\frac{1}{2}}-q^{-\frac{1}{2}}  }   z^dQ^{\b} .
\end{align}
In the fourth line we introduce a shorter notation $\PE\Big[z^dQ^\b, N_{(d,\b)}^{(s,r)},s,r \Big]_{*\text{-brane}}$ to denote $\PE[f(z^dQ^\b, q,t)]$ for an appropriate brane type. Note that for a $t$-brane, the contribution from each open BPS state takes form
\begin{align}
\PE\Big[z^dQ^\b, N_{(,\b)}^{(s,r)},s,r \Big]_{\tbrane}=\left( q^{-s +\frac{1}{2}  }  t^{r+\frac{1}{2}} z^dQ^\b,q  \r)_\inf^{ (-1)^{2s}N_{(d,\b)}^{(s,r)}}.
\end{align}

Once we presented an integral expansion for a $t$-brane, we can take advantage of (\ref{qbarteql}) and (\ref{qtoqbar}) to write down analogous expansions for other types of branes. Note that for all these branes refined open BPS invariants $N_{(d,\b)}^{(s,r)}$ are the same.  
For a $\bar{q}$-brane, using the exchange transformation $q\rightarrow 1/t, t\rightarrow 1/q$ and (\ref{qbarteql}), we get
\begin{align}\label{tbarbranePE}
\begin{split}
Z_{\qbarbrane} &=    \prod\limits_{(d,\b) \in H_2(X,L,\mathbb{Z})}
\prod\limits_{s,r \in \mathbb{Z}/2  }  \prod\limits_{ n=0}^{\inf}
\left(  1-t^{s-n-\frac{1}{2}  } q^{-r-\frac{1}{2}} ~
z^dQ^{\b}             \r)^{ (-1)^{2 s} N_{(d,\b)}^{(s,r)}} = 
\\
&=	\prod\limits_{(d,\b) \in H_2(X,L,\mathbb{Z})}
\prod\limits_{s,r \in \mathbb{Z}/2  }  
\PE\Bigg[  \frac{ (-1)^{2s+1} N_{(d,\b)}^{(s,r)} t^{s}  q^{-r-\frac{1}{2}}    } { t^{
		\frac{1}{2}}-t^{-\frac{1}{2}}  }   z^dQ^{\b}\Bigg] \equiv \\ 
&\equiv \prod\limits_{(d,\b) \in H_2(X,L,\mathbb{Z})}
\prod\limits_{s,r \in \mathbb{Z}/2  }  
\PE\Big[z^dQ^\b, N_{(d,\b)}^{(s,r)},s,r \Big]_{\qbarbrane} 
\end{split}
\end{align}
where
\begin{align}
\PE\Big[z^dQ^\b, N_{(d,\b)}^{(s,r)},s,r \Big]_{\qbarbrane} =\left( t^{s -\frac{1}{2}  }  q^{-r-\frac{1}{2}} z^dQ^\b, 1/t  \r)_\inf^{ (-1)^{2s}N_{(d,\b)}^{(s,r)}  }.
\end{align}
Analogously we can write down partition functions for a $\bar{t}$-brane and $q$-brane.

In what follows, we read off open BPS invariants $N_{(d,\b)}^{(s,r)}$ from the above product expansions of open string partition functions; for example, a factor $( z\sqrt{qt},q)_\inf$ encodes the open BPS invariant $N_z^{(0,0)}=1$. Moreover, in some cases it is convenient to redefine the labels $r$ and $s$ as follows
\begin{align}
s\rightarrow r-\frac{j}{2}+1 ,~~ r\rightarrow r-\frac{1}{2}  \,,
\end{align}
which turns \eqref{openGV} into
\begin{align}
\begin{split}
Z_{\tbrane}&=
\prod\limits_{(d,\b) \in H_2(X,L,\mathbb{Z})}
\prod\limits_{ r\in\mathbb{Z}/2, j \in\mathbb{Z}  }  \prod\limits_{ n=0}^{\inf}
\Big(  1-q^{n+\frac{j-1}{2}  } \Big(\frac{t}{q}  \Big)^r ~
z^dQ^{\b}             \Big)^{ (-1)^{-j+2r} \widetilde{N}_{(d,\b)}^{(j,r)}} \label{refDTgen} = \\
&=\prod\limits_{(d,\b) \in H_2(X,L,\mathbb{Z})}
\prod\limits_{ r \in\mathbb{Z}/2, j \in\mathbb{Z}  }   
\Big( q^{\frac{j-1}{2}  }   \Big( \frac{t}{q} \Big)^r z^dQ^{\b};q    \Big)_\inf^{ (-1)^{2r-j } \widetilde{N}_{(d,\b)}^{(j,r)}   },     
\end{split}
\end{align}
with refined open BPS invariants $\widetilde{N}_{(d,\b)}^{(j,r)} $. In this form we can immediately take the unrefined limit $t=q$ 
\begin{align}
\begin{split}
Z_{\tbrane}\big{|}_{t=q}&=
\prod\limits_{(d,\b) \in H_2(X,L,\mathbb{Z})}
\prod\limits_{j\in \mathbb{Z}  }  \prod\limits_{ n=0}^{\inf}
\Big(  1-q^{n+\frac{j-1}{2}  } 
z^dQ^{\b}             \Big)^{ (-1)^{j+1}~  \sum\limits_r(-1)^{2 r+1} \widetilde{N}_{(d,\b)}^{(j,r)}}  = \label{unreftbrane}\\
&=\prod\limits_{(d,\b) \in H_2(X,L,\mathbb{Z})}
\prod\limits_{j \in \mathbb{Z}  }  
\big( q^{\frac{j}{2}  }   z^dQ^{\b};q    \big)_\inf^{ (-1)^{j } N_{(d,\b)}^{j}   },
\end{split}
\end{align}
which captures unrefined open BPS invariants $N_{(d,\b)}^{j}$, related to refined ones by  
\begin{align}
N_{(d,\b)}^j=\sum\limits_{r\in\mathbb{Z}/2}(-1)^{2 r} \widetilde{N}_{(d,\b)}^{(j,r)}.
\end{align}
Analogously to the closed string case, while refined invariants $ \widetilde{N}_{(d,\b)}^{(j,r)}$ are non-negative integers, the unrefined invariants $N_{(d,\b)}^j$ are integer however not necessarily non-negative.

It is also interesting to consider the classical limit 
\begin{align}
y(z,Q,t)= \lim_{q \to 1}  \frac{Z_{\tbrane}( q z,Q,t,q)}{Z_{\tbrane}(z,Q,t,q)} 
= \prod\limits_{ d=0}^{\inf}  \prod\limits_{\b}
\prod\limits_{r \in \mathbb{Z}/2  } 
\left(  1- t^r 
z^d Q^{\b}             \r)^{ (-1)^{2 r} 
d \, n^r_{(d,\b)}}	,     \label{class-OV}
\end{align}
which defines refined open BPS states
\begin{align}
n^r_{(d,\b)} = \sum_{j} (-1)^{j} \widetilde{N}_{(d,\b)}^{(j,r)}.
\end{align}
The function $y=y(z,Q,t)$ is a solution of a refined mirror curve equation $A(y,z;t)=0$, which for $t=1$ reduces to the standard (unrefined) mirror curve. Furthermore, for $t=q\to 1$, we get unrefined classical open BPS invariants  $\sum_{j,r} (-1)^{j+2r} \widetilde{N}^{(j,r)}_{(d,\beta)}$, which are therefore captured by the unrefined mirror curve.

Finally, we stress that all formulas presented above are relevant in the assignment of $q$ and $t$ to toric legs presented in fig. \ref{fig:convenassign}. In the alternative assignment (with opposite assignment of $q$ and $t$ to toric legs) the above formulas are modified. Consider first the partition function (\ref{openGV}). The transformation $q\leftrightarrow t$ changes the Higgsing value of a $t$-brane to $q \sqqt$ and makes it natural to use $s'=-s,  r'=-r-1$, so that (\ref{openGV}) is transformed into
\begin{align}
Z_{q\text{-brane}}(z,Q,t,q) ^{\text{alt.}}=
\prod\limits_{(d,\b) \in H_2(X,L,\mathbb{Z})}
\prod\limits_{s',r' \in \mathbb{Z}/2  }  
\PE\Bigg[- \frac{ (-1)^{2s'+1} N_{(d, \b)}^{(s',r')}  t^{s'}  q^{-r'-\frac{1}{2}}    } { t^{
		\frac{1}{2}}-t^{-\frac{1}{2}}  } z^d  Q^{\b}\Bigg],
\end{align}
which we identify with a partition functon of a $q$-brane in the alternative assignment (we stress that it is not equal to the partition function for a $q\text{-brane}$ in the standard assignment). Similarly, after the transformation $q\leftrightarrow t$, the formula (\ref{tbarbranePE}) is turned into the refined Ooguri-Vafa formula for $\bar{t}\text{-brane}$ in the alternative assignment
\begin{align}
Z_{\bar{t}\text{-brane}}(z,Q,t,q) ^{\text{alt.}}=
\prod\limits_{(d,\b) \in H_2(X,L,\mathbb{Z})}
\prod\limits_{s',r' \in \mathbb{Z}/2  }  
\PE\Bigg[- \frac{ (-1)^{2s'} N_{(d, \b)}^{(s',r')}  q^{-s'}  t^{r'+\frac{1}{2}}    } { q^{
		\frac{1}{2}}-q^{-\frac{1}{2}}  } z^d  Q^{\b}\Bigg]
\end{align}
where $s'=-s,\,  r'=-r-1$. We emphasize that refined BPS invariants are always the same non-negative integers in both standard and alternative assignments, up to a shift of indices $s$ and $r$. In what follows, to avoid confusion, we always choose the versions of Ooguri-Vafa formulas consistent with the assignment presented in fig. \ref{fig:convenassign}.

\subsection{Quivers and refined topological strings}

The last general feature of topological string theory that we discuss is its recently discovered relation to quivers. It turns out that open topological string partition function can be written in the form of quiver motivic generating series. This implies that open BPS states are bound states of a finite number of elementary states, and integrality of open BPS invariants automatically follows once a quiver corresponding to a given setup is found. Originally this relation was found in the context of knots-quivers correspondence, which naturally makes contact with topological strings once the knots are engineered in appropriate brane systems \cite{Kucharski:2017poe,Kucharski:2017ogk}. The relations between quivers and topological string theory was further elucidated in \cite{Ekholm:2018eee,Ekholm:2019lmb}, and analogous relations to quivers were found for Aganagic-Vafa branes in strip geometries \cite{Panfil:2018faz,Kimura:2020qns}; for related results see also \cite{Bousseau:2020fus,Bousseau:2020ryp}. In this work we generalize these results to refined topological strings.

In quiver representation theory, the motivic generating series associated to a symmetric quiver with $C_{ij}=C_{ji}$ arrows from vertex $i$ to vertex $j$ takes form
\begin{align}\label{quivergen}
P_C(q;x_1,x_2,\ldots,x_m) =\sum_{d_1,...,d_m=0}^{\inf}  
(-q^{1/2})^{ \sum_{i,j=1}^m C_{ij}d_i d_j} 
\frac{ x_1^{d_1}\cdots x_m^{d_m}} { (q,q)_{d_1}\cdots (q,q)_{d_m} },
\end{align}
where $x_i$ are generating parameters. It is then shown that this generating series has the following product decomposition
\begin{align}\label{DTgen}
P_C(q;x_1,x_2,\ldots,x_m) =\prod_{d_1,...,d_m=0}\prod_{ j \in \mathbb{Z}} \prod_{n=0}^{\inf} \left( 1- q^{n+\frac{j-1}{2} }  x_1^{d_1}\cdots x_m^{d_m}  \r) ^{(-1)^{j+1} \Omega_{d_1,...,d_m;j}   }  \,,
\end{align}
where $\Omega_{d_1,...,d_m;n} $ are non-negative integer motivic Donaldson-Thomas invariants \cite{Kontsevich:2010px,efimov2012}.  

In analogy to (\ref{class-OV}), we can also consider the classical limit
\begin{align}
\begin{split}
y(x_1,\ldots,x_m) &= \lim_{q\to 1} \frac{P_C(q;qx_1,qx_2,\ldots,qx_m)}{P_C(q;x_1,x_2,\ldots,x_m)} \equiv \sum_{l_1,\ldots,l_m} b_{l_1,\ldots,l_m} x_1^{l_1}\cdots x_m^{l_m} = \\
&= \prod_{(d_1,\ldots,d_m)\neq 0} \Big(1 - x_1^{d_1}\cdots x_m^{d_m}\Big)^{(d_1+\ldots+d_m)\Omega_{d_1,\ldots,d_m}},
\end{split}
\end{align}
which encodes classical Donaldson-Thomas invariants
\begin{align}
\Omega_{d_1,\ldots,d_m} = \sum_j (-1)^j \Omega_{d_1,\ldots,d_m;j}.
\end{align}
Explicit formulae for coefficients $b_{l_1,\ldots,l_m}$ and classical Donaldson-Thomas invariants $\Omega_{d_1,\ldots,d_m}$ have been found in \cite{Panfil:2018sis}.

Note that the product decomposition (\ref{DTgen}) is analogous to the product decomposition of topological string partition function (\ref{openGV}) or (\ref{unreftbrane}). As shown in \cite{Panfil:2018faz} for strip geometries and in the unrefined case (\ref{unreftbrane}), appropriate identification of generating parameters $x_i$ leads to exact identification of these partition functions, which implies that topological string partition function can be written in the form (\ref{quivergen}). In this work we generalize this observation to the refined case: in what follows we show that open string partition functions for strip geometries and the closed topological vertex can be written in the form (\ref{quivergen}), with appropriate identification of $x_i$ with closed and open K\"ahler parameters, and we find corresponding quivers. Among others, this implies that refined open BPS numbers are identified with motivic Donaldson-Thomas invariants. Moreover, as it is proved that the latter invariants are non-negative integers, it follows that all refined open BPS numbers associated to a given lagrangian brane are also non-negative integers, as we expect. For this reason non-negativity of motivic Donaldson-Thoms invariants is crucial in the refined case (contrary to the unrefined case and the original knots-quivers correspondence, where BPS invariants are not necessarily positive).



\section{Strip geometries}   \label{sec-strip}

In section \ref{sec-refined} we summarized formalism of refined topological string theory and properties of quiver generating series. In the rest of the paper we analyze various classes of toric threefolds with branes and their refined open partition functions. In this section we discuss a class of toric manifolds without compact four-cycles, which are called strip geometries or generalized conifolds \cite{Iqbal:2004ne,Panfil:2018faz}. These examples are interesting for several reasons. First, they enable us to illustrate how the method of geometric transition can be used to determine partition functions for branes. Second, we show that refined open BPS invariants for these manifolds are non-negative integers, thereby confirming consistency of refined topological string theory. Third, we represent refined open partition functions for strip geometries in terms of quiver generating series, generalizing to the refined realm the results of \cite{Panfil:2018faz}. We also show that refined open BPS invariants for strip geometries have some specific structure that we illustrate in various explicit examples. More details concerning interpretation of such results in 3d $\mathcal{N}=2$ theories can be found e.g. in \cite{DGH,Cheng:2020zbh}


Recall that toric diagrams for strip geometries are dual to a triangulation (with triangles of unit height) of a rectangle of unit height. Such diagrams consist of a chain of finite segments that represent local $\mathbb{P}^1$'s, as well as vertical and horizontal legs extending to infinity. They can be represented in type II superstring theory upon identification of these segments and legs with a system of NS5-branes ending on a D5-brane. In this work we consider an extra brane, which is also represented by an additional leg extending to infinity, and which can be attached either to a vertical leg representing NS5-brane, see fig. \ref{fig:strip1}, or to a horizontal leg representing D5-brane, see fig. \ref{fig:strip2}. These two cases are qualitatively different. In what follows we discuss first these two cases in general, and then present various explicit results for some particular strip geometries. We stress that fig.  \ref{fig:strip1} and \ref{fig:strip2} are schematic; more properly, the horizontal axis should be replaced by a chain of segments representing local $\mathbb{P}^1$'s, whose slopes encode types of those local $\mathbb{P}^1$'s. 

In principle we could compute open amplitudes in the above setting directly using the formalism of refined topological vertex. However, such computations are quite subtle; for example one needs to take into account refinement of holonomies discussed in section \ref{ssec-holonomies}. For this reason we compute refined open partition functions using the method of geometric transition presented in sections \ref{ssec-geomtran}. These computations, for strip geometries and also for other manifolds analyzed in subsequent sections, illustrate that this method is indeed powerful  and can be effectively used in quite complicated setting. Moreover, we also determine refined open BPS invariants encoded in these partition functions and show that they are non-negative integers, which provides an independent, non-trivial confirmation of correctness and consistency of the geometric transition method.


\subsection{Branes on a vertical leg of a strip geometry and refined quiver invariants}

\begin{figure}[]
	\centering
	\includegraphics[width=3.5in]{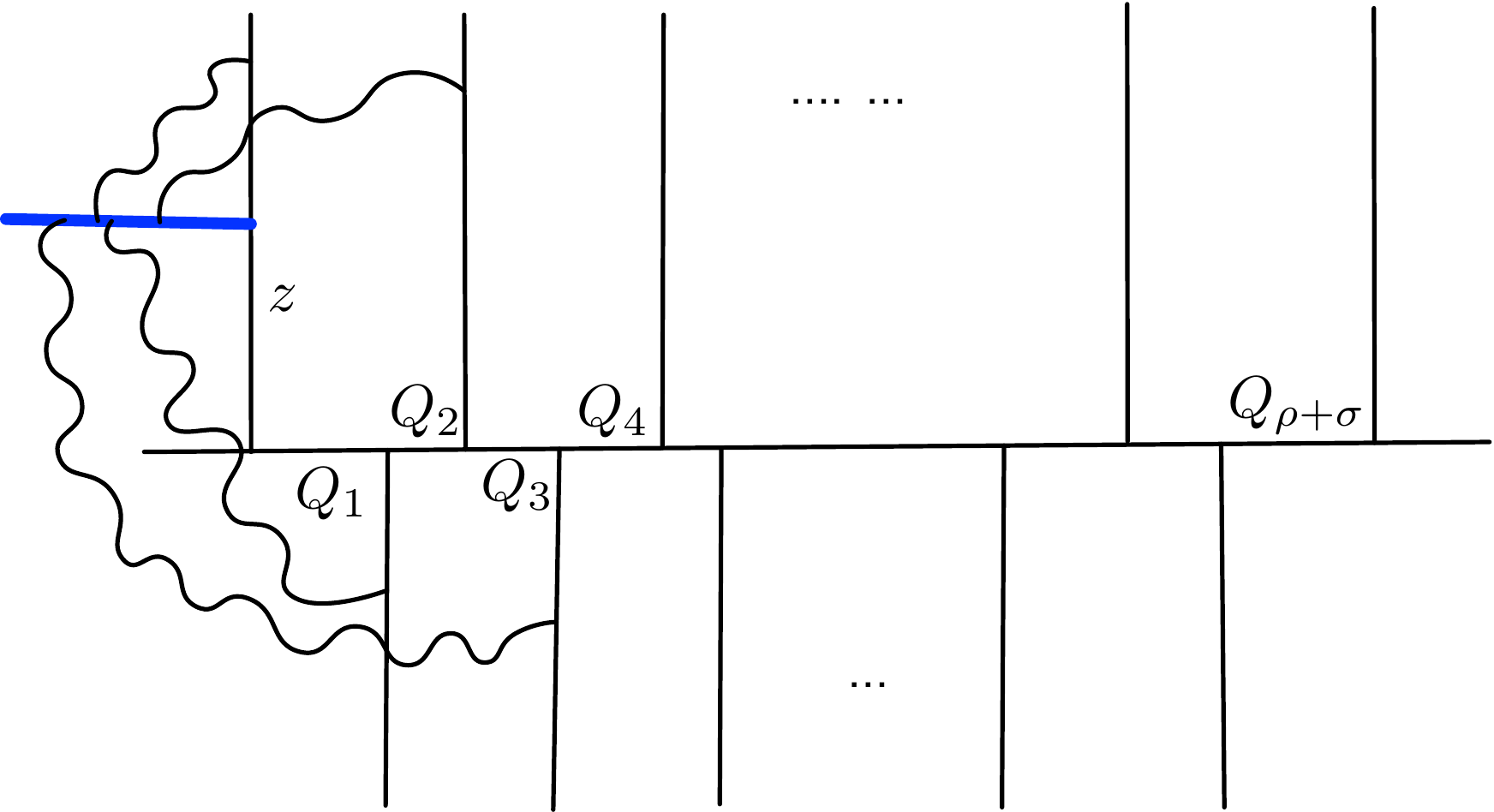}
	\caption{A strip geometry can be represented in type II superstring theory by a system of NS5-branes (vertical legs) attached to a D5-brane (horizontal line). A lagrangian brane (in blue) is attached to a vertical leg. Open strings stretched between this brane and NS5-branes are shown as wavy lines. }
	\label{fig:strip1}
\end{figure}

To start with, we compute refined open partition functions for various types of branes attached to a vertical leg, as shown in fig. \ref{fig:strip1}. We denote the open modulus by $z$ (which is identified with certain closed K\"ahler parameter $Q$ before the transition, as in fig. \ref{fig:geotrans}), and closed K\"ahler parameters of the strip geometry by $Q_{i}$ for $i=1,\ldots,\rho+\sigma$ (labeling segments in a toric diagram from left to right), where $\rho$ is the number of half-infinite legs pointing downwards, and $\sigma$ is the number of half-infinite legs pointing upwards (ignoring the first leg to which the brane is attached). The total number of vertices in such a geometry is $1+\rho+\sigma$. As in \cite{Panfil:2018faz}, we introduce $\alpha_i=Q_{1}Q_{2}\cdots Q_{i}$ where $i=1,\ldots,\rho$ labels one of the legs pointing downwards, and $\beta_j=Q_{1}Q_{2}\cdots Q_{j}$ where $j=1,\ldots,\sigma$ labels one of the legs pointing upwards. 

As mentioned above, to determine brane partition functions we use the method of geometric transition presented in section \ref{ssec-geomtran}. To this end, we first need to engineer an appropriate diagram without a brane, which will produce a brane configuration of our interest after the transition. One example of this process is shown in fig. \ref{fig:geotrans}. For a general strip geometry, we simply need to generalize the setup of that figure. In particular, instead of the middle panel in fig. \ref{fig:geotrans}, for a general strip geometry we consider a horizontal half-infinite leg attached to the left-most vertical leg, which intersects all other vertical legs that point upwards. Each of these intersections arises from a local conifold with  K\"ahler parameter $Q_{m_i}$, whose value is set to $\sqqt $ or $\sqtq$ (depending on flops and assignment of $q$ and $t$) upon the geometric transition. In addition we set $Q_{m_1}$ to one of the values in (\ref{defqqtt}), which results in a strip geometry with an appropriate type of brane. This brane is labeled by a symmetric $\{n\}$ or antisymmetric representation $\{1,1,...,1\}$, depending on the direction of arrows in the diagram; we can always adjust these directions in order to choose e.g. a symmetric labeling $\{n\}$. Note that $Q_{m_j}$ should not be confused with $Q_{i}$ mentioned in the previous paragraph; the latter ones are K\"ahler parameters of the strip geometry of our interest (after the geometric transition), while the former ones characterize local conifolds in the initial geometry that are absent in the strip geometry after the transition.

In particular, engineering appropriate more general diagrams without branes, following (\ref{defqqtt}) and setting $Q_{m_1}= \frac{1}{q} \sqtq$, and adjusting all other $Q_{m_i}$ as mentioned above, we obtain refined partition functions for a $\bar{q}$-brane in the form of a $q$-hypergeometric function. From such computations for various examples discussed in what follows we deduce that for general strip geometry such a partition function takes form
%
\begin{align}\label{qbarbranehypergeometry} 
Z_{\bar{q}\text{-brane}}
(z,\a_i,\b_j)=\sum\limits_{n=0}^{\inf} \frac{ \left(z \sqtq \r)^n}{(t,t)_n} 
\frac{
	\left(\a_1 \sqtq,t \r)_n \left(\a_2\sqtq,t\r)_n \left(\a_3\sqtq,t \r)_n \cdots \left(\a_{\rho}\sqtq,t \r)_n 
}
{ 
	\left(\b_1 \frac{t}{q} ,t \r)_n \left(\b_2 \frac{t}{q} ,t \r)_n \left(\b_3 \frac{t}{q} ,t\r)_n \cdots \left(\b_{\sigma} \frac{t}{q} ,t \r)_n 	
}, 
\end{align}
in agreement with \cite{Kozcaz:2010af}. 
Similarly, the $t$-brane partition function can be obtained by setting $Q_{m_1} = t \sqtq$, 
or by using equivalence relations in fig. \eqref{branes-types} and substituting $t \leftrightarrow q^{-1}$ in the above result. For illustration, let us write two equivalent expressions for $t$-brane partition function
\begin{align}\label{tbranehypergeometry}
\begin{split}
& Z_{t\text{-brane}}(z,\a_i,\b_j)=
\sum\limits_{n=0}^{\inf} \frac{ \left(z \sqtq\r)^n 
}{   \left(  \frac{1}{q},\frac{1}{q}\r)_n } 
\frac{
	\left(\a_1\sqtq,\frac{1}{q}  \r)_n \left(\a_2\sqtq,\frac{1}{q} \r)_n \left(\a_3\sqtq,\frac{1}{q} \r)_n \cdots \left(\a_{\rho}  \sqtq,\frac{1}{q} \r)_n 
}
{ 
	\left(  \b_1  \frac{t}{q}  ,\frac{1}{q} \r)_n \left( \b_2 \frac{t}{q}  , \frac{1}{q}  \r)_n \left(\b_3 \frac{t}{q}  , \frac{1}{q} \r)_n \cdots \left(\b_{\sigma} \frac{t}{q} , \frac{1}{q}  \r)_n} 	= \\
& \qquad =   \sum\limits_{n=0}^{\inf} 
(-\sqrt{q})^{(1+r-s ) \, n^2}
\frac{ \tilde{z} ^n 
}{   \left(  q, q\r)_n } 
\frac{
	\left(\a_1^{-1}\sqqt, q  \r)_n \left(\a_2^{-1}\sqqt, q \r)_n \cdots \left(\a_{\rho}^{-1}  \sqqt,q \r)_n 
}
{ 
	\left(  \b_1^{-1}  \frac{q}{t}  , q \r)_n \left( \b_2^{-1} \frac{q}{t}  ,  q	 \r)_n \cdots \left(\b_{\sigma}^{-1} \frac{q}{t} , q  \r)_n 	},
	\end{split}
\end{align}
where to get the second line we used identities from the appendix and we set
\begin{align}
\tilde{z}= - z ~t^{(1+s-r)/2}\Big(\frac{q}{t}\Big)^{r/2} \frac{\prod_{i=1}^{\rho}  \a_i }{ \prod_{j=1}^{\sigma}  \b_j} \,.
\end{align}
For $q=t$ the above partition functions reduce to unrefined partition functions in \cite{Panfil:2018faz}.


Having found the above open partition functions, we can determine associated open BPS degeneracies. In the case of strip geometries we consider two perspectives: on one hand we analyze the refined Ooguri-Vafa decomposition such as (\ref{refDTgen}), and on the other hand motivic Donaldson-Thomas invariants associated to the quiver representation of partition functions. 

First, we consider refined Ooguri-Vafa form of partition function, and show that for the class of branes we are considering here, the corresponding refined BPS invariants have some particular structure. Let us write a general product decomposition (\ref{refDTgen}) in the notation appropriate to the current situation, with $Q^{\beta}\equiv \a^{\mathbf{l}} \b^{ \mathbf{k}} = \alpha_1^{l_1}\cdots \alpha_{\rho}^{l_{\rho}} \beta_1^{k_1}\cdots \beta_{\sigma}^{k_{\sigma}}$
\begin{align}\label{refinedtbraneOV}
Z_{\tbrane}(z,t,q) &=\prod\limits_{ d=1}^{\inf}  \prod\limits_{\bf{l},\bf{k}=0}^{\inf}
\prod\limits_{r \in \mathbb{Z}/2 , j\in\mathbb{Z} }  \prod\limits_{ n=0}^{\inf}
\Big(  1-q^{n+\frac{j-1}{2}  } \Big(\frac{t}{q}\Big)^{r} ~
z^d \a^{\mathbf{l}} \b^{ \mathbf{k}}             \Big)^{ (-1)^{2r-j} \widetilde{N}_{d, \mathbf{l}, \mathbf{k}}^{(j,r)}} .
\end{align} 
On the other hand, note that the whole dependence on the refined parameter $t$ in (\ref{tbranehypergeometry}) can be encoded by shifting parameters $\a_i$ and $\b_j$ in  the unrefined version. Indeed, by setting
$\tilde{z} = z \sqrt{\frac{t }{q}} \,, \tilde{\a}_i = \a_i \sqrt{\frac{t }{q}} \,,  \tilde{\b}_j = \b_j {\frac{t }{q}}$, the expression (\ref{tbranehypergeometry}) takes form
\begin{align}
Z_{t\text{-brane}}(z,t,q)=
\sum\limits_{n=0}^{\inf} \frac{ \tilde{z} ^n 
}{   \left(  \frac{1}{q},\frac{1}{q}\r)_n } 
\frac{
	\left(\tilde{\a}_1,\frac{1}{q}  \r)_n \left( \tilde{\a}_2,\frac{1}{q} \r)_n \left(\tilde{\a}_3,\frac{1}{q} \r)_n \cdots \left(\tilde{\a}_\rho,\frac{1}{q} \r)_n 
}
{ 
	\left(  \tilde{\b}_1,\frac{1}{q} \r)_n \left( \tilde{\b}_2,   \frac{1}{q}  \r)_n \left( \tilde{\b}_3 , \frac{1}{q} \r)_n \cdots \left(\tilde{\b}_\sigma , \frac{1}{q}  \r)_n 	
}\,,
\end{align}
which is the same as the unrefined open partition function for a strip geometry, which then has the unrefined Ooguri-Vafa decomposition
\begin{align}\label{shiftabsorb}
\begin{split}
Z_{\tbrane}(z,t,q) &=\prod\limits_{ d=1}^{\inf}  \prod\limits_{\bf{l},\bf{k}=0}^{\inf}
\prod\limits_{j\in \mathbb{Z}  }  \prod\limits_{ n=0}^{\inf}
\left(  1-q^{n+\frac{j-1}{2}  }  ~
\tilde{z}^d \tilde{\a}^{\mathbf{l}} \tilde{\b}^{ \mathbf{k}}             \r)^{ (-1)^{j} N_{d, \mathbf{l}, \mathbf{k}}^{j}}  = \\
& = \prod\limits_{ d=1}^{\inf}  \prod\limits_{\bf{l},\bf{k}=0}^{\inf}
\prod\limits_{j\in \mathbb{Z}  }  \prod\limits_{ n=0}^{\inf}
\Big(  1-q^{n+\frac{j-1}{2}  }    \Big(\frac{t}{q}   \Big)^{ (d+\sum\limits_i l_i)/2 + \sum\limits_j k_j}  
z^d \a^{\mathbf{l}} \b^{ \mathbf{k}}             \Big)^{ (-1)^{j} N_{d, \mathbf{l}, \mathbf{k}}^{j}}  
\end{split}
\end{align} 
where $N_{d,\mathbf{l}, \mathbf{k}}^j= \sum\limits_r (-1)^{2 r} N_{d,\mathbf{l}, \mathbf{k}}^{(j,r)}  $, and in the second line we return to the original parameters $\a,\b$ and $z$. Comparing now (\ref{refinedtbraneOV}) and the second line of (\ref{shiftabsorb}), whose dependence on $q$ and $t$ must be the same, we see that for fixed $d$, $\mathbf{l}$, $\mathbf{k}$ and $j$, there is only one value of $r$ 
\begin{align}
r = (n+\sum\limits_i l_i)/2 + \sum\limits_j k_j ,
\end{align}
for which $\widetilde{N}_{d, \mathbf{l}, \mathbf{k}}^{(j,r)}$ is non-zero. Furthermore, we then have
\begin{align}
N_{n,\mathbf{l}, \mathbf{k}}^j=  (-1)^{2 r} \widetilde{N}_{n,\mathbf{l}, \mathbf{k}}^{(j,r)} 
\,.
\end{align}
We will confirm in various examples below that, for fixed d,$\mathbf{l}$, $\mathbf{k}$ and $j$, refined invariants $\widetilde{N}_{n,\mathbf{l}, \mathbf{k}}^{(j,r)}$ are non-zero indeed for only one particular value of $r$.


In turn, we consider the relation to quivers. It arises from rewriting brane partition functions in the form of quiver generating series (\ref{quivergen}). Unrefined amplitudes were rewritten in this form in \cite{Panfil:2018faz}. We can use the same strategy now and rewrite partition functions (\ref{qbarbranehypergeometry}) and (\ref{tbranehypergeometry}) in the quiver form using the relations
\begin{align}
\begin{split}\label{factors}
 (\a,q)_n &= \frac{ (\a ,q)_\inf    }{(\a q^n,q)_\inf   }  = (\a ,q)_\inf \sum_{d=0}^{\infty}  \frac{\a^d q^{d n}}{(q,q)_d}, \\
\frac{1}{ (\b,q)}_n   &= \frac{ (\b q^n,q)_\inf}{ (\b,q)_\inf  }  = \frac{ 1 }{ (\b,q)_\inf  } \sum_{d=0}^{\infty}  \frac{(-1)^d q^{dn+d(d-1)/2} \b^d   }{(q,q)_d}.
\end{split}
\end{align}
Together with one factor of $(t,t)_{n}$ or $(q,q)_{n}$ already present in (\ref{qbarbranehypergeometry}) and (\ref{tbranehypergeometry}), these brane partition functions take form (\ref{quivergen}), with an underlying quiver of size $1+\rho+\sigma$, and with overall $z$-independent prefactor. This prefactor does not encode open BPS states associated to a lagrangian brane, so it is not quite relevant for our analysis. Nonetheless, we could rewrite (\ref{factors}) further by expanding prefactors $(\a ,q)_\inf$ and $(\b,q)_\inf^{-1}$ into the summation form. This would produce a larger quiver, of size $1+2\rho+2\sigma$, such that each of the extra $\rho+\sigma$ vertices would not be connected by arrows to any other vertex.

In particular, after such a rewriting, the $\bar{q}$-brane partition function (\ref{qbarbranehypergeometry}) takes form
\begin{align}
Z_{\bar{q}\text{-brane}}(z,\a_i,\b_j)= Z_{extra}\cdot P_C\left(t; z\sqtq, \a_1 \sqtq\,,  \cdots\,, \a_\rho \sqtq\,,~ \b_1 \frac{\sqrt{t} }{ q}\,,\cdots\,, \b_\sigma \frac{\sqrt{t} }{ q}  \right) ,
\end{align}
where the $z$-independent prefactor reads
\begin{align}
Z_{extra}= \frac{
	\left(\a_1 \sqtq,t \r)_\inf \left(\a_2\sqtq,t\r)_\inf \left(\a_3\sqtq,t \r)_\inf \cdots \left(\a_\rho\sqtq,t \r)_\inf 
} 
	{ 
\left(\b_1 \frac{t}{q} ,t \r)_\inf \left(\b_2 \frac{t}{q} ,t \r)_\inf \left(\b_3 \frac{t}{q} ,t\r)_\inf \cdots \left(\b_\sigma \frac{t}{q} ,t \right)_\inf }	   \label{Zextra1}
\end{align}
and the quiver matrix takes form
$$
C= \left[\begin{array}{c|ccc|ccc}
{0}& 1 & \dots &  1 & 1 & \dots & 1  \\ \hline
1 & 0 & \dots & 0 & 0 & \dots & 0 \\
\vdots & & \ddots & & & \ddots &  \\
1 & 0 & \dots & 0 & 0 & \dots & 0 \\ \hline
1 & 0 & \dots & 0 & 1 & \dots & 0 \\
\vdots & & \ddots & & & \ddots &  \\
1 & 0 & \dots & 0 & 0 & \dots & 1 \\
\end{array}\right]   \,.
$$

Similarly, the partition function (\ref{tbranehypergeometry}) for $t$-brane can be written in the quiver form as 
\begin{align}
Z_{t\text{-brane}}(z,\a_i,\b_j) = Z_{extra} \cdot P_C\left(q; z\sqrt{t},   \a_1 \sqrt{t}\,,  \cdots\,, \a_\rho \sqrt{t}\,,~ \b_1\,t \,,\cdots\,, \b_\sigma\,t\right) \,,
\end{align}
where the extra $z$-independent factor is
\begin{align}
Z_{extra} = \frac{
	\left(\a_1 \sqtq,\frac{1}{q}  \r)_\inf \left(\a_2\sqtq,\frac{1}{q} \r)_\inf \left(\a_3\sqtq,\frac{1}{q}  \r)_\inf \cdots \left(\a_\rho\sqtq,\frac{1}{q}  \r)_\inf 
}{ 
	\left(\b_1 \frac{t}{q} ,\frac{1}{q} \r)_\inf \left(\b_2 \frac{t}{q} ,\frac{1}{q}  \r)_\inf \left(\b_3 \frac{t}{q} ,\frac{1}{q} \r)_\inf \cdots \left(\b_\sigma \frac{t}{q} ,\frac{1}{q}  \right)_\inf }	   \label{Zextra2}
\end{align}
and the quiver matrix takes form
\begin{equation}
C   = \left[\begin{array}{c|ccc|ccc}
1& -1 & \dots &  -1 & -1 & \dots & -1  \\ \hline
-1 & 1 & \dots & 0 & 0 & \dots & 0 \\
\vdots & & \ddots & & & \ddots &  \\
-1 & 0 & \dots & 1 & 0 & \dots & 0 \\ \hline
-1 & 0 & \dots & 0 & 0 & \dots & 0 \\
\vdots & & \ddots & & & \ddots &  \\
-1 & 0 & \dots & 0 & 0 & \dots & 0 \\
\end{array}\right]\,.  
\end{equation}

Having determined the above quivers, we can then identify refined BPS numbers $\widetilde{N}_{n,\mathbf{l}, \mathbf{k}}^{(j,r)}$ in (\ref{refinedtbraneOV}) with motivic Donaldson-Thomas invariants defined via (\ref{DTgen}), analogously to the  unrefined case \cite{Panfil:2018faz}. The latter invariants are non-negative integers \cite{Kontsevich:2010px,efimov2012}, which implies that all refined open BPS numbers must be non-negative integers too, as we intended to show. From the perspective of refined topological strings it is thus crucial that motivic Donaldson-Thomas invariants are non-negative.


\subsection{Branes on a horizontal leg}

\begin{figure}[]
	\centering
	\includegraphics[width=3.5in]{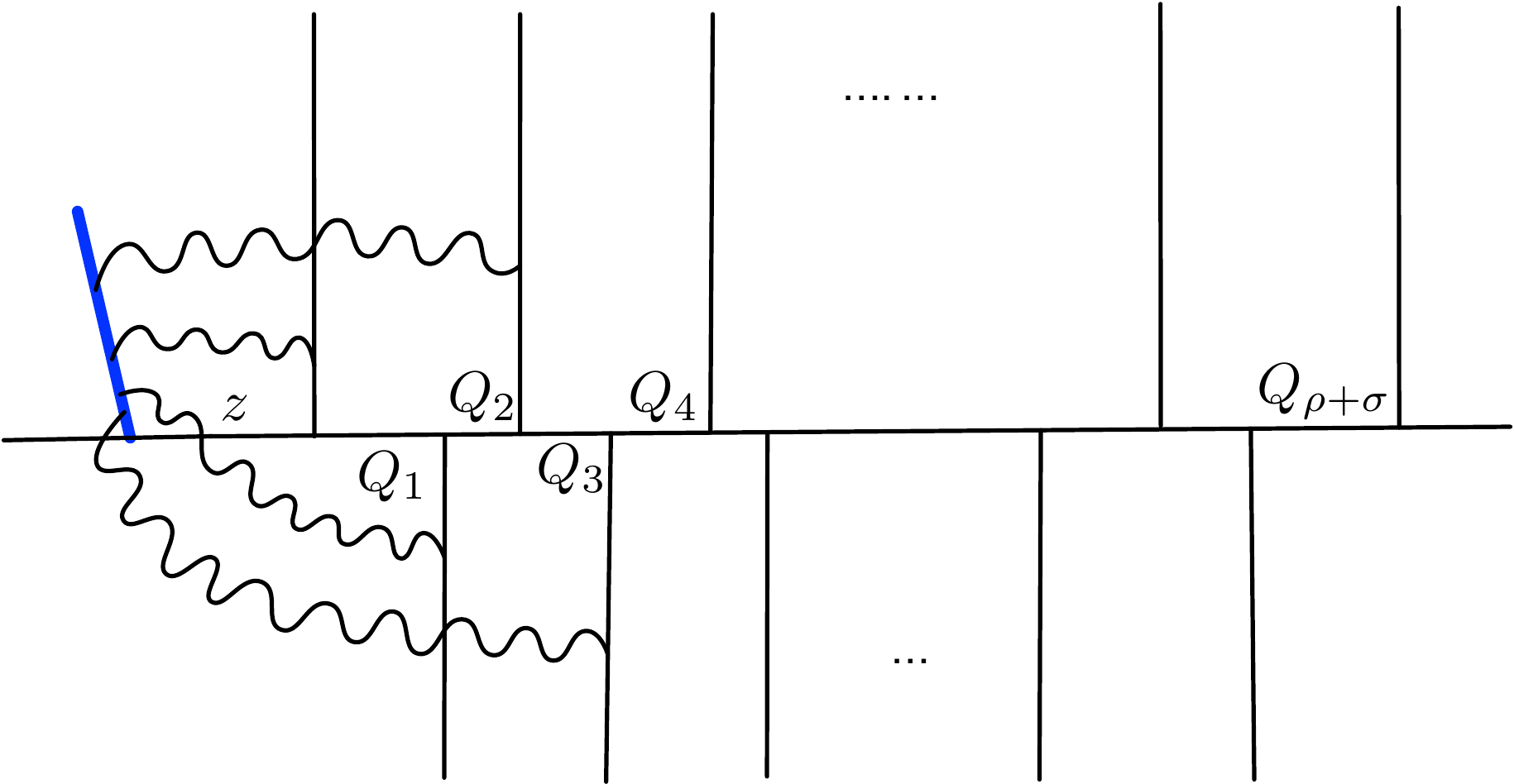}
	\caption{Open strings connect the lagrangian brane and vertical legs. All external legs are associated with empty Young diagrams. }
	\label{fig:strip2}
\end{figure}

Consider now a different system, with a lagrangian brane attached to the horizontal leg, as shown in fig. \ref{fig:strip2}. In this case the partition function takes form of a product of a finite number of $q$-Pochhammer symbols
\begin{align}
\begin{split}
Z_{\tbrane}  (z,\a_i,\b_j)&=
 \frac{ 	\left(z \frac{t}{q} ,\frac{1}{q} \r)_\inf
	\left(z\b_1 \frac{t}{q} ,\frac{1}{q} \r)_\inf \left(z\b_2 \frac{t}{q} ,\frac{1}{q}  \r)_\inf  \cdots \left(z\b_\sigma \frac{t}{q} ,\frac{1}{q}  \right)_\inf }	{
	\left(z \a_1 \sqtq,\frac{1}{q}  \r)_\inf \left(z \a_2\sqtq,\frac{1}{q} \r)_n \cdots \left(z\a_\rho\sqtq,\frac{1}{q}  \r)_\inf 
} = \\
&=\frac{
	(z\a_1\sqrt{qt} ,q)_\inf   (z\a_2  \sqrt{qt} ,q)_\inf\cdots (z\a_\rho \sqrt{qt} ,q)_\inf
}{   
	(z t,q)_\inf (z\b_1 t,q)_\inf   (z\b_2 t,q)_\inf\cdots (z\b_\sigma t,q)_\inf
}\,,\\
Z_{\qbarbrane}  (z,\a_i,\b_j)&=\frac{   
	\left(z \frac{t}{q} ,t\r)_\inf \left(z\b_1  \frac{t}{q},t \r)_\inf   \left(z\b_2 \frac{t}{q},t \r)_\inf\cdots \left(z\b_\sigma \frac{t}{q},t \r)_\inf   }
{  \left(z\a_1 \sqtq,t\r)_\inf   \left(z\a_2 \sqtq, t\r)_\inf\cdots \left(z\a_\rho  \sqtq,t\r)_\inf
}.
\label{10brane}
\end{split}
\end{align}
These results are already in the refined Ooguri-Vafa form. Each quantum dilogarithm corresponds to one open BPS invariant and in total there are finitely many open BPS invariants $N_\b^{(s,r)}$
\begin{align}
N_{z}^{(1/2,1/2)}=1,\quad N_{z\a_i }^{ (0,0)}=1,\quad N_{z\b_j}^{(1/2,1/2)}=1,
\end{align}
which are clearly non-negative integers. Expanding all factors in (\ref{10brane}) using (\ref{factors}) we can also determine the associated quiver, which consists of a finite number of disconnected nodes. Furthermore, open partition functions (\ref{10brane}), from the viewpoint of 3d $\mathcal{N}=2$ theories, represent just a bunch of free chiral multiplets.




\subsection{$\mathbb{C}^3$ geometry}


\begin{figure}[]
	\centering
	\includegraphics[width=3.5in]{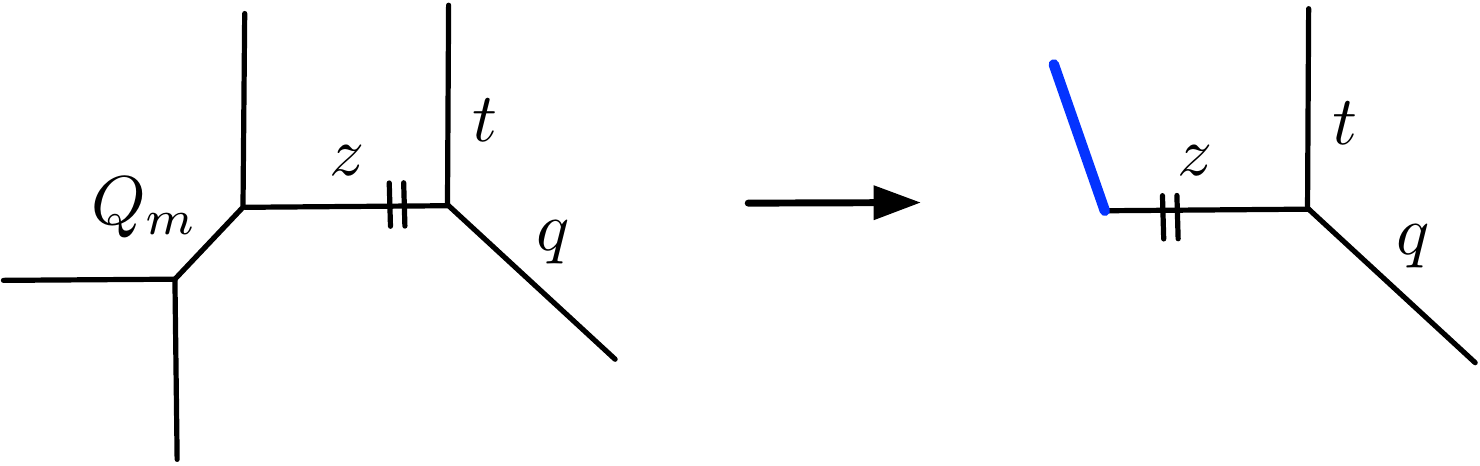}
	\caption{A geometric transition that transforms a double-$\mathbb{P}^1$ geometry with K\"ahler parameters $Q_m$ and $z$ into $\mathbb{C}^3$ with a brane. After the transition, the modulus $z$ plays role of an open parameter. The assignment of $q$ and $t$ and the preferred direction are shown in the diagram.}   \label{fig:C3higgsing}
\end{figure}

In what follows we illustrate the above analysis in several specific examples, starting from $\mathbb{C}^3$ geometry. Let us illustrate the method of geometric transition and compute the refined partition function for a brane in the configuration shown in  fig. \ref{fig:C3higgsing} (right). First, we engineer a double-$\mathbb{P}^1$ geometry with K\"ahler parameters $Q_m$ and $z$, shown in fig. \ref{fig:C3higgsing} (left). Its closed partition function, before the transition, can be computed using topological vertex rules and takes form
\begin{align}
Z^{\text{closed}} (z,Q_m)= Z^{\text{closed}}=\frac{ M\left( z Q_m  \sqrt{\frac{t}{q}}, t,q \right)   }{ M\left( z \frac{t}{q},t,q \right)  },
\end{align}
where refined MacMahon function $M(z,q,t)$ is defined in (\ref{M-Qtq}). By comparing the diagram in fig. \ref{fig:C3higgsing} (left) with  \eqref{ttqq-brane} we find that the types of branes that can be introduced in this case are $\tbrane$ and $\qbarbrane$. Imposing $Q_m = t\sqtq$ and reinterpreting $z$ as an open parameter we get
\begin{align}
Z_{t\text{-brane}} = Z^{\text{closed}} \left( z, Q_m = t \sqrt{\frac{t}{q} }  \right) =\frac{1}{  (zt,q )_\inf  }.
\end{align}
Similarly, setting $Q_m=\frac{1}{q}\sqtq$, we obtain
 \begin{align}
Z_{\bar{q}\text{-brane}} = Z^{\text{closed}} \left( z, Q_m = \frac{1}{q} \sqrt{\frac{t}{q} }  \right) =   \left(z \frac{t}{q},t \right)_\inf . 
 \end{align}
Using the identities for  $q$-Pochhammer symbols from the appendix we confirm that the exchange symmetry holds
\begin{align}
	 Z_{\tbrane}  \xleftrightarrow{t \leftrightarrow q^{-1}} 	 Z_{\qbarbrane}  \,.
\end{align}
We can also write the above open partition functions in the Ooguri-Vafa form, as well as in the quiver form (\ref{quivergen})
\begin{align}
\begin{split}
&Z_{t\text{-brane}} = \frac{1}{  (zt,q )_\inf  } = \text{PE}[ z,1, 1/2,1/2]_{  t\text{-brane} }  = P_{C_t} (q; z t ) \,, \quad C_t = [0]  \,, \\
&Z_{\bar{q}\text{-brane}} =  \left(z \frac{t}{q},t \right)_\inf   = \text{PE}[ z,1, 1/2,1/2]_{  \bar{q}\text{-brane} }  = P_{C_{\bar{q} }} \left(q; z \frac{\sqrt{t}}{q} \right) \,, \quad C_{\bar{q} }= [1]  \,, 
\end{split}
\end{align}
with quiver matrices $C$ of size 1. It follows from (\ref{openGV}) and (\ref{tbarbranePE}) that in these cases there is one open BPS number $N_{z}^{(1/2,1/2)}=1$. Furthermore, the open partition functions for $\tbarbrane$ and $\qbrane$ can be obtained by using the relations (\ref{qtoqbar}), which yields 
\begin{align}
Z_{\bar{t}\text{-brane}} = ( z \sqrt{tq},t )_\inf  \,,\qquad Z_{q\text{-brane}} = \frac{1} {\left( z \sqrt{\frac{q}{t}  } ,q \right)_\inf   },
\end{align}
with the same open BPS invariant.




\subsection{Resolved conifold}

\begin{figure}[]
	\centering
	\includegraphics[width=4.5in]{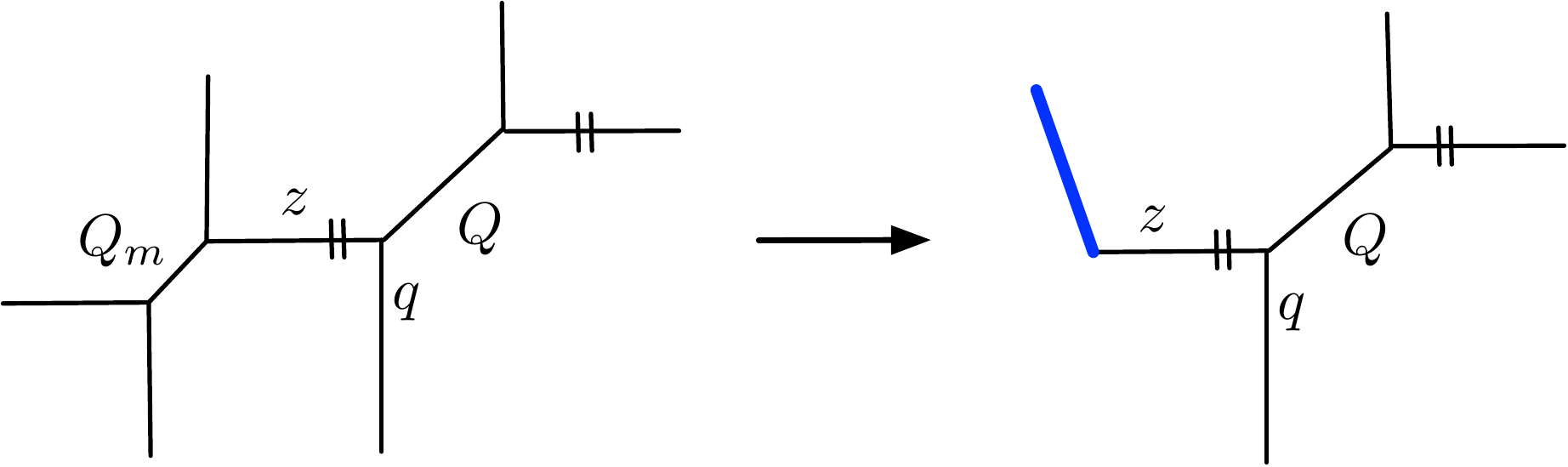}
	\caption{Geometric transition that transforms a triple-$\mathbb{P}^1$ geometry into resolved conifold with a lagrangian brane. In this process the value of $Q_m$ is fixed, while $z$ plays role of the open K\"ahler parameter after the transition.} 
	\label{fig:conifoldtwocases}
\end{figure}

The next example we consider is resolved conifold, which also enables to illustrate the method of geometric transition in a non-trivial setup. As the first step in this method we engineer a triple-$\mathbb{P}^1$ manifold without branes, with closed K\"ahler parameters $Q_m$, $Q$ and $z$, see fig. \ref{fig:conifoldtwocases} (left). The value of $Q_m$ is fixed during the transition, $Q$ becomes the K\"ahler parameter of the resulting conifold, while $z$ is interpreted as the open parameter after the transition. Following the strategy presented in section \ref{ssec-geomtran}, we consider the manifold represented by the diagram in fig. \ref{fig:conifoldtwocases} (left), whose closed partition function takes form of a product of a finite number of refined MacMahon functions (raised to integer powers that encode Gopakumar-Vafa invariants)
\begin{align}
Z^{ \text{closed}} \sim  \frac{   M\left(  z \sqrt{\frac{t}{q}} ,t,q \right)M\left(  z Q Q_m\sqrt{\frac{t}{q}} ,t,q \right)   }{ M\left(  z Q_m ,t,q \right)  M\left(  z Q {\frac{t}{q}} ,t,q \right)   },
\end{align}
where $\sim$ means that we have ignored the closed string contributions that do not depend on K\"ahler parameter $z$ (and thus do not contribute to the resulting open amplitudes). After the transition, in the case of $t$-brane we set $Q_m=t\sqtq$, which yields the open refined partition function
\begin{align}  
\begin{split}
Z_{t\text{-brane}}&=Z^{ \text{closed}} \left(z, Q_m = t \sqrt{ \frac{t}{q}}\right) = \frac{ \left(z \sqrt{tq} ,q  \right)_\inf  }{ \left(  z Q t,q \right)_\inf   } = \sum\limits_{n=0}^{\inf}
\frac{  \left( z \sqrt{ \frac{t}{q}} \right)^n   \left( Q \sqrt{ \frac{t}{q} } , \frac{1}{q} \right)_n }{   \left( \frac{1}{q}, \frac{1}{q} \right)_n   } = \\
&= \PE[ z,1,0,0  ]_{t} \PE[z Q,1/2,1/2 ]_t =P _{C_{t}}\left( q; z \sqrt{t}, z Q t \right) .
\end{split}
\end{align}
Similarly, for $\bar{q}$-brane we substitute $Q_m=\frac{1}{q}\sqtq$, which yields the following open refined partition function 
\begin{align}
\begin{split}
Z_{\qbarbrane}&=Z^{ \text{closed}} \left(z, Q_m = \frac{1}{q}\sqrt{ \frac{t}{q}}\right) = \frac{ \left(z Q \frac{t}{q} ,t  \right)_\inf  }{ \left(  z \sqrt{\frac{t}{q}  },t \right)_\inf   } = \sum\limits_{n=0}^{\inf}
\frac{  \left( z \sqrt{ \frac{t}{q}} \right)^n   \left( Q \sqrt{ \frac{t}{q} } , t \right)_n }{   \left( t, t \right)_n   } = \\
&= \PE[ z,1,0,0  ]_{\bar{q}} \PE[z Q,1/2,1/2 ]_{\bar{q}} = P _{C_{\bar{q}}}\left( t; z \frac{\sqrt{t}}{q}, z Q \sqrt{\frac{t}{q}  } \right).
\end{split}
\end{align}
Above we have also provided quiver forms (\ref{quivergen}), this time with quiver matrices of size 2 of the same form
\begin{align}
C_{\bar{q}}= \left[ 
\begin{array}{cc}
1 & 0 \\
0 & 0\\
\end{array}
\right] \qquad\qquad  C_t= \left[ 
	\begin{array}{cc}
1 & 0 \\
0 & 0\\
\end{array}
\right] 
\end{align}
Moreover, for these branes, by comparing with (\ref{openGV}) and (\ref{tbarbranePE}), we can easily read off refined Ooguri-Vafa invariants $N_{z}^{(0,0)}=1,~N_{z Q}^{(1/2,1/2)}=1$.


\subsection{Resolution of $\mathbb{C}^3/\mathbb{Z}_2$}

\begin{figure}
	\centering
	\includegraphics[width=3.5in]{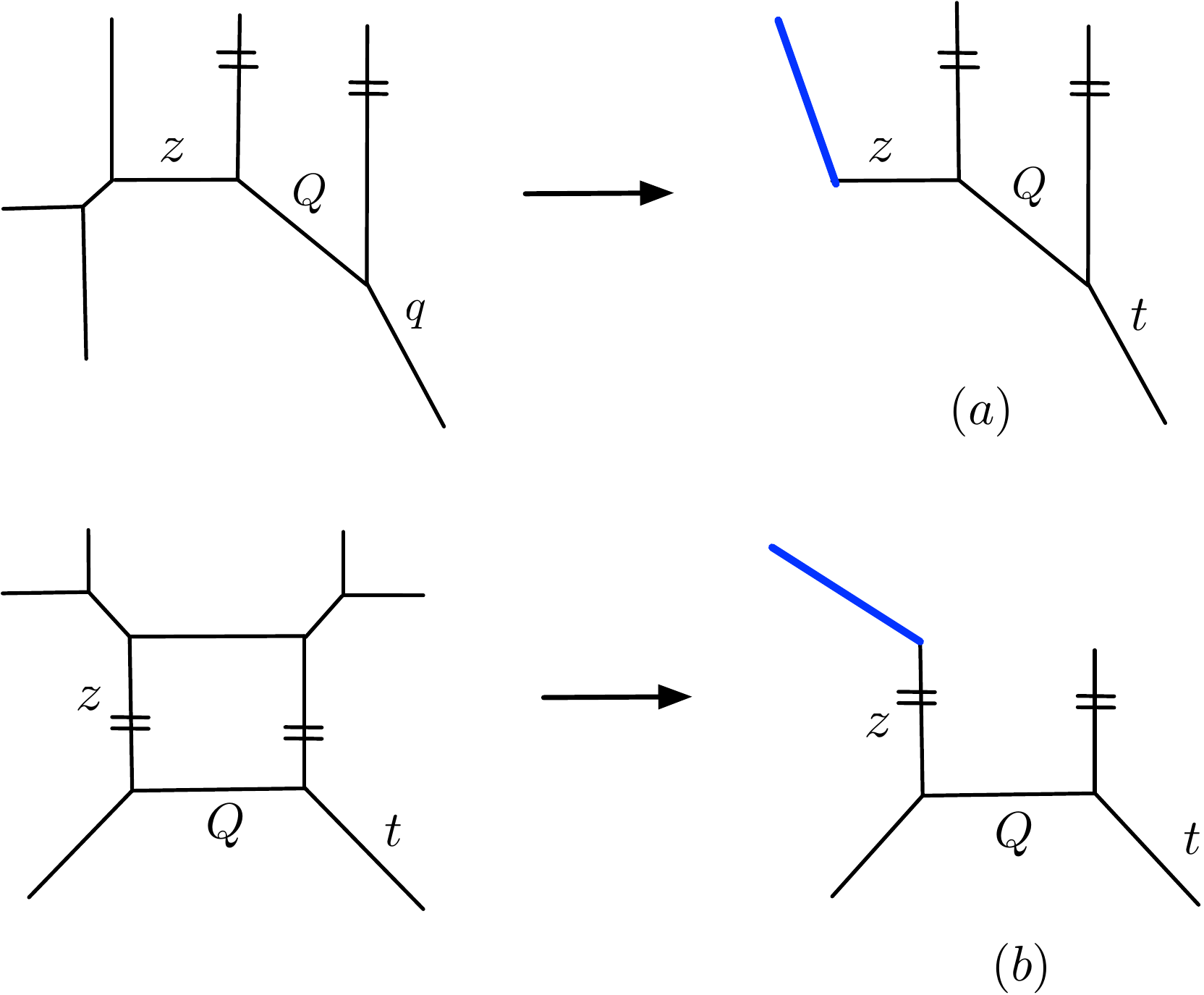}
	\caption{Geometric transitions that produce branes on the horizontal $(a)$ or vertical $(b)$ leg in the resolution of $\mathbb{C}^3/\mathbb{Z}_2$.  }   \label{fig:c2overz2}
\end{figure}

Another example of a strip geometry with one local $\mathbb{P}^1$ is $\mathcal{O}(0)\oplus\mathcal{O}(-2)\to\mathbb{P}^1$, or equivalently a resolution of $\mathbb{C}^3/\mathbb{Z}_2$. In this case we consider two brane locations, either on a horizontal or a vertical leg (in our earlier conventions), as shown in fig. \ref{fig:c2overz2}. Partition functions for these branes can be obtained by two different geometric transitions. For branes on a horizontal leg the partition function takes form of a product of a finite number of quantum dilogarithms, which arise from a ratio of a finite number of MacMahon functions (representing the closed partition before the transition). Simplifying such MacMahon functions, for $t$-brane and $\bar{q}$-brane we obtain respectively
\begin{align}
\begin{split}
Z_\tbrane^{(a)} &= \frac{1}{ (z t, q )_\inf  ( z Qt,q  )_\inf  } , \\
Z_\qbarbrane^{(a)} &=  \left(z \frac{t}{q}, t \right)_\inf   \left(z Q \frac{t}{q}, t \right)_\inf .
\end{split}
\end{align}
The corresponding open BPS invariants are $N_z^{(1/2, 1/2)} =1$ and $N_{zQ}^{(1/2,1/2)} =1$.

\begin{table}[h!]
	{
		\begin{tabular}{|l|c|c|}\hline
			$d=1,~(d_0,d_1)$ & $2 r$& $2 s=1$   \\\hline
			$(1,0)$&1&1
			\\	\hline 
		\end{tabular}
		\begin{tabular}{|l|c|c|}\hline
			$d=2,~(d_0,d_1)$ & $2 r$& $2 s=3$   \\\hline
			$(1,1)$&3&1
			\\	\hline 
		\end{tabular}
		\begin{tabular}{|l|c|c|}\hline
			$d=3,~(d_0,d_1)$ & $2 r$& $2 s=5$   \\\hline
			$(2,1),(1,2)$&5&1
			\\	\hline 
		\end{tabular}
		\begin{tabular}{|l|c|cc|}\hline
			$d=4,~(d_0,d_1)$ & $2 r$& $2 s=7$&9   \\\hline
			$(3,1),(1,3)$&7&1&
			\\	
			$(2,2)$&7&1&1\\\hline
		\end{tabular}
		\begin{tabular}{|l|c|c cc|}\hline
			$d=5,~(d_0,d_1)$ & $2 r$& $2 s=9$&11&13   \\\hline
			$(4,1),(1,4)$&9&1&&
			\\	
			$(2,3),(3,2)$&9&1&1&1
			\\\hline 
		\end{tabular}
		\begin{tabular}{|l|c|ccccc|}\hline
			$d=6,~(d_0,d_1)$ & $2 r$& $2 s=11$&13&15&17&19   \\\hline
			$(5,1),(1,5)$&11&1&&&&
			\\
			$(4,2),(2,4)$&11&2&2&1&1&\\
			$(3,3)$&11&3&3&3&1&1 \\
			\hline 
		\end{tabular}
		\begin{tabular}{|l|c|ccccccc|}\hline
			$d=7,~(d_0,d_1)$ & $2 r$& $2 s=13$&15&17&19&21&23&25   \\\hline
			$(6,1),(1,6)$&13&1&&&&&&
			\\
			$(5,2),(2,5)$&13&2&2&1&1&&&\\
			$(4,3),(3,4)$&13&5&5&6&4&3&1&1 \\
			\hline 
		\end{tabular}
		\begin{tabular}{|l|c|cccccccccc|}\hline
			$d=8,~(d_0,d_1)$ & $2 r$& $2 s=15$&17&19&21&23&25&27&29&31&33   \\\hline
			$(7,1),(1,7)$&15&1&&&&&&&&&
			\\
			$(6,2),(2,6)$&15&3&3&2&2&1&1&&&&\\
			$(5,3),(3,5)$&15&7&8&10&8&7&4&3&1&1& \\
			$(4,4)$&15&8&12&14&14&10&8&5&3&1&1  \\
			\hline 
		\end{tabular}
		\begin{tabular}{|l|c|ccccccccccccc|}\hline
			$d=9,~(d_0,d_1)$ & $2 r$& $2 s=17$&19&21&23&25&27&29&31&33&35&37&39&41   \\\hline
			$(8,1),(1,8)$&17&1&&&&&&&&&&&&\\
			$(7,2),(2,7)$&17&4&3&3&2&2&1&1&&&&&&   \\ 
			$(6,3), (3,6)$&17&9&12&15&13&13&9&7&4&3&1&1&&
			\\
			$(5,4),(4,5)$&17&14&21&30&30&29&22&19&12&9&5&3&1&1\\
			\hline 
		\end{tabular}
	} 
	\caption{Refined open BPS invariants for the resolution of $\mathbb{C}^3/\mathbb{Z}_2$ in fig. \ref{fig:c2overz2} $(b)$. Degrees $(d_0,d_1)$ correspond to the term $Q^{d_0} Q_1^{d_1} $. }\label{tb:c2overz2}
\end{table}

On the other hand, the partition function for $t$-brane in panel $(b)$ in fig. \ref{fig:c2overz2} takes form
\begin{align}\label{quiver=|}
\begin{split}
Z_\tbrane^{(b)} &=\sum\limits_{n=0}^{\inf} \frac{ Q^n t^n }{ \left( q,q \r)_n  \left(  Q_1 \frac{t}{q}, \frac{1}{q}  \r)_n  }= 
\frac{1}{ ( Q \frac{t}{q}, \frac{1}{q} )_\inf  } \sum\limits_{n,d=0}^{\inf}    ( -\sqrt{q} )^{-2nd} \frac{ (Q t)^n  (Q_1t)^d }{ \left( q,q \right)_n \left( q,q \right)_d } =   \\
& =\frac{1}{ ( Q_1 \frac{t}{q}, \frac{1}{q} )_\inf  } P_C(q;Qt,Q_1t ),  
\end{split}
\end{align}
which is a framed version of (\ref{tbranehypergeometry}); the quiver matrix in the representation (\ref{quivergen}) in this case reads
\begin{align}
C = \left[ 
\begin{array}{cc}
0 &-1   \\
-1 & 0\\
\end{array}
\right] \,.
\end{align}
In this case there is an infinite number of open BPS invariants, see table \ref{tb:c2overz2}. Note that for fixed $d_0$ and $d_1$, non-zero invariants arise only for one particular value of $r$, in agreement with our earlier prediction. As usual, the partition function for a $\qbarbrane$ can be obtained by substituting $q\rightarrow 1/t,~t\rightarrow 1/q$ in \eqref{quiver=|} 
\begin{align}\label{quiverII}
	Z_\qbarbrane^{(b)} &=\sum\limits_{n=0}^{\inf} \frac{ 
	(-\sqrt{t})^{n^2}   \left( Q \sqtq\r)^n
	}
	{   (t,t)_n \left( Q_1 \frac{t}{q},t \r)_n  }
	=\frac{1}{ ( Q \frac{t}{q}, t )_\inf  } P_C\left(t; Q\sqtq ,Q_1 \frac{ \sqrt{t}}{q} \right)
\end{align} 
with quiver matrix in the representation (\ref{quivergen})
\begin{align}
C = \left[ 
\begin{array}{cc}
1 &1   \\
1 & 1\\
\end{array}
\right] .
\end{align}


\subsection{Double-$\mathbb{P}^1$}

\begin{figure}[]
	\centering
	\includegraphics[width=4.5in]{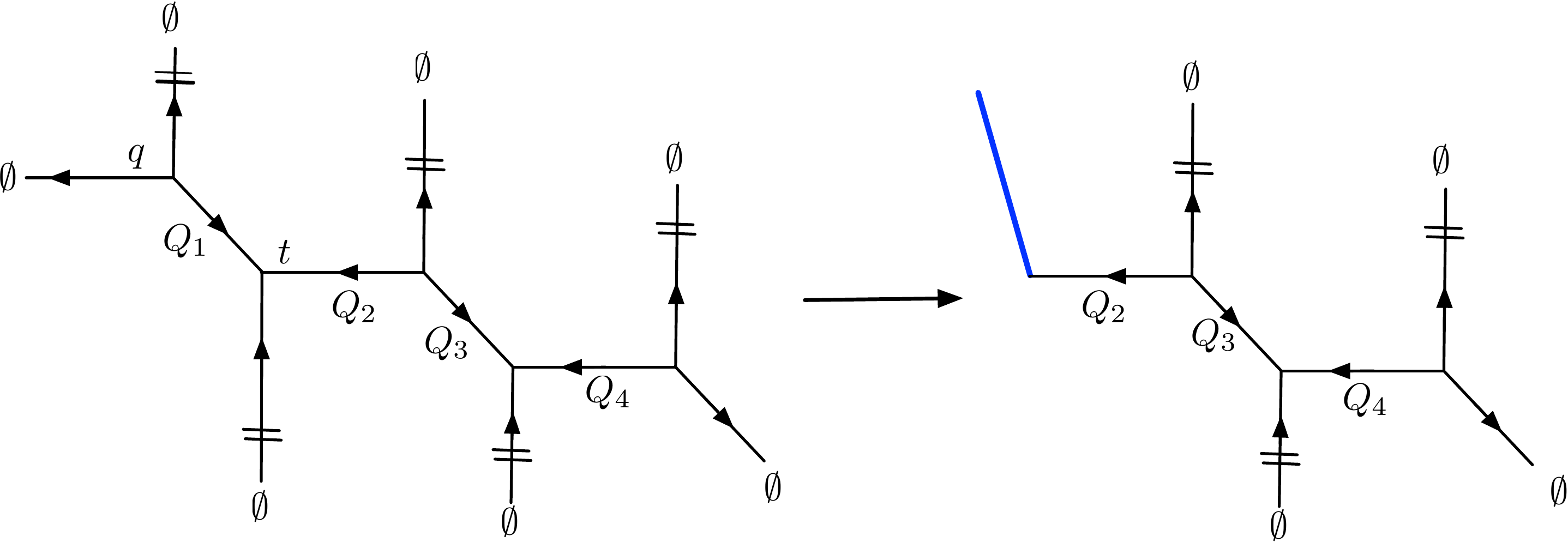}
	\includegraphics[width=4.5in]{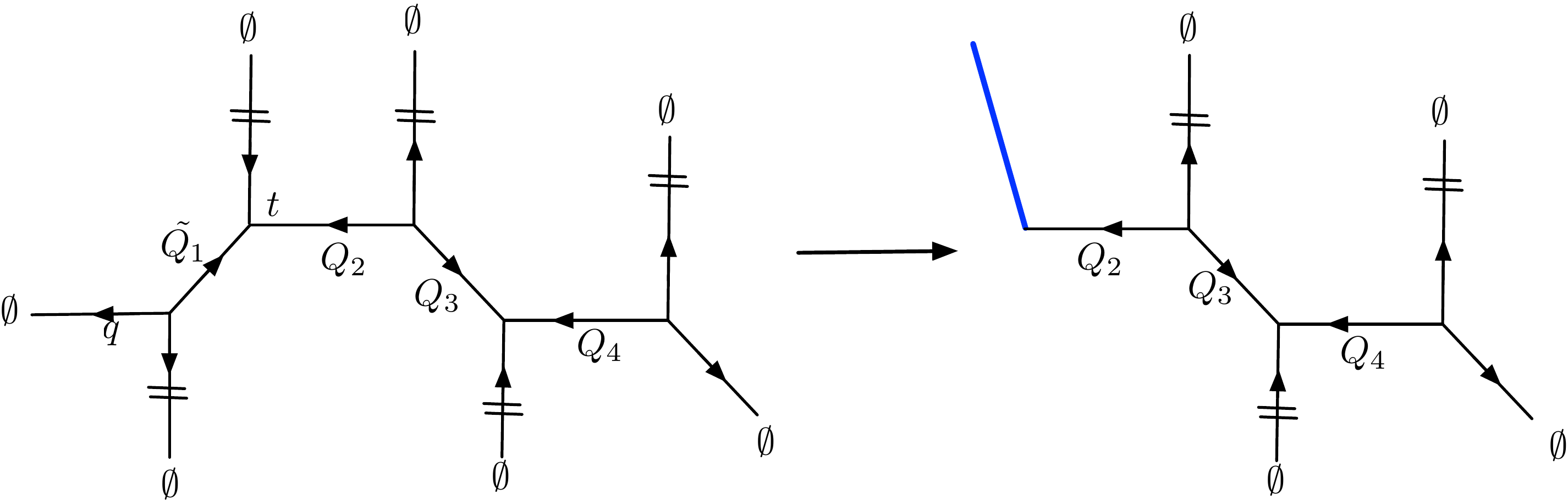}
	\caption{Geometric transition for two strip geometries related by a flop transition (left) produces the same double-$\mathbb{P}^1$ geometry, however with an additional lagrangian brane of various types (right). }  \label{fig:qqttexample1}
\end{figure}

In turn, we consider branes in a double-$\mathbb{P}^1$ geometry (that involves two local $\mathbb{P}^1$'s), either on a horizontal or vertical leg. First, by considering branes attached to a horizontal leg, as shown in fig. \ref{fig:qqttexample1}, we illustrate the effect of a flop transition on the resulting brane. Second, we consider branes on a vertical leg, and show that corresponding  open BPS invariants are non-negative integers.

\begin{figure}[]
	\centering
	\includegraphics[width=3in]{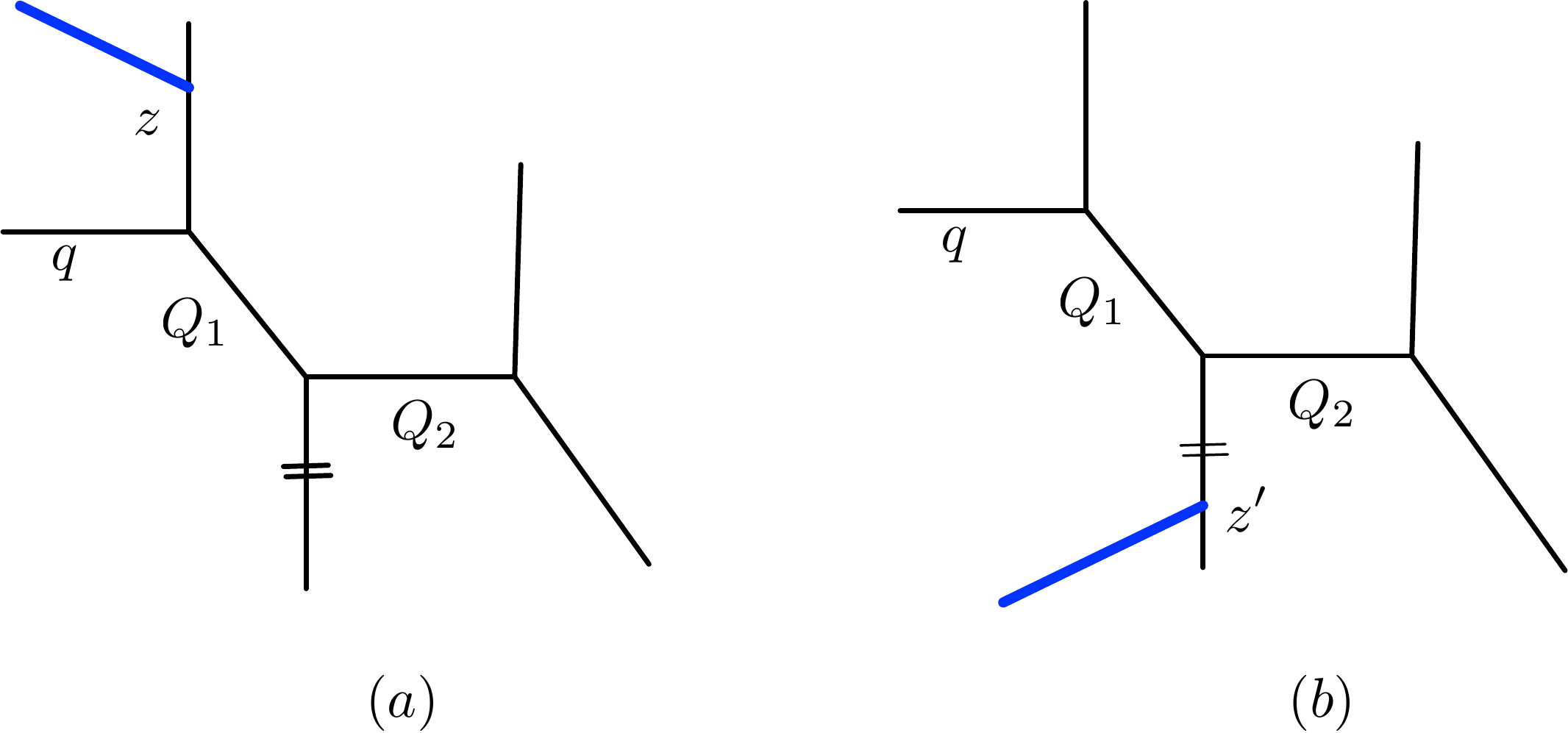}
	\caption{Branes on various vertical legs of a double-$\mathbb{P}^1$ geometry.  }
	\label{fig:updown3p1}
\end{figure}

To start with, consider the geometry in the top left in fig. \ref{fig:qqttexample1}. This is also a strip geometry, so its closed partition function is expressed in terms of a finite number of refined MacMahon functions
\begin{align}
\begin{split}
Z^{\text{closed}}=
&\frac{ 
	M \big( Q_1 \sqtq,t,q \big) 
	M \big( Q_2 \sqtq,t,q \big) 
	M \big( Q_3 \sqtq,t,q \big) 
	M \big( Q_4 \sqtq,t,q \big) 
}
{ 
	M \big( Q_1 Q_2 ,q,t \big) 
	M \big( Q_2 Q_3 ,t,q \big) 
	M \big( Q_3 Q_4 ,q,t \big) 
	M \big( Q_1 Q_2 Q_3 Q_4,q,t \big) 
}	\times \\
&\times
M \big( Q_1 Q_2 Q_3 \sqtq,t,q \big) 
M \big( Q_2 Q_3 Q_4 \sqtq,t,q \big) .
\end{split}
\end{align}
In order to introduce a brane we perform a geometric transition at the cycle of size $Q_1$. In this process we can ignore the terms $\frac{M(Q_3\sqtq,t,q)M(Q_4\sqtq,t,q) }{M(Q_3Q_4,q,t)}$ that capture only closed string contributions. Furthermore, following \eqref{qqtthorizontal}, we substitute $Q_1=q \sqqt$ or $Q_1=\frac{1}{t}  \sqtq$ respectively to obtain partition functions for $q$-brane or $\bar{t}$-brane with the open K\"ahler parameter $z\equiv Q_2$ 
\begin{align}
\begin{split}
Z_{q\text{-brane}}&=\frac{ 
	M\left( Q_2 \sqtq,t,q \r)
	M\left(  Q_2 Q_3 q,t,q \r)
	M\left(  Q_2 Q_3 Q_4\sqtq,t,q \r)
}
{M\left( Q_2 \sqrt{q t},t,q \r)
	M\left( Q_2 Q_3,t,q \r)
	M\left( Q_2 Q_3 Q_4 \sqrt{q t},t,q \r)
} = \\
	&=   \frac{  ( Q_2 \sqrt{tq},t  )_\inf    (Q_2 Q_3 Q_4 \sqrt{tq},t   )_\inf  }{  (Q_2 Q_3 q,t)_\inf }  = P_{C_{q}}\left(t; Q_2 \sqrt{q}, Q_2Q_3Q_4 \sqrt{q} ,Q_2Q_3 q \right) ,  \label{qbraneA11}\\
Z_{\bar{t}\text{-brane}}&=\frac{ 
	M\left( Q_2 \sqtq,t,q \r)
	M\left(  Q_2 Q_3/t,t,q \r)
	M\left(  Q_2 Q_3 Q_4\sqtq,t,q \r)
}
{M\left( Q_2/ \sqrt{q t},t,q \r)
	M\left( Q_2 Q_3,t,q \r)
	M\left( Q_2 Q_3 Q_4/ \sqrt{q t},t,q \r) 
} = \\
& =\frac{ (  Q_2 Q_3 \frac{q}{t}  , q )_\inf  }{( Q_2 \sqqt , q )_\inf   ( Q_2Q_3Q_4 \sqqt,q )_\inf } = P_{C_{ \bar{t}}}\left( q; Q_2 Q_3 \frac{\sqrt{q}}{t}, Q_2 \sqqt, Q_2 Q_3 Q_4 \sqqt   \right ),
\end{split}
\end{align}
	where 
	\begin{align}
	C_q=\left[ 
	\begin{array}{ccc}
	1 &0  &0 \\
	0& 1 &0\\
	0 &0  &0\\
	\end{array}
	\right]  \qquad \qquad
	C_{\bar{t}}=\left[ 
	\begin{array}{ccc}
	1 &0  &0 \\
	0& 0 &0\\
	0 &0  &0\\
	\end{array}
	\right]  
	\end{align}  
As a check, note that for the above partition functions the relation \eqref{qbarteql} holds.

On the other hand, we consider a geometry represented by the diagram in the bottom left in fig.  \ref{fig:qqttexample1}, related to the previous geometry by a flop transition on a cycle of size $Q_1$. Upon the geometric transition it leads to the same double-$\mathbb{P}^1$ geometry, however with different types of branes: substituting $\tilde{Q}_1=t \sqtq$ or $\tilde{Q}_1=\frac{1}{q} \sqqt$ we respectively obtain $t$-brane or $\bar{q}$-brane, whose partition functions take form 
\begin{align}
\begin{split}
Z_{t\text{-brane}}&=\frac{ 
	M\left( Q_2 t,q,t \r)
	M\left(  Q_2 Q_3 \sqtq,t,q \r)
	M\left(  Q_2 Q_3 Q_4 t,q,t \r)
}
{M\left( Q_2 ,q,t \r)
	M\left(  Q_2 Q_3  \sqrt{q t},q,t \r)
	M\left( Q_2 Q_3 Q_4 ,q,t \r)
}   = \\
& = \frac{(Q_2 Q_3 \sqrt{tq} ,q  )_\inf  }{ (Q_2 t ,q )_\inf  ( Q_2Q_3Q_4 t,q  )_\inf  }  = P_{C_t}\left(  q; Q_2Q_3\sqrt{t},Q_2 t, Q_2Q_3Q_4 t      \right) , \\
Z_{\bar{q}\text{-brane}}&=\frac{ 
	M\left( Q_2/q ,q,t \r)
	M\left(  Q_2 Q_3 \sqtq,t,q \r)
	M\left(  Q_2 Q_3 Q_4 /q,q,t \r)
}
{M\left( Q_2, q,t \r)
	M\left( Q_2 Q_3/{\sqrt{q t}},q,t \r)
	M\left( Q_2 Q_3 Q_4,q,t\r)
} = \\
&= \frac{  (Q_2 \frac{t}{q} ,t )_\inf    ( Q_2Q_3Q_4 \frac{t}{q},t  )_\inf }{  ( Q_2Q_3 \sqtq,t )_\inf   }  =P_{C_{\bar{q}}}  \left(    t ; Q_2 \frac{\sqrt{t}}{q}, Q_2Q_3Q_4\sqtq, Q_2Q_3\sqtq  \right)
\end{split}
\end{align}
where quiver matrices in the representation (\ref{quivergen}) are
\begin{align}
C_t=\left[ 
\begin{array}{ccc}
1 &0  &0 \\
0& 0 &0\\
0 &0  &0\\
\end{array}
\right]  \qquad \quad 
C_{\bar{q}}=\left[ 
\begin{array}{ccc}
1 &0  &0 \\
0& 1 &0\\
0 &0  &0\\
\end{array}
\right]  
\end{align}
There are just 3 open BPS numbers
\begin{align}
N_{Q_2}^{(1/2,1/2)}=1, \quad N_{Q_2 Q_3}^{(0,0)}=1, \quad N_{Q_2 Q_3 Q_4}^{(1/2,1/2)}=1  \,.
\end{align}

To sum up, the geometric transition may produce only two particular types of branes. A flop of a cycle that undergoes the geometric transition leads then to another two types of branes. As a check, all relations from fig. \ref{branes-types} hold for the partition functions determined above.


\begin{figure}[]
	\centering
	\includegraphics[width=5.5in]{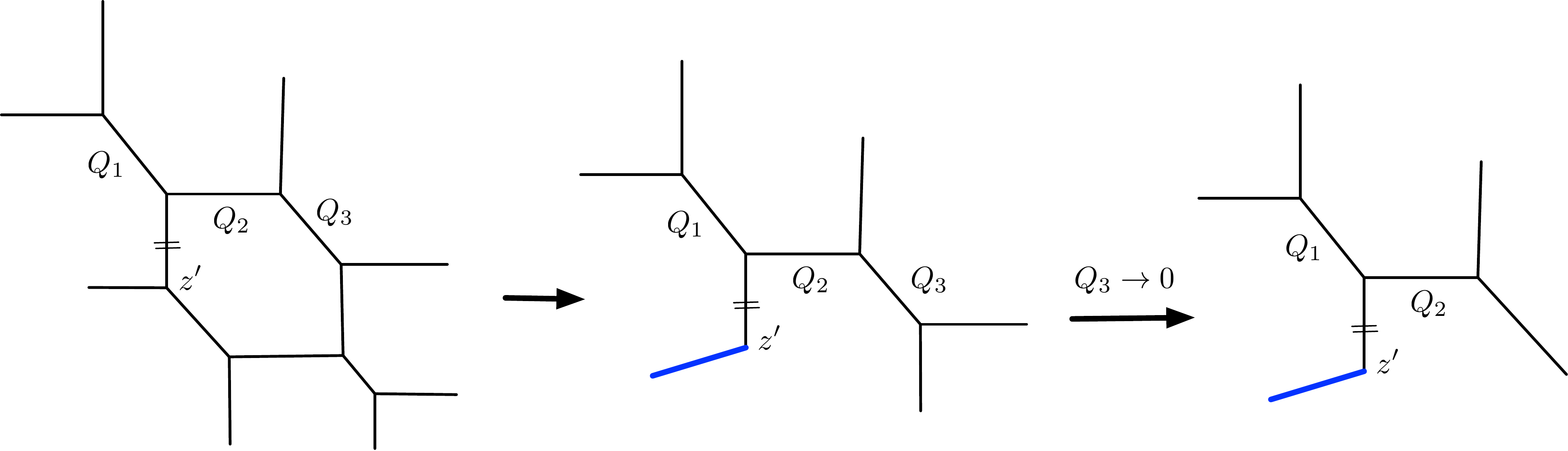}
	\caption{The geometric transition that engineers the diagram $(b)$ in fig. \ref{fig:updown3p1}.  }
	\label{fig:higgsdecoup}
\end{figure}

Furthermore, we consider branes on vertical legs. Partition functions for a $\bar{q}$-brane and $t$-brane in the panel $(a)$ in fig. \ref{fig:updown3p1} take form
\begin{align}\label{2p1tbrane}
\begin{split}
Z^{(a)}_{\bar{q}\text{-brane}} &= \sum\limits_{n=0}^{\inf}
\frac{
	\left(z\sqtq\r)^n \left( Q_1 \sqtq, t\r)_n
}
{ (t,t)_n  \left(  Q_1Q_2\frac{t}{q},t\r)_n
} = \frac{
	\left( Q_1 \sqtq, t\r)_\inf
}
{   \left(  Q_1Q_2\frac{t}{q},t\r)_\inf
}P_{C_{\bar{q}}} \left(t; z \sqtq, Q_1 \sqtq  , Q_1Q_2 \frac{ \sqrt{t} }{q} \right) ,  \\
Z^{(a)}_{\tbrane}&= \sum\limits_{n=0}^{\inf}
\frac{ (-z)^n q^{ \frac{n^2}{2}}t^{\frac{n}{2}}
	\left( Q_1 \sqtq,  \frac{1}{q} \r)_n
}
{ (q,q)_n  \left(  Q_1Q_2\frac{t}{q}, \frac{1}{q}\r)_n
} = \frac{\left( Q_1 \sqtq,  \frac{1}{q} \r)_\inf}{\left(  Q_1Q_2\frac{t}{q}, \frac{1}{q}\r)_\inf}
P_{C_t} \left(  q; z \sqrt{t}, Q_1Q_2 t    \right) .
\end{split}
\end{align}
The first summation formulae above are special cases of (\ref{qbarbranehypergeometry}) and  (\ref{tbranehypergeometry}), while in the quiver form (\ref{quivergen}) quiver matrices read
\begin{align}
C_{\bar{q}} = \left[ 
\begin{array}{ccc}
0 &1  &1 \\
1 & 0 &0\\
1 &0  &1\\
\end{array}
\right] \qquad \quad
C_{t} = \left[ 
\begin{array}{ccc}
1&-1  &-1 \\
-1 & 1 &0\\
-1 &0  &0\\
\end{array}
\right] 
\end{align}
We verify that corresponding refined open BPS invariants are indeed non-negative integers, as shown in table \ref{tb:2P1openBPS}. Note that for fixed $(d_0,d_1,d_2)$, non-zero BPS invariants arise only for one particular value of $r$, as predicted above.

On the other hand, partition functions for branes in a diagram $(b)$ in fig. \ref{fig:updown3p1} can  be obtained by performing a geometric transition and then blowing down the cycle of size $Q_3\rightarrow 0$, as shown in fig. \ref{fig:higgsdecoup}. In this way we obtain $\qbrane$ and $\tbarbrane$ partition functions
\begin{align}
\begin{split}
Z^{(b)}_{\qbrane}&=\sum\limits_{n=0}^{\inf} \frac{ q^n (z' Q_1)^n  \left(  \frac{1}{Q_1} \sqtq ,t \r)_n \left( Q_2 \sqtq,t \r) }{ (t,t)_n   }\,,\quad \\
Z^{(b)}_{\tbarbrane} &=\sum\limits_{n=0}^{\inf}
\frac{  (-z' Q_2)^n q^{ \frac{ n-n^2 }{2}} 
	\left(Q_1 \sqqt,q\r)_n  \left( \frac{1}{Q_2} \sqqt,q \r)_n  
}{  (q,q)_n }.
\end{split}
\end{align}
One can now apply \eqref{qtoqbar} to get $\qbarbrane$ and $\tbrane$ partition functions
\begin{align}
\begin{split}
Z^{(b)}_{\qbarbrane}&=\sum\limits_{n=0}^{\inf}
\frac{ \left( \tilde{z} \sqtq \r)^n  \left( \tilde{Q}_1 \sqtq,t \r)_n  \left( Q_2\sqtq, t \r)_n    }{ (t,t)_n  } = \\
&=   \left( \tilde{Q}_1 \sqtq,t \r)_\inf \left( Q_2\sqtq, t \r)_\inf ~ P_{C_q} \left(  t; \tilde{z} \sqtq,\tilde{Q}_1\sqtq, {Q}_2\sqtq   \right) 
,  \\
Z^{(b)}_{\tbrane}&=\sum\limits_{n=0}^{\inf}
\frac{ \left(\tilde{z} \sqtq\r)^n \left( \tilde{Q}_1\sqtq,\frac{1}{q}  \r)_n  \left(  Q_2 \sqtq, \frac{1}{q}\r)_n    }{\left( \frac{1}{q},  \frac{1}{q} \r)_n    } = P_{C_{\bar{t}}} (q;~ \tilde{z}\sqrt{t}, \tilde{Q}_1 \sqrt{t},  Q_2 \sqrt{t}   )   ,\label{2p1downt}
\end{split}
\end{align}
where $\tilde{Q}_1=1/Q_1, \tilde{z}=z' Q_1 q\sqqt $. These formulas have a standard form for open partition functions on a strip geometry, thus the corresponding open BPS invariants are non-negative integers. The corresponding quiver matrices in the representation (\ref{quivergen}) are
\begin{align}
C_{q} = \left[ 
\begin{array}{ccc}
0 &1  &1 \\
1 & 0 &0\\
1 &0  &0\\
\end{array}
\right] \qquad \quad
C_{\bar{t}} = \left[ 
\begin{array}{ccc}
1 &-1  &-1 \\
-1 & 1 &0\\
-1 &0  &1\\
\end{array}
\right]
\end{align}

\begin{table}[h!]
	\centering
	{
		
		\begin{tabular}{|l|c|c|}\hline
			$d=1,~(d_0,d_1,d_2)$ & $2 r$& $2 s=0$   \\\hline
			$(1,0,0)$&0&1
			\\	\hline 
		\end{tabular}
		\begin{tabular}{|l|c|c|}\hline
			$d=2,~(d_0,d_1,d_2)$ & $2 r$& $2 s=1$   \\\hline
			$(1,1,0)$&1&1
			\\	\hline 
		\end{tabular}
		\begin{tabular}{|l|c|c|}\hline
			$d=3,~(d_0,d_1,d_2)$ & $2 r$& $2 s=2$   \\\hline
			$(1,1,1)$&2&1
			\\	\hline 
		\end{tabular}
		\begin{tabular}{|l|c|c|}\hline
			$d=4,~(d_0,d_1,d_2)$ & $2 r$& $2 s=3$   \\\hline
			$(1,2,1)$&3&1
			\\	\hline 
		\end{tabular}
		\begin{tabular}{|l|c|c|}\hline
			$d=5,~(d_0,d_1,d_2)$ & $2 r$& $2 s=4$   \\\hline
			$(2,2,1),(1,2,2)$&4&1
			\\	\hline 
		\end{tabular}
		\begin{tabular}{|l|c|c|}\hline
			$d=6,~(d_0,d_1,d_2)$ & $2 r$& $2 s=5$   \\\hline
			$(2,3,1),(2,2,2),(1,3,2)$&5&1
			\\	\hline 
		\end{tabular}
		\begin{tabular}{|l|c|c c|}\hline
			$d=7,~(d_0,d_1,d_2)$ & $2 r$& $2 s=6$&8   \\\hline
			$(3,3,1),(1,3,3)$&6&1&\\
			$(2,3,2)$&6&2&1
			\\	\hline 
		\end{tabular}
		\begin{tabular}{|l|c|c c|}\hline
			$d=8,~(d_0,d_1,d_2)$ & $2 r$& $2 s=7$&9   \\\hline
			$(3,4,1),(1,4,3)$&7&1&\\
			$(3,3,2),(2,4,2),(2,3,3)$&7&1&1
			\\	\hline 
		\end{tabular}
		\begin{tabular}{|l|c|c c c|}\hline
			$d=9,~(d_0,d_1,d_2)$ & $2 r$& $2 s=8$&10&12   \\\hline
			$(4,4,1),(1,4,4)$&8&1&&\\
			$(3,3,3)$&8&&1&\\
			$(3,4,2),(2,4,3)$&8& 3& 2&1
			\\	\hline 
		\end{tabular}
		\begin{tabular}{|l|c|c c c c|}\hline
			$d=10,~(d_0,d_1,d_2)$ & $2 r$& $2 s=9$&11&13&15   \\\hline
			$(4,5,1),(1,5,4)$&9&1&&&\\
			$(4,4,2),(3,5,2),(2,5,3),(2,4,4)$&9&2&1&1&\\
			$(3,4,3)$&9&3& 4&2&1
			\\	\hline 
		\end{tabular}
		\begin{tabular}{|l|c|c c c c c|}\hline
			$d=11,~(d_0,d_1,d_2)$ & $2 r$& $2 s=10$&12&14&16&18   \\\hline
			$(5,5,1),(1,5,5)$&10&1&&&& \\
			$(4,4,3),(3,4,4)$&10&1&2&1&1& \\
			$(4,5,2),(2,5,4)$&10&4& 3&2&1&  \\
			$(3,5,3)$&10& 6&6&5&2&1
			\\	\hline 
		\end{tabular}
		\begin{tabular}{|l|c|c c c c c c|}\hline
			$d=12,~(d_0,d_1,d_2)$ & $2 r$& $2 s=11$&13&15&17&19&21   \\\hline
			$(5,5,1),(1,5,5)$&11&1&&&&& \\
			$(4,4,4)$&11&&1&&1&&   \\
			$(5,5,2),(4,6,2),(2,6,4),(2,5,5)$&11&2&2&1&1& &\\
			$(3,6,3)$&11&3& 3&3&1&1&  \\
			$(4,5,3),(3,5,4)$&11& 6&9&7&5&2&1
			\\	\hline 
		\end{tabular}
	} 
	\caption{Refined open BPS invariants for a double-$\mathbb{P}^1$ strip diagram in fig. \ref{fig:updown3p1} $(a)$. $(d_0,d_1,d_2)$ are degrees corresponding to the term $z^{d_0} Q_1^{d_1} Q_2^{d_2}$. }\label{tb:2P1openBPS}
\end{table}


\subsection{Triple-$\mathbb{P}^1$ }    \label{ssec-3P1}

\begin{figure}[]
	\centering
	\includegraphics[width=2in]{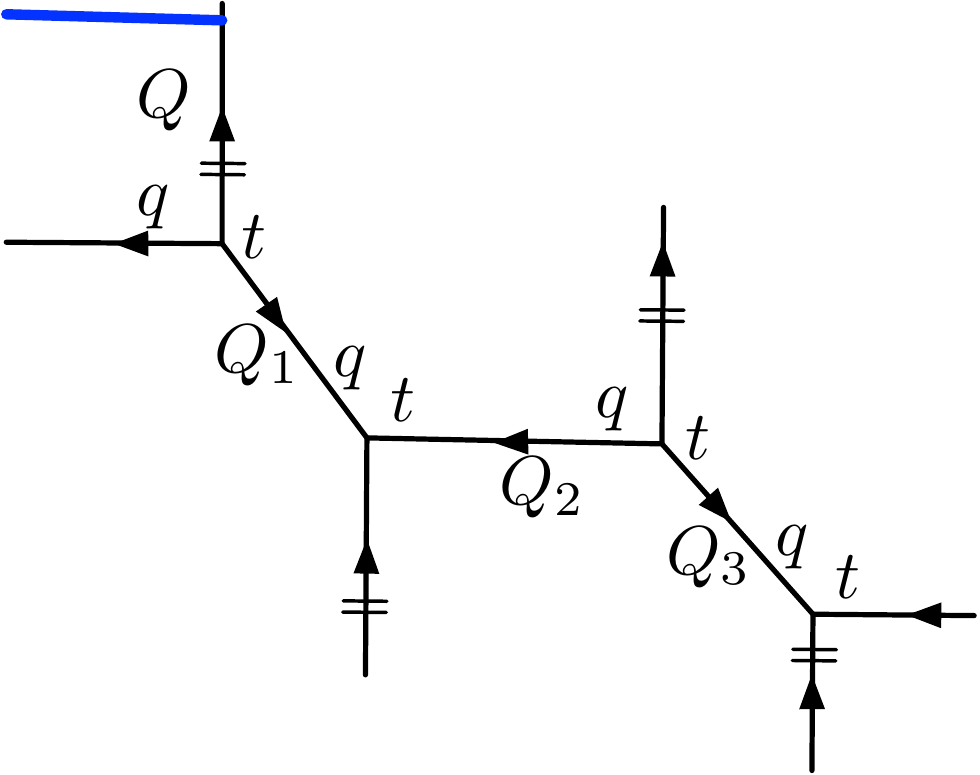}
	\caption{A brane on a vertical leg in a triple-$\mathbb{P}^1$. }
	\label{fig:3P1}
\end{figure}

In turn, we consider branes on a vertical leg in a triple-$\mathbb{P}^1$ geometry, as shown in fig.  \ref{fig:3P1}.  
After the geometric transition (from an appropriate more complicated threefold), denoting the resulting open K\"ahler  parameter by $Q$, for a $\bar{q}$-brane and $t$-brane we find respectively
\begin{align}
\begin{split}
Z_{\qbarbrane}&= \sum\limits_{n=0}^{\inf} \frac{  
	\left(Q \sqtq \r)^n }
{ (t,t)_n  } 
\frac{ \left(   Q_1 \sqtq, t\r)_n  \left(  Q_1 Q_2 Q_3 \sqtq, t\r)_n   }
{   \left(   Q_1 Q_2 \frac{t}{q}  , t\r)_n     } =\\
& = \frac{ \left(   Q_1 \sqtq, t\r)_\inf  \left(  Q_1 Q_2 Q_3 \sqtq, t\r)_\inf   }
		{   \left(   Q_1 Q_2 \frac{t}{q}  , t\r)_\inf   }   P_{C_{\bar{q}}}\left(t;  Q \sqtq,Q_1 \sqtq, Q_1Q_2Q_3 \frac{ \sqrt{t}}{q} \right) 
\end{split}
\end{align}
and
\begin{align}
\begin{split}
Z_{\tbrane}&= \sum\limits_{n=0}^{\inf} \frac{  
			\left(Q \sqtq \right)^n }
		{ \left(\frac{1}{q},\frac{1}{q}\right)_n  } 
		\frac{ \left(   Q_1 \sqtq, \frac{1}{q}\r)_n  \left(  Q_1 Q_2 Q_3 \sqtq, \frac{1}{q}\r)_n   }
		{   \left(   Q_1 Q_2 \frac{t}{q}  , \frac{1}{q}\r)_n     } = \\
&=\frac{ \left(   Q_1 \sqtq, \frac{1}{q}\r)_\inf  \left(  Q_1 Q_2 Q_3 \sqtq, \frac{1}{q}\r)_\inf   }
		{   \left(   Q_1 Q_2 \frac{t}{q}  , \frac{1}{q}\r)_\inf   }   P_{C_t}\left(q;  Q \sqrt{t},Q_1 \sqrt{t}, Q_1Q_2Q_3 t \right)  
\end{split}
\end{align}
where
\begin{align}
C_{\bar{q}} =  \left[ 
		\begin{array}{cccc}
			0 &1 &1 &1 \\
			1 & 0&0  &0\\
			1 & 0 & 0&0\\
			1 &0 &0  &1\\
		\end{array}
		\right]\qquad\quad
C_{t} =  \left[ 
		\begin{array}{cccc}
			1 &-1 &-1 &-1 \\
			-1 & 1&0  &0\\
			-1 & 0 & 1&0\\
			-1 &0 &0  &0\\
		\end{array}
		\right]		
\end{align}
We verify that refined open BPS invariants $N_{(d,d_1,d_2,d_3)}^{(s,r)}$ are non-negative integers as expected, as shown in table \ref{tb:3P1openBPS}. Note that for fixed $(d,d_1,d_2,d_3)$, non-zero invariants arise only for one particular value of $r$, as predicted earlier. We also conjecture that, for a given $d$, non-zero open BPS invariants arise for indices $(s,r)$ that are in the range
\begin{align}
r \leq s \leq  d+1, \qquad \frac{1}{2}\Big[\frac{d}{2}\Big] \leq r \leq \frac{d-1}{2},
\end{align}
and open BPS invariants with the maximal spin $s=d+1$ are equal to one.


\begin{table}[h!]
	\centering
	{\small
		\begin{tabular}{|l|c|c |}\hline
			$d=1,~(d_0,d_1,d_2,d_3)$ & $2 r$& $2 s=0$   \\\hline
			$(1,0,0,0)$&0&1   
			\\	\hline 
		\end{tabular}
		\begin{tabular}{|l|c|c |}\hline
			$d=2,~(d_0,d_1,d_2,d_3)$ & $2 r$& $2 s=1$   \\\hline
			$(1,1,0,0)$&1&1   
			\\	\hline 
		\end{tabular}
		\begin{tabular}{|l|c|c |}\hline
			$d=3,~(d_0,d_1,d_2,d_3)$ & $2 r$& $2 s=2$   \\\hline
			$(1,1,1,0)$&2&1   
			\\	\hline 
		\end{tabular}
		\begin{tabular}{|l|c|c c |}\hline
			$d=4,~(d_0,d_1,d_2,d_3)$ & $2 r$& $2 s=1$  & 3 \\\hline
			$(1,1,1,1)$&1&1&\\
			$(1,2,1,0)$&3&&1
			\\	\hline 
		\end{tabular}
		\begin{tabular}{|l|c|c c |}\hline
			$d=5,~(d_0,d_1,d_2,d_3)$ & $2 r$& $2 s=2$  & 4 \\\hline
			$(1,2,1,1)$&2&1&\\
			$(2,2,1,0),(1,2,2,0)$&4&&1
			\\	\hline 
		\end{tabular}
		\begin{tabular}{|l|c|c c |}\hline
			$d=6,~(d_0,d_1,d_2,d_3)$ & $2 r$& $2 s=3$  & 5 \\\hline
			$(2,2,1,1),(1,2,2,1)$&1&1&\\
			$(2,3,2,0),(2,2,2,0),(1,3,2,0)$&1&&1
			\\	\hline 
		\end{tabular}
		\begin{tabular}{|l|c|c c c|}\hline
			$d=7,~(d_0,d_1,d_2,d_3)$ & $2 r$& $2 s=4$  & 6&8 \\\hline
			$(2,3,1,1),(2,2,2,1),(1,3,2,1)$& 4&1&&
			\\
			$(3,3,1,0),(1,3,3,0)$&6&&1&\\
			$(2,3,2,0)$&6&&1&1
			\\	\hline 
		\end{tabular}
		\begin{tabular}{|l|c|c c c|}\hline
			$d=8,~(d_0,d_1,d_2,d_3)$ & $2 r$& $2 s=5$  & 7 & 9  \\\hline
			$(1,3,3,1),(3,3,1,1)$&5&1&& \\
			$(2,3,2,1)$ & 5& 3&1& \\
			$(1,4,3,0),(3,4,1,0)$ & 7& & 1 &  \\
			$(2,3,3,0),( 2,4,2,0),(3,3,2,0)$ & 7&&1&1
			\\	\hline 
		\end{tabular}
		\begin{tabular}{|l|c|c c c c c|}\hline
			$d=9,~(d_0,d_1,d_2,d_3)$ & $2 r$& $2 s=4$  & 6 & 8 & 10 &12  \\\hline
			$(2,3,2,2)$&4&1&&&& \\
			$(1,4,3,1),(3,4,1,1)$&6&&1&&&\\
			$(2,3,3,1),(2,4,2,1),(3,3,2,1)$&6&&2&1&&
			\\ $(1,4,4,0),(4,4,1,0)$&8&&&1&&  \\
			$(3,3,3,0)$&8&&&&1& \\
			$(2,4,3,0),(3,4,2,0)$&8&&&3&2&1
			\\	\hline 
		\end{tabular}
		\begin{tabular}{|l|c|c c c c c c|}\hline
			$d=10,~(d_0,d_1,d_2,d_3)$ & $2 r$& $2 s=5$ & 7 & 9 & 11 & 13 & 15 \\\hline
			$(2,3,3,2)$,$(2,4,2,2)$,$(3,3,2,2)$  & 5&1 &&&&&\\
			$ (1,4,4,1) , (4,4,1,1)$  & 7 & & 1  &&&&             \\
			$ (3,3,3,1) $  & 7 & & 1  &1&&&             \\
			$ (2,4,3,1),(3,4,2,1) $  & 7 & & 5 &3&1&&             \\
			$(1,5,4,0 ),(4,5,1,0 )$ &9 &&&1        &&&\\
			$(2,4,4,0 ),(2,5,3,0 ),(3,5,2,0),(4,4,2,0)$ &9 &&&2&1&1  &      \\
			$(3,4,3,0)$&9 &&&3&4&2&1\\
			\hline 
		\end{tabular}
		\begin{tabular}{|l|c|c c c c c c c|}\hline
			$d=11,~(d_0,d_1,d_2,d_3)$ & $2 r$& $2 s=6$  &  8 & 10 &12 & 14 & 16 &18 \\\hline
			$(3,3,3,2)$&6&1&&&&&&\\
			$(2,4,3,2),(3,4,2,2)$&6&2&1&&&&&\\
			$(1,5,4,1),(4,5,1,1)$&8&&1&&&&&  \\
			$(2,4,4,1 ),(2,5,3,1),(3,5,2,1),(4,4,2,1)$&8&&3&2&1&&&\\
			$(3,4,3,1)$&8&&7&7&3&1&&\\
			$(1,5,5,0),(5,5,1,0)$&10&&&1&&&&\\
			$(3,4,4,0),(4,4,3,0)$&10&&&1&2&1&1&\\
			$(2,5,4,0),(4,5,2,0)$&10&&&4&3&2&1&    \\
			$(3,5,2,0)$&10&&&6&6&5&2&1     \\
			\hline 
		\end{tabular}
	} 
	\caption{Open refined BPS invariants for triple-$\mathbb{P}^1$ strip diagram in fig. \ref{fig:3P1}. }\label{tb:3P1openBPS}
\end{table}

\clearpage



\subsection{$5 \mathbb{P}^1$ geometry}

\begin{figure}
	\centering
	\includegraphics[width=5.3in]{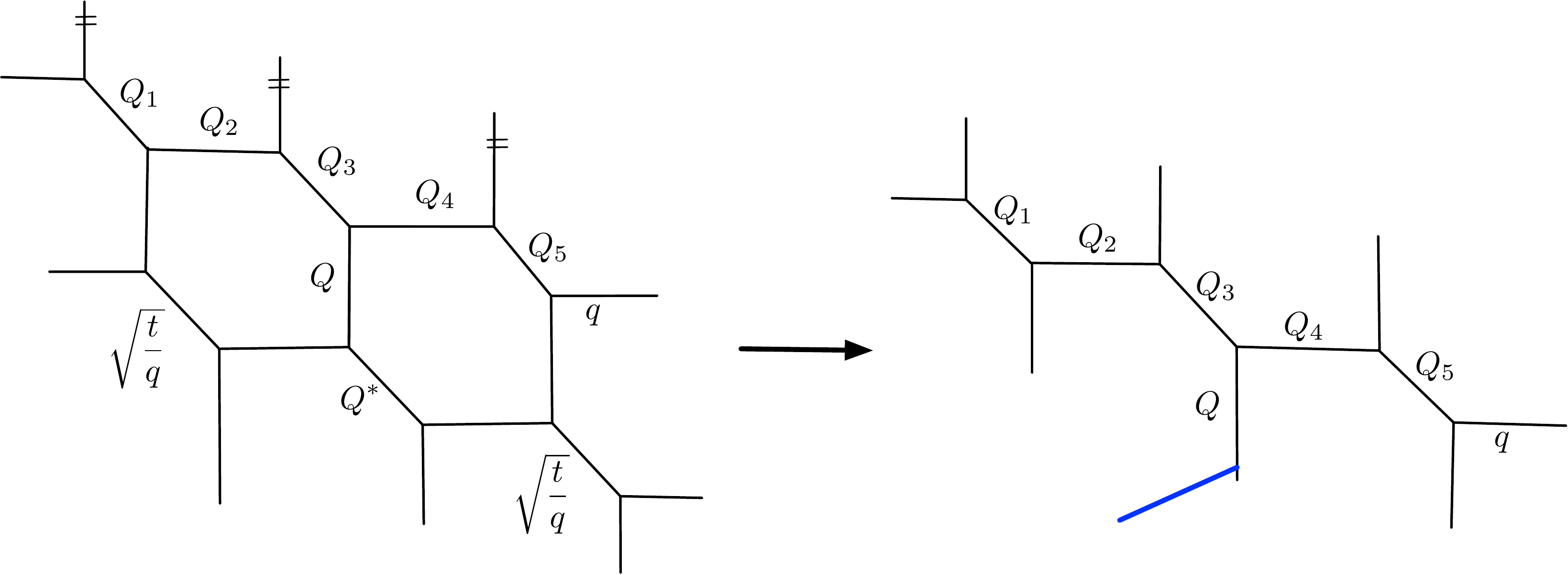}
	\caption{A geometric transition that produces $5 \mathbb{P}^1$ geometry with a lagrangian brane. }
	\label{fig:middle5p1}
\end{figure}

The last example of a strip geometry that we consider is a configuration with 5 local $\mathbb{P}^1$'s, which we also refer to simply as $5 \mathbb{P}^1$ geometry. We consider a geometric transition shown in fig. \ref{fig:middle5p1}; by 
 tuning $Q^*\rightarrow t \sqtq$ we get the $t$-brane partition function for brane configuration in fig. \ref{fig:middle5p1} (right)
\begin{align}
\begin{split}
Z_{\tbrane}&=\sum\limits_{n=0}^{\inf}  
\frac{ \left(\frac{Q}{Q_5} \sqtq\r)^n 
	\left( Q_3\sqtq,\frac{1}{q} \r)_n 
	\left(Q_1 Q_2 Q_3\sqtq,\frac{1}{q} \r)_n
	\left( \frac{1}{Q_4}  \sqtq,\frac{1}{q} \r)_n
}
{    \left( \frac{1}{q},\frac{1}{q} \r)_n
	\left( Q_2 Q_3\frac{t}{q},\frac{1}{q} \r)_n
	\left( \frac{1}{Q_4Q_5} \frac{t}{q},\frac{1}{q} \r)_n
} = \\
&=	\frac{ 
		\left( Q_3\sqtq,\frac{1}{q} \r)_\inf
	\left(Q_1 Q_2 Q_3\sqtq,\frac{1}{q} \r)_\inf
	\left( \frac{1}{Q_4}  \sqtq,\frac{1}{q} \r)_\inf
}
{ 
	\left( Q_2 Q_3\frac{t}{q},\frac{1}{q} \r)_\inf
	\left( \frac{1}{Q_4Q_5} \frac{t}{q},\frac{1}{q} \r)_\inf
} \times \\
&\quad \times P_C \left( Q Q_5^{-1} \sqtq \,,
Q_1 Q_2 Q_3\sqtq\,,  Q_4^{-1}\sqtq\,, Q_2Q_3 \frac{\sqrt{t}}{q} \,,  Q_4^{-1}Q_5^{-1} \frac{\sqrt{t}}{q}
\right)  
\end{split}
\end{align}
with a  quiver matrix in the representation (\ref{quivergen})
\begin{align}
C = \left[ 
\begin{array}{cccccc}
0 &1  &1 &1&1 & 1 \\
1 & 0 &0 & 0&0 &0 \\
1 &0  &0 &0&0 &0 \\
1 &0&0& 0&0 &0\\
1 &0&0& 0&1&0\\
1 &0&0& 0&0 &1\\
\end{array}
\right] .
\end{align}
The fact that this partition function can be presented in the quiver form automatically assures that corresponding refined open BPS numbers are non-negative integers, as we discussed earlier. We also note that the above result is a special case of (\ref{tbranehypergeometry}); to see this directly one can make a redefinition
$\tilde{Q}_4=\frac{1}{Q_4},\tilde{Q}_5=\frac{1}{Q_5},\tilde{Q}=\frac{Q}{Q_5} \sqtq$, which turns the above partition function into the form
\begin{align}
Z_{\tbrane}=\sum\limits_{n=0}^{\inf}  
\frac{ \left(\tilde{Q} \sqtq\r)^n \left( Q_3\sqtq,\frac{1}{q} \r)_n 
	\left(Q_1 Q_2 Q_3\sqtq,\frac{1}{q} \r)_n
	\left( \tilde{Q}_4  \sqtq,\frac{1}{q} \r)_n
}
{    \left( \frac{1}{q},\frac{1}{q} \r)_n
	\left( Q_2 Q_3\frac{t}{q},\frac{1}{q} \r)_n
	\left( \tilde{Q}_4\tilde{Q}_5 \frac{t}{q},\frac{1}{q} \r)_n
} .
\end{align}


\section{Closed topological vertex geometry and Hanany-Witten transitions}   \label{sec-T2}

\begin{figure}[h]
	\centering
	\includegraphics[width=2.in]{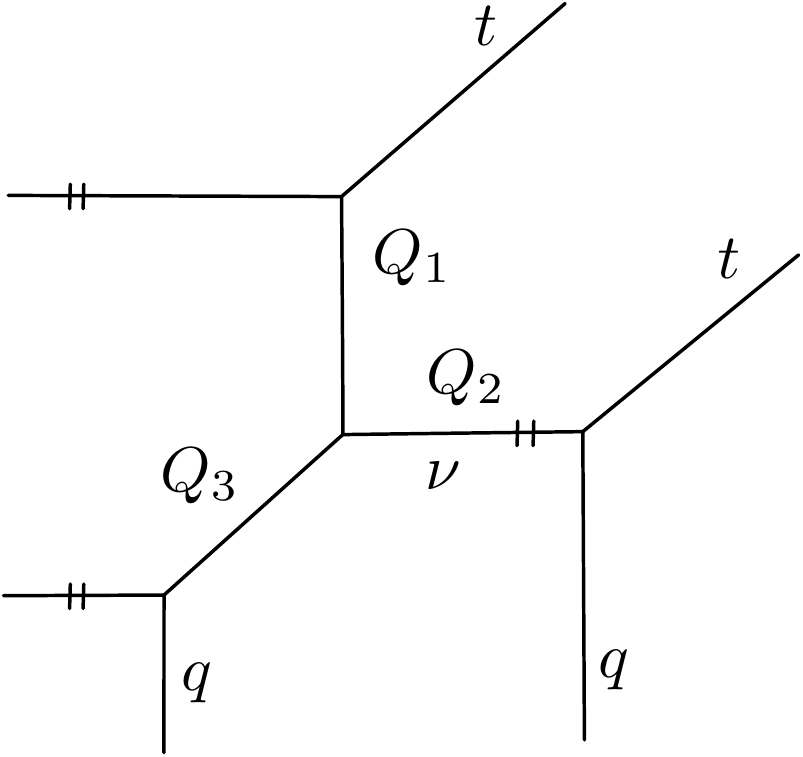}
	\caption{Closed topological vertex geometry ($T_2$-geometry).}
	\label{fig:T2geom}
\end{figure}

In this section we consider another important example of a Calabi-Yau threefold without compact four-cycles, which is called the closed topological vertex or $T_2$-geometry \cite{Bryan:2003yd,Sulkowski:2006jp,Kozcaz:2010af,Brini:2014fea}, see fig. \ref{fig:T2geom}. It can be identified as a resolution of $\mathbb{C}^3/\mathbb{Z}_2\times\mathbb{Z}_2$; its geometry includes three local $\mathbb{P}^1$'s that meet in one point. $T_2$-geometry is a particular example of $T_N$-geometries that engineer non-lagrangian theories \cite{Benini:2009gi,Hayashi:2013qwa} -- for this reason, analysis of all these geometries is particularly important. Some unrefined open amplitudes for the closed topological vertex were computed in \cite{Takasaki:2015raa}.

One aim of this section is to show that refined open BPS invariants for $T_2$-geometry are non-negative integers. As we will show, this statement follows from a simple relation of the open partition function of $T_2$-geometry to the partition function of a triple-$\mathbb{P}^1$ geometry discussed in section \ref{ssec-3P1}. Using this relation one can also easily write the open partition function for a brane in $T_2$-geometry in a quiver form, which implies that all corresponding refined open BPS numbers are non-negative integers.

Another aim of this section is to confirm consistency of refined topological string theory in processes that involve Hanany-Witten transitions. After reinterpreting toric diagrams as webs of five-branes \cite{Hanany:1996ie,Benini:2009gi} we introduce D7-branes that can move around a given diagram. When such a D7-brane crosses a five-brane (some leg of a toric diagram), some other five-branes can be created or annihilated, which is the process referred to as Hanany-Witten transition. On the level of topological strings, partition functions before and after such a transition should be the same, possibly up to some simple factors that represent some extra strings that appear in such processes. In this section we verify that this is indeed the case for refined partition functions for various threefolds that involve $T_2$-geometry. Analysis of Hannay-Witten transitions in the context of closed topological strings has been conducted before, e.g. in \cite{Hayashi:2013qwa,Cheng:2018aa}. On the other hand, the configurations that we consider in this section involve additional lagrangian branes -- to our knowledge, apart from an example in the unrefined case in \cite{Kim:2020npz}, Hanany-Witten transitions in such systems have not been analyzed before. We show that open partition functions for such systems posses the expected properties, which is yet another confirmation of consistency of refined topological strings. 

Furthermore, yet another interesting phenomenon that we discuss in the context of refined open amplitudes for threefolds that involve $T_2$-geometry is $T_2$-tuning. Our results generalize the earlier analysis of this phenomenon that involved only closed topological strings \cite{Cheng:2018aa}.

\begin{figure}
	\centering
	\includegraphics[width=4in]{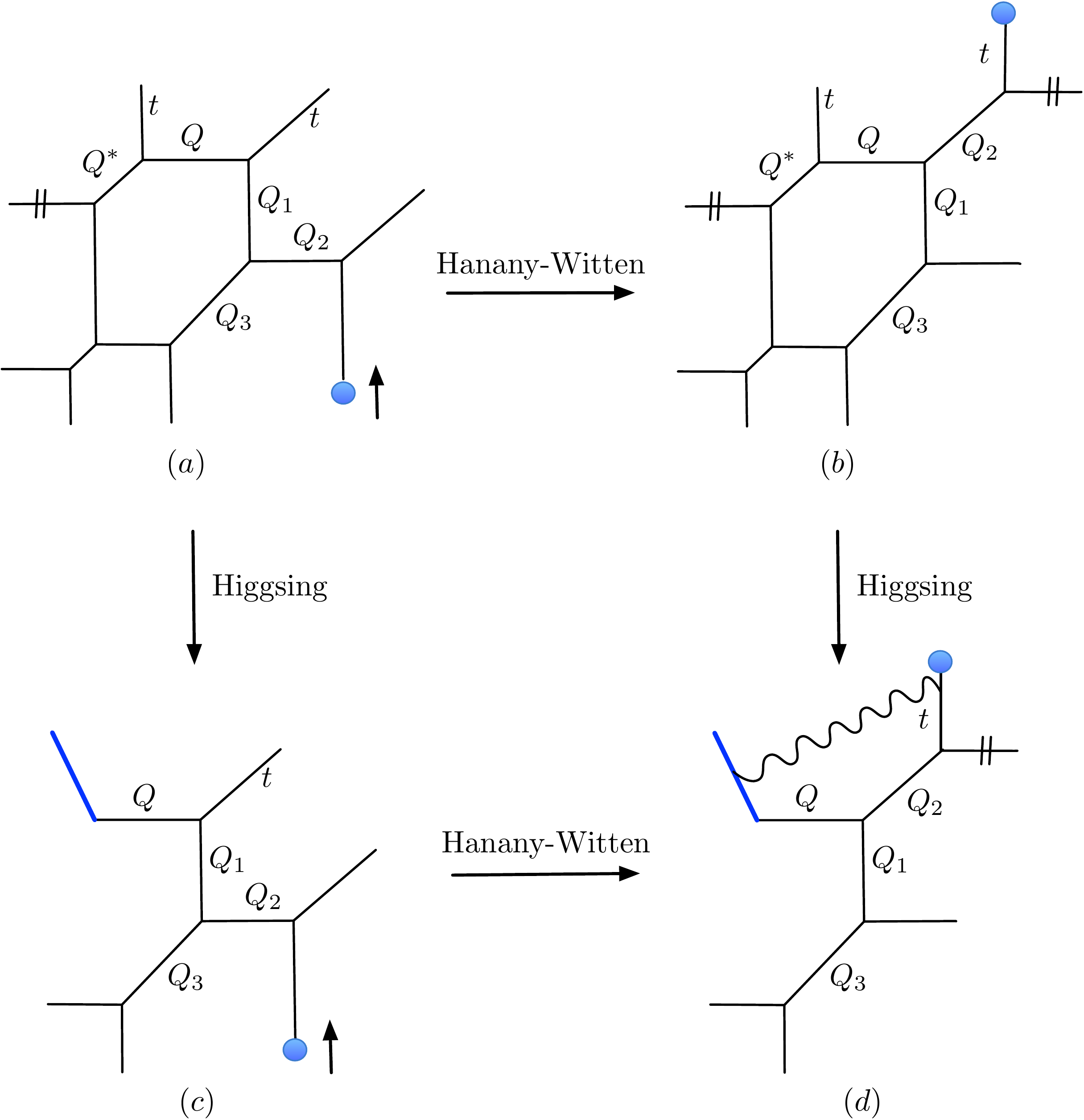}
	\caption{Geometric transitions (Higgsing) and Hanany-Witten transitions that relate the closed topological vertex geometry ($T_2$-geometry) with a lagrangian brane (diagram $(c)$) to other Calabi-Yau threefolds.}
	\label{fig:T2HW}
\end{figure}

To start with, we show how various geometric transitions and Hanany-Witten transitions, shown in fig. \ref{fig:T2HW}, enable to compute a partition function for the closed topological vertex with a brane, and relate it to a triple-$\mathbb{P}^1$ geometry analyzed in section \ref{ssec-3P1}. The closed topological vertex geometry with one lagrangian brane (represented by a blue segment) is shown in diagram $(c)$. This geometry can be obtained by Higgsing (geometric transition) the diagram $(a)$. Furthermore, we can make the Hanany-Witten transition in these two geometries, which we interpret as introducing a D7-brane from infinity from the bottom (as represented by a blue dot) and moving it up. When applied to the $T_2$-geometry, this operation produces a triple-$\mathbb{P}^1$ geometry with extra open strings (diagram $(d)$). This is why (open) partition functions for triple-$\mathbb{P}^1$ and $T_2$-geometry are simply related. Furthermore, this triple-$\mathbb{P}^1$ geometry can be obtained by Higgsing yet another geometry $(b)$ that completes the whole diagram; the geometry in diagram $(b)$ itself can be obtained from the geometry $(a)$ upon the  Hanany-Witten transition.

We stress that geometric transitions and Hanany-Witten transitions in fig. \ref{fig:T2HW}, and in the rest of this section, have very different status. Geometric transitions enable us to determine explicitly open partition functions of our interest, upon specialization of certain K\"{a}hler parameters for appropriately engineered more complex geometries (whose partition functions can be computed using refined topological vertex without difficulties that arise for open amplitudes). For example, partition functions for diagrams $(c)$ and $(d)$ in fig. \ref{fig:T2HW} arise respectively from diagrams $(a)$ and $(b)$ upon geometric transitions. On the other hand, Hanany-Witten transitions predict that partition functions for certain threefolds should be related, however we do not have direct means how to represent such transitions quantitatively on the level of topological strings. For this reason we calculate partition functions before and after the Hanany-Witten transition independently (e.g. using geometric transitions) and check whether that they agree. We find that they indeed do, which is a non-trivial manifestation of Hanany-Witten transition.

Let us discuss various elements of fig. \ref{fig:T2HW} quantitatively. Partition function for a brane in $T_2$-geometry represented by the diagram $(c)$ (ignoring here the blue circle, as well as overall contributions of the form (\ref{Z^M-factor})) takes form
\begin{align}\label{T2partfuns}
\begin{split}
&Z^{T_2}_{\text{brane}}=
\sum\limits_{\u,\v   }  
(-1)^{|\u|+|\v|} q^{ \frac{ ||\u||^2+||\v||^2 }{2}}  Q^{ |\u|} Q_2^{|\v|}        
||Z_\u(q,t)||^2   ||Z_{\v}(q,t)||^2  \times \\
&\qquad \qquad\qquad
\times \frac{   N^{\half,- } _\u (  Q^*,t^{-1},q^{-1})  
	N^{\half,- } _\v (  Q_3,t^{-1},q^{-1})
	N _{\u\v} \left(  Q_1\sqtq,t^{-1},q^{-1} \r)
}  
{   N ^{\half,-}_\u \left(  Q_1Q_3\sqqt,t^{-1},q^{-1} \r)  }.
\end{split}
\end{align}
Note that this is the parameter $Q$ that plays a role of an open string modulus. This partition function is obtained through the geometric transition represented by the left vertical arrow in fig. \ref{fig:T2HW},  so that  the values of $Q^*$ in the term $N^{\half,- } _\u (  Q^*,t^{-1},q^{-1})$ are adjusted appropriately, following the rules from section \ref{ssec-geomtran}. For $Q^*=t\sqtq$ we get a $t$-brane and for $Q^*=\frac{1}{q}\sqtq$ we get $\bar{q}$-brane, whose partition functions we denote respectively by $Z^{T_2}_{\tbrane}$ and $Z^{T_2}_{\qbarbrane}$. 

On the other hand, consider the partition function for a brane in a triple-$\mathbb{P}^1$ geometry, shown in diagram $(d)$ in fig.  \ref{fig:T2HW}. It can be computed using the geometric transition for the diagram $(b)$ analogously as in section ref. \ref{ssec-3P1}, and for $t$-brane and $\bar{q}$-brane respectively takes form (again, for this result the blue circle does not play a role)
\begin{align}
\begin{split}   \label{HW1}
Z^{3\mathbb{P}^1}_{\tbrane} &=\sum\limits_{n=0}^{\inf} \frac{ ( Q Q_2 t)^n \left(  Q_1 \sqqt,q\r)_n \left( \frac{1}{ Q_2} \sqqt,q \r)_n  }{  (q,q)_n  ( Q_1 Q_3 q/t,q)_n   } =     \\
& = P_{C_t}   \left( q; Q Q_2 t , Q_1 \sqqt, Q_2^{-1} \sqqt, Q_1 Q_3 \frac{\sqrt{q} }{t}    \right) ,\\
Z^{3\mathbb{P}^1}_{\qbarbrane} &= \sum\limits_{n=0}^{\inf} 
\frac{  \left( \frac{ Q} {Q_3 q} \r)^n \left( \frac{1}{Q_1}\sqtq , t\r)_n  \left(  Q_2 \sqtq,t\r)_n
}
{ (t,t)_n   \left( \frac{t}{Q_1Q_3 q},t \r)_n   } = \\
& = P_{C_{\bar{q}} }   \left( t; Q Q_2 \frac{\sqrt{t}}{q} , Q_1 \sqrt{q}, Q_2^{-1} \sqrt{q}, Q_1 Q_3 q  \right), 
\end{split}
\end{align}
where we also provide quiver forms (\ref{quivergen}) (ignoring $Q$-independent prefactors analogous to (\ref{Zextra1}) and (\ref{Zextra2})), for which quiver matrices read
\begin{align}  \label{HW1-quivers}
C_t=\left[ 
\begin{array}{cccc}
0 &1  &1 &1  \\
1& 0 &0 &0  \\
1 &0  &0 & 0 \\
1 &0  &0 & 1 \\
\end{array}
\right]
  \qquad \qquad
  C_{\bar{q} }=\left[ 
  \begin{array}{cccc}
  -1 &-1  &-1 &-1  \\
 -1& 1 &0 &0  \\
  -1 &0  &1 & 0 \\
  1 &0  &0 & 0\\
  \end{array}
  \right] 
\end{align}
We can now compare partition functions in (\ref{T2partfuns}) with (\ref{HW1}). We find order by order that they agree up to a simple factor
\begin{align}    \label{T2partfun}
Z^{T_2}_{\tbrane} =\frac{Z^{3\mathbb{P}^1}_{\tbrane} }{ (Q Q_2t,q)^{-1}_{\inf}},\qquad  \quad Z^{T_2}_{\qbarbrane} =\frac{Z^{3\mathbb{P}^1}_{\qbarbrane} }{ (Q Q_2\frac{t}{q},t)_{\inf}}.
\end{align}
The factor $(Q Q_2t,q)^{-1}_{\inf}=  (Q Q_2\frac{t}{q},\frac{1}{q})_{\inf} = \PE[Q Q_2, 1, 1/2,1/2 ]_{\tbrane}$ in the denominator for the $t$-brane encodes a single open BPS invariant $N_{QQ_2}^{(1/2,1/2)}=1$, and represents open strings of length $QQ_2$, denoted in diagram $(d)$ by a wavy line. The factor $(Q Q_2\frac{t}{q},t)_{\inf}$ for the $\bar{q}$-brane has analogous interpretation. The open strings corresponding to these factors can be interpreted as arising during the Hanany-Witten transition represented by the bottom arrow in fig. \ref{fig:T2HW}. Altogether, the relations (\ref{T2partfun}) confirm that refined topological strings are consistent with Hanany-Witten transitions. Note that in the unrefined case, an analogous relation between open amplitudes for the closed topological vertex and a triple-$\mathbb{P}^1$ geometry was observed in \cite{Takasaki:2015raa}.

Furthermore, the relations (\ref{T2partfun}) immediately imply that partition functions for branes in the closed topological vertex geometry can be also presented in the quiver form -- in this case quivers look like the quivers for triple-$\mathbb{P}^1$ geometry in (\ref{HW1-quivers}), and in addition have one extra node (disconnected from other nodes) that represents the denominators on the right side of (\ref{T2partfun}). It follows that the quiver forms (\ref{quivergen}) of the open partition functions for branes in the closed topological vertex (again ignoring $Q$-independent prefactors analogous to (\ref{Zextra1}) and (\ref{Zextra2})) read
\begin{align}
\begin{split}
Z^{T_2}_{\tbrane} &  = P_{C_t}   \left( q; Q Q_2\frac{t}{\sqrt{q}} ,   Q Q_2 t , Q_1 \sqqt, Q_2^{-1} \sqqt, Q_1 Q_3 \frac{\sqrt{q} }{t}    \right) , \\
Z^{T_2}_{\qbarbrane} &=  P_{C_{\bar{q} }}   \left( t ; Q Q_2\frac{t}{q} , Q Q_2 \frac{\sqrt{t}}{q} , Q_1 \sqrt{q}, Q_2^{-1} \sqrt{q}, Q_1 Q_3 q    \right),
\end{split}
\end{align}
with quiver matrices
\begin{align}
C_t=\left[ 
\begin{array}{ccccc}
1 &0 &0 &0 &0  \\
0&0 &1  &1 &1  \\
0&1& 0 &0 &0  \\
0&1 &0  &0 & 0 \\
0&1 &0  &0 & 1 \\
\end{array}
\right] \qquad \quad
 C_{\bar{q}}=\left[ 
\begin{array}{ccccc}
0 &0 &0 &0 &0  \\
0&1 &-1  &-1 &-1  \\
0&-1& 1 &0 &0  \\
0&-1 &0  &1 & 0 \\
0&-1 &0  &0 & 0 \\
\end{array}
\right]
\end{align}
As motivic Donaldson-Thomas invariants for any symmetric quiver are non-negative integers, the above relation also immediately implies that all refined open BPS invariants for closed topological vertex are non-negative integers.

\bigskip

\begin{figure}
	\centering
	\includegraphics[width=6in]{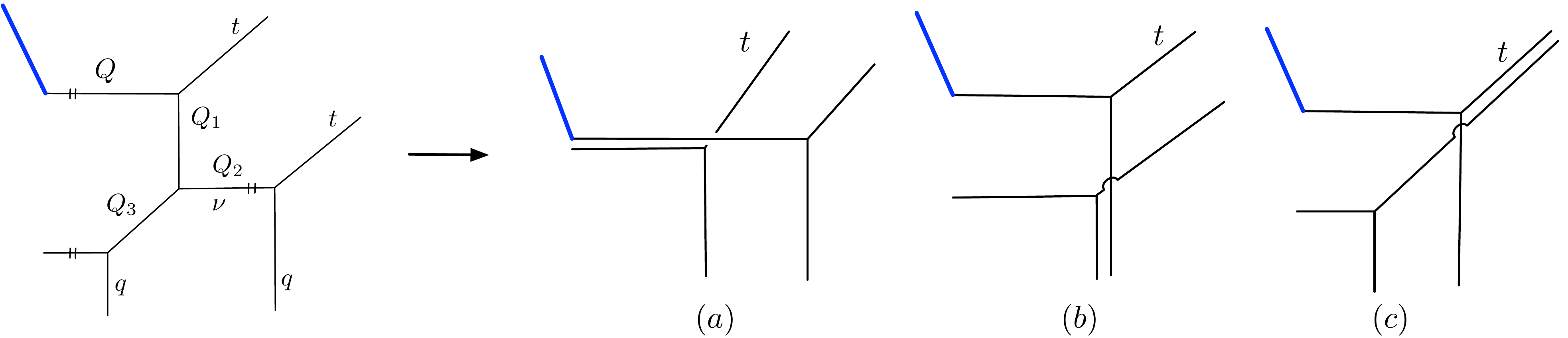}
	\caption{$T_2$-tuning with a lagrangian brane: tuning appropriately K\"{a}hler parameters of $T_2$-geometry reduces it to two copies of $\mathbb{C}^3$ in a non-toric configuration (i.e. with two of external legs intersecting).}
	\label{fig:T2higgsing}
\end{figure}

Having determined the partition function for a brane in the closed topological vertex geometry, let us discuss some of its properties. One of them is referred to as $T_2$-tuning, which for closed strings was discussed in \cite{Cheng:2018aa}. In the context of closed strings this the statement that after adjusting two of the three K\"ahler parameters of the closed topological vertex, either as $Q_1=Q_3=\sqtq$ or $\sqqt$, or $Q_2=Q_3=\sqtq$,  or $Q_1=Q_2=\sqqt$, appropriate two parallel external legs of $T_2$-geometry overlap, and the geometry itself reduces effectively to two copies of $\mathbb{C}^3$ in a non-toric configuration (i.e. which involves an intersection of toric legs). Such a process is shown in fig. \ref{fig:T2higgsing} if we ignore the lagrangian brane. Note that for closed strings, in case $(a)$, $Q_1$ and $Q_3$ can be fixed to two different values $\sqtq$ or $\sqqt$; this appears to be a feature of two legs overlapping along the preferred direction.

Let us consider now $T_2$-tuning for $T_2$-geometry with a lagrangian brane, as in fig. \ref{fig:T2higgsing}. Similarly as in the closed string case, in situations $(a)$, $(b)$ and $(c)$ we set respectively $Q_1=Q_3=\sqtq$, or $Q_2=Q_3=\sqtq$,  or $Q_1=Q_2=\sqqt$. Note that now the presence of a brane breaks the symmetry, so in case $(a)$ the parameters $Q_1$ and $Q_3$ can be fixed to only one value $\sqtq$. Similarly as in the closed string case, once we fix values of two K\"ahler parameters as above, the dependence on third parameter drops out. 
For example, fixing $Q_1=Q_3=\sqtq$ in the partition function (\ref{T2partfuns}), it follows from (\ref{constraint1}) that $\nu=\emptyset$ and hence $Q_2^{|\v|}=1$. Ultimately, in all configurations $(a), (b)$ and $(c)$ the $\tbrane$ partition function (\ref{T2partfuns}) reduces to $(Q\sqrt{q t},q)_\inf$, which is indeed the same as $\tbrane$ partition function in $\mathbb{C}^3$ (ignoring overall closed string contributions of the form (\ref{Z^M-factor})). This shows that the phenomenon of $T_2$-tuning arises consistently in the open string case too.

\bigskip

\begin{figure}
	\centering
	\includegraphics[width=4in]{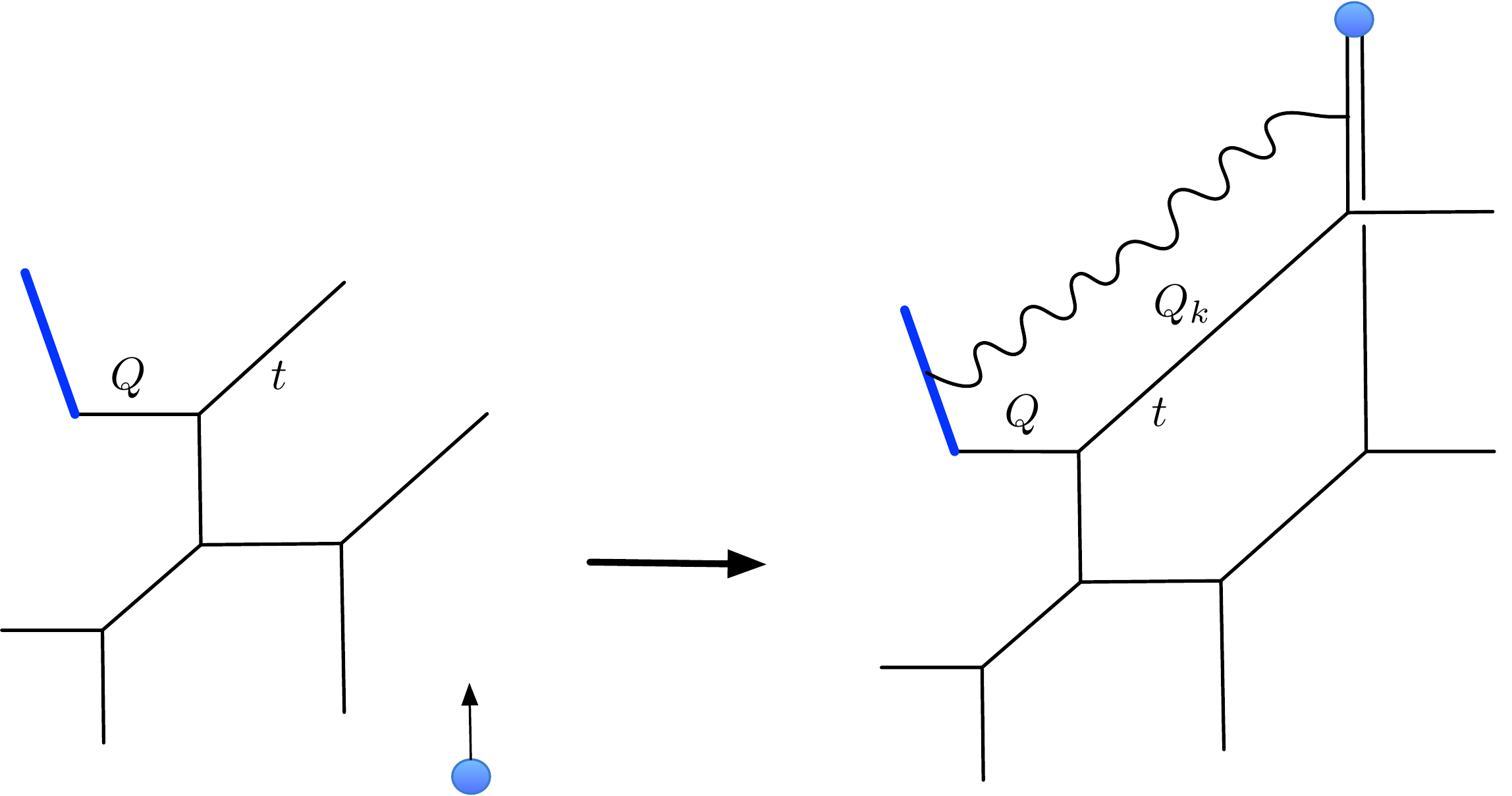}
	\caption{Hanany-Witten transition that moves a D7-brane (denoted by a blue dot) from infinity to the $T_2$-geometry results in a non-toric structure with one extra set of open strings of length $Q Q_k$.}
	\label{fig:HWexample1}
\end{figure}

\begin{figure}[ht]
	\centering
	\includegraphics[width=5.5in]{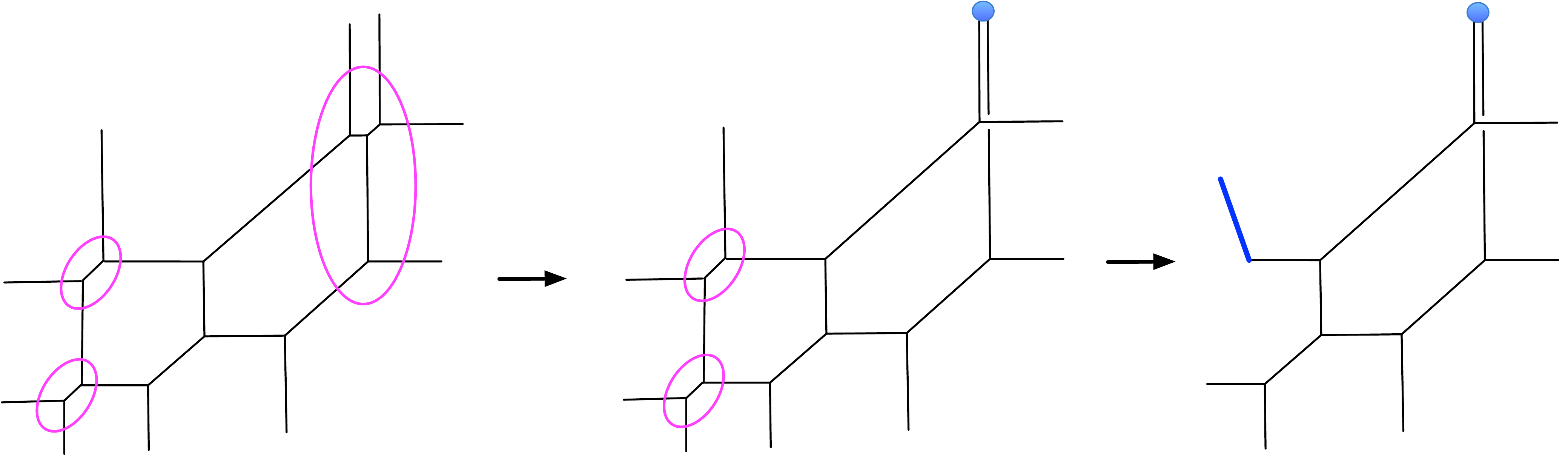}
	\caption{$T_2$-tuning (in the first step) followed by a geometric transition on two local $\mathbb{P}^1$'s (in the second step) engineers the geometry with a lagrangian brane shown in fig. \ref{fig:HWexample1} (right).}  \label{T2oneD7up}
\end{figure}

Finally, let us consider more complicated examples of Hanany-Witten transitions in $T_2$-geometry and related systems. First, consider the process in fig. \ref{fig:HWexample1}, where moving a D7-brane from infinity to the $T_2$-geometry (left) results in a non-toric structure that involves one extra set of open strings (right). The partition function for the $T_2$-geometry (left) is given in (\ref{T2partfuns}). To determine the partition function for the geometry on the right we engineer yet more complicated threefold shown in fig. \ref{T2oneD7up} (left), compute its partition function using refined topological vertex, and then fix appropriate parameters according to the rules of $T_2$-tuning, and fix other parameters in the process of geometric transition that introduces a lagrangian brane. This process produces the geometry shown in fig. \ref{T2oneD7up} (right), which is the same as in \ref{fig:HWexample1} (right), and whose partition function we denote by $Z^{\text{right}}_{\tbrane}$. Then, comparing partition functions of both geometries in fig. \ref{fig:HWexample1} we find
\begin{align}
&Z^{T_2}_{\tbrane} =\frac{Z^{\text{right}}_{\tbrane} }{ (Q Q_k t,q)^{-1}_{\inf}},
\end{align}
where the factor $(Q Q_k t,q)_{\inf}^{-1}  = \PE[Q Q_k, 1, 1/2,1/2 ]_{\tbrane}$ represents extra open strings of length $QQ_k$ that are produced during the Hanany-Witten transition. The above relation confirms that refined topological strings are consistent with Hanany-Witten transitions.

\begin{figure}
	\centering
	\includegraphics[width=4in]{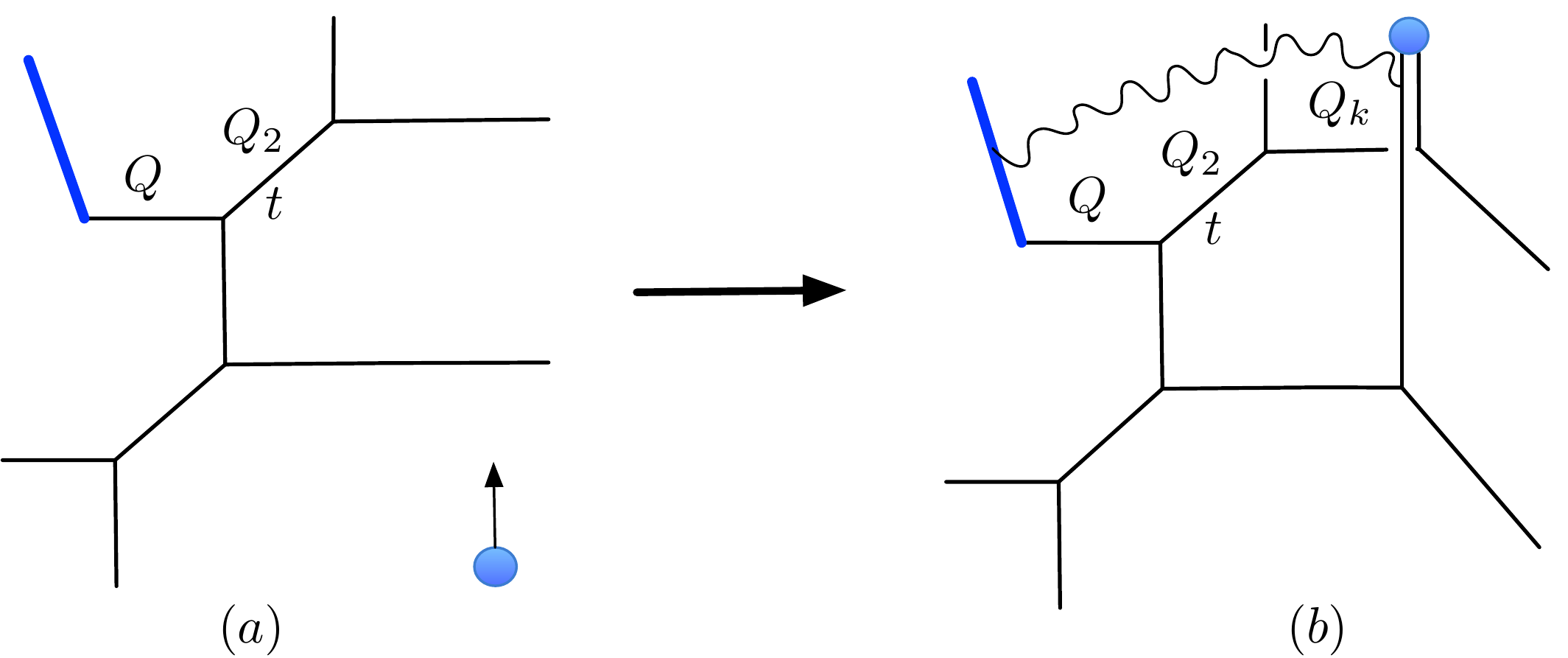}
	\caption{Hanany-Witten transition that moves a D7-brane (denoted by a blue dot) from infinity to the triple-$\mathbb{P}^1$ geometry results in a non-toric structure with one extra set of open strings of length $Q Q_2 Q_k$.}
	\label{fig:HWexample2}
\end{figure}

An analogous example, however involving the triple-$\mathbb{P}^1$ geometry, is shown in fig. \ref{fig:HWexample2}. We have computed the partition function $Z^{3\mathbb{P}^1}_{\tbrane}$ for a $t$-brane in triple-$\mathbb{P}^1$ geometry in (\ref{HW1}). We determine the partition function $Z^{(b)}_{\tbrane} $ for the geometry shown in \ref{fig:HWexample2} in diagram $(b)$ following fig. \ref{3P1oneD7up}, and again applying $T_2$-tuning and geometric transition to an appropriately engineered more complicated geometry. We then find that
\begin{align}
Z^{3\mathbb{P}^1}_{\tbrane}=\frac{Z^{(b)}_{\tbrane}   } {(Q Q_2 Q_k t^2 q^{-1},q  )_\inf^{-1}   },
\end{align}
where the denominator $(Q Q_2 Q_k t^2 q^{-1},q  )_\inf^{-1}   =  \PE[Q Q_2 Q_k, 1, 3/2,3/2 ]_{\tbrane}$ represents open strings of length $Q Q_2 Q_k$ introduced by the Hanany-Witten transition in fig. \ref{fig:HWexample2}.

\begin{figure}[ht]
	\centering
	\includegraphics[width=5.5in]{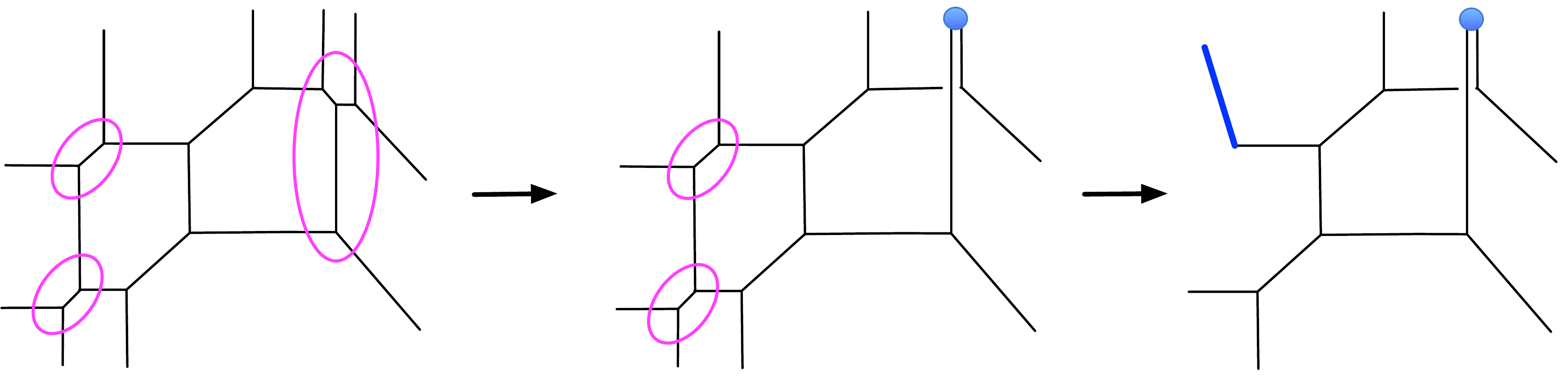}
	\caption{$T_2$-tuning (in the first step) followed by a geometric transition on two local $\mathbb{P}^1$'s (in the second step) engineers the geometry with a lagrangian brane shown in fig. \ref{fig:HWexample2} (right).}  \label{3P1oneD7up}
\end{figure}

\begin{figure}
	\centering
	\includegraphics[width=3.5in]{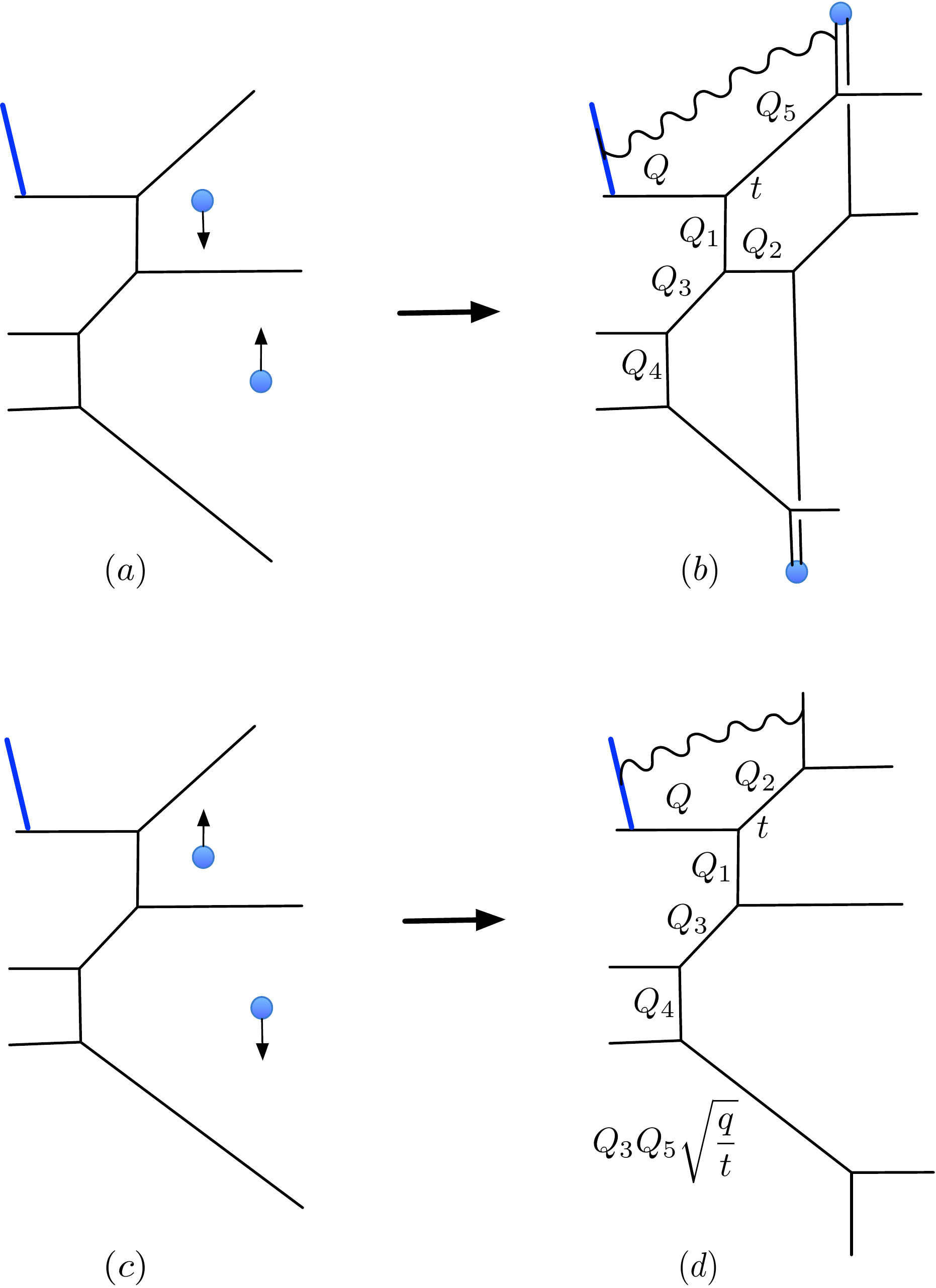}
	\caption{Hanany-Witten transition that involves moving two D7-branes to infinity, and as usual produces extra open strings. The shift $\sqqt$ in $Q_3 Q_5 \sqqt$ is caused by Higgsing.}
	\label{fig:HWexample3}
\end{figure}

\begin{figure}[ht]
	\centering
	\includegraphics[width=5in]{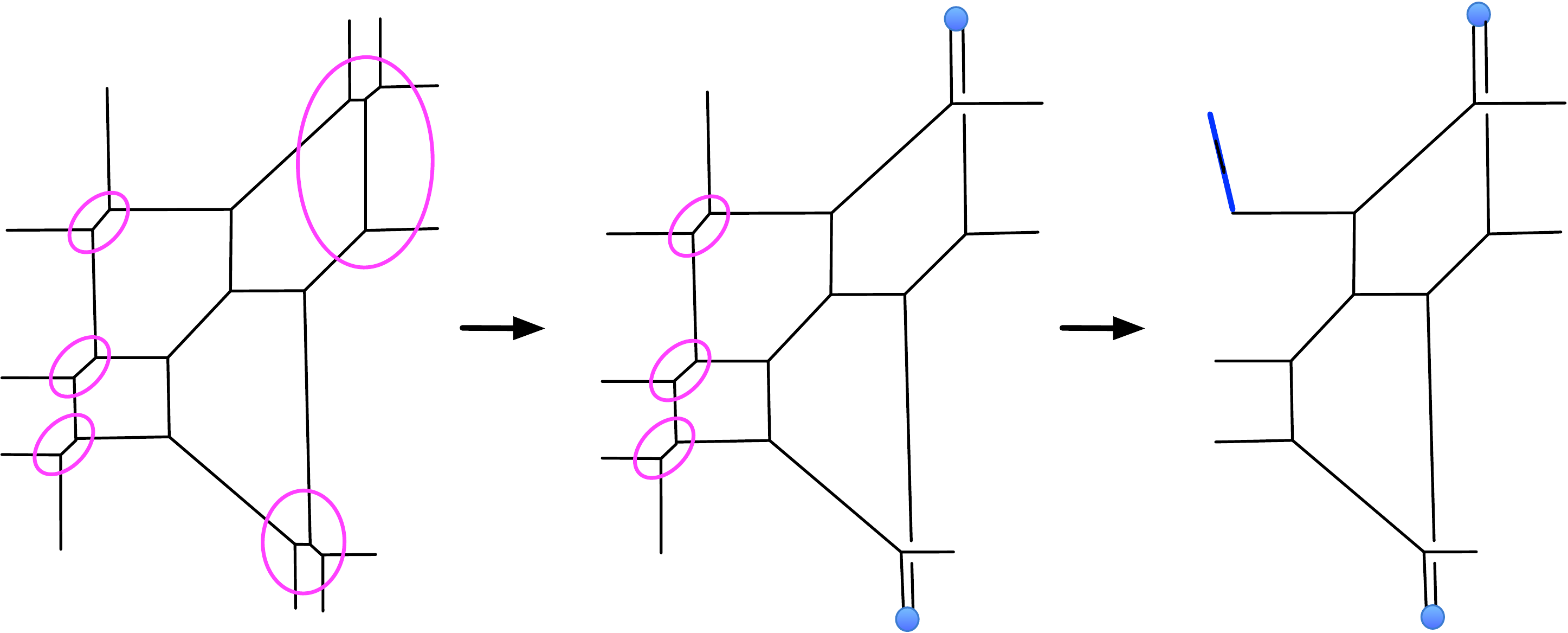}
	\caption{$T_2$-tuning on two pairs of legs (in the first step) followed by a geometric transition on three local $\mathbb{P}^1$'s (in the second step) engineers the geometry with a lagrangian brane shown in fig. \ref{fig:HWexample3} $(b)$.}    \label{3p12D7b}
\end{figure}

\begin{figure}[ht]
	\centering
	\includegraphics[width=3in]{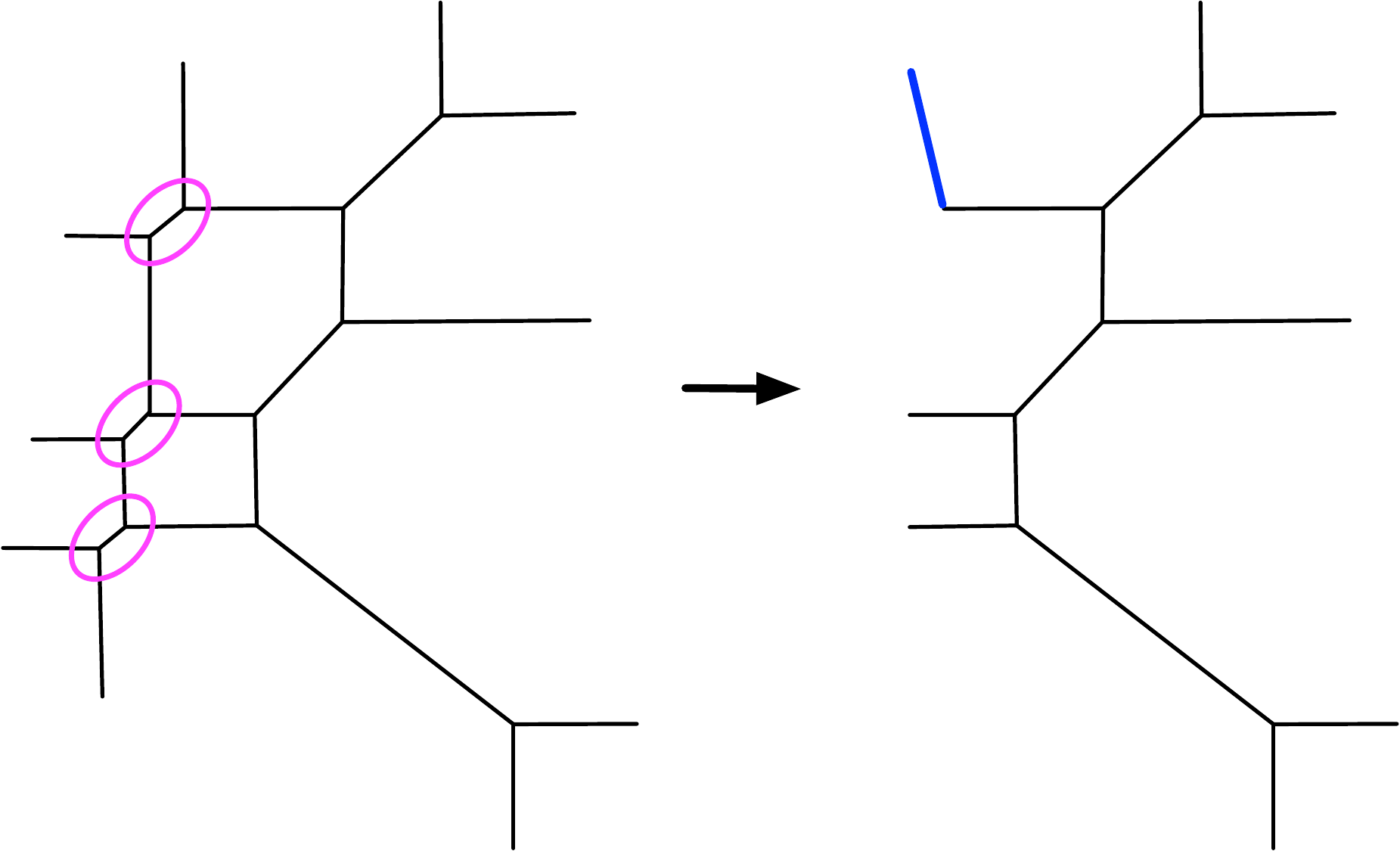}
	\caption{	
	Geometric transition on three local $\mathbb{P}^1$'s engineers the geometry with a lagrangian brane shown in fig. \ref{fig:HWexample3} $(d)$.}    \label{3p12D7d}
\end{figure}

Furthermore, let us consider a more complicated process, where anti-fundamental chiral hypermultiplets are introduced by sandwiching D7-branes between 5-branes, see fig. \ref{fig:HWexample3}. We choose either to move one D7-brane down to infinity and another D7-brane up to infinity, see diagram $(a)$ in fig. \ref{fig:HWexample3}, or to move these branes in opposite directions, see diagram $(c)$. These operations result respectively in geometries represented by diagrams $(b)$ and $(d)$. We expect that their partition functions should be equal, up to some factors representing open strings that arise in these different Hanay-Witten transitions. To verify this claim, we compute refined partition functions for diagrams $(b)$ and $(d)$ following respectively fig. \ref{3p12D7b} and fig. \ref{3p12D7d}. Denoting these partition functions by $Z^{(b)}_{\tbrane}$ and $Z^{(d)}_{\tbrane}$, we then find
\begin{align}
\frac{Z^{(b)}_{\tbrane}}{(Q Q_5 t,q)_\inf^{-1}   }   =\frac{
	Z^{(d)}_{\tbrane}
} { (Q Q_2 t,q)_\inf^{-1}  }.
\end{align}
This indeed the expected results, where extra open strings arising during Hanany-Witten transitions are are represented by $(Q Q_5 t,q)_\inf^{-1} =\PE[Q Q_5, 1, 1/2,1/2 ]_{\tbrane}$ and $(Q Q_2 t,q)_\inf^{-1} = \PE[Q Q_2, 1, 1/2,1/2 ]_{\tbrane}$. This result again confirms consistency of refined topological strings with Hanany-Witten transitions.

\section{Toric manifolds with compact four-cycles}   \label{sec-4}

In this section we consider examples of toric manifolds with compact four-cycles --  we focus on lagrangian branes in Hirzebruch surfaces with blown-up points. As before, our primary aim is to show that in this case refined open BPS degeneracies are also non-negative integers. To determine relevant partition functions that encode these BPS degeneracies, at the same time we further develop the method of geometric transition, as these partition functions are difficult to compute by other means. The fact that the resulting BPS numbers are integer assures that our approach based on the geometric transition is correct. 

\begin{figure}[h]
	\centering
	\includegraphics[width=2.5in]{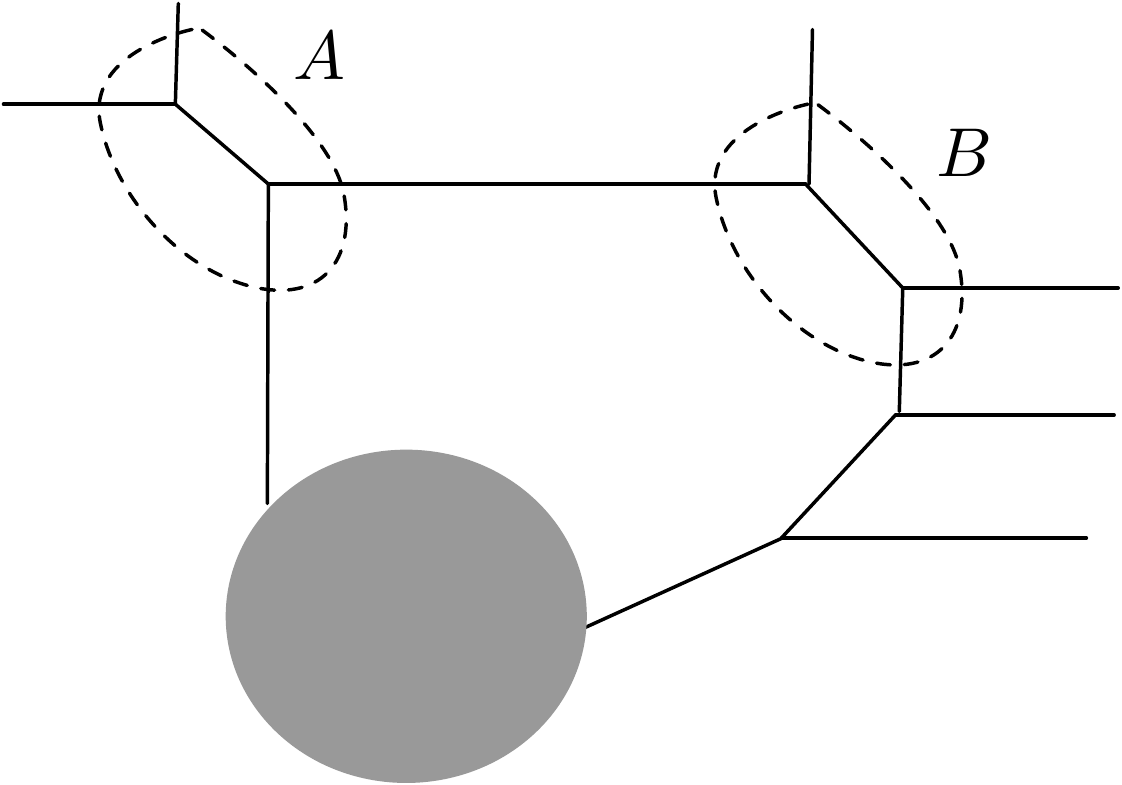}
	\caption{Introducing local $\mathbb{P}^1$'s may produce parallel external legs. In this example, lagrangian branes can be engineered at positions $A$ or $B$ through the geometric transition.}
	\label{fig:higgspara}
\end{figure}

Let us first make the following remark. Our approach seems to work only for 5-branes that are introduced through the geometric transition between external parallel legs. In other situations one can blow up one or more local $\mathbb{P}^1$'s, see fig. \ref{fig:higgspara}, which effectively changes the direction of external legs, and then decouple them at the end of calculations. Engineering carefully such blow-ups may introduce parallel legs, thereby enabling to use our method. This is also a reason why in this section we consider Hirzebruch surfacs with blown-up points.


\begin{figure}
	\centering
	\includegraphics[width=4.5in]{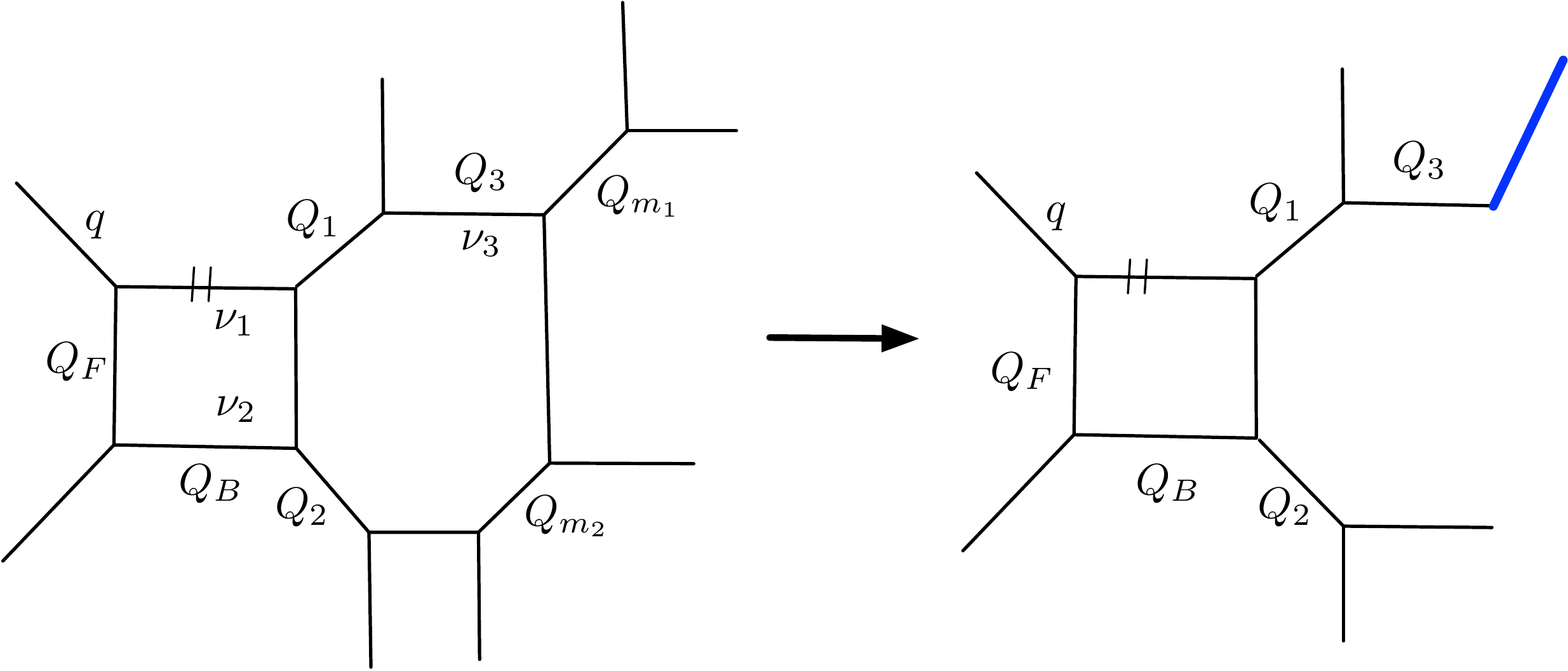}
	\caption{Geometric transition at $Q_{m_1}$ and $Q_{m_2}$ introduces a lagrangian brane at one external leg of $\mathbb{F}_0^2$. One can also switch the values of $Q_{m_1}$ and $Q_{m_2}$ to introduce the lagrangian brane on the bottom external leg. }
	\label{fig:SU2+2F}
\end{figure}

As the first example we consider the Hirzebruch surface $\mathbb{F}_0^2$ with a lagrangian brane. Its toric diagram is shown in fig. \ref{fig:SU2+2F} (right); it can be obtained through the geometric transition of the geometry shown on the left. Recall that the Hirzebruch surface $\mathbb{F}_0^2$ engineers 5-dimensional $\mathcal{N}=1$ $SU(2)$ gauge theory with two flavors, and with a surface defect on an external leg. We just calculate $t$-brane partition function, as other types of branes give rise to analogous results. To this end, in the diagram in fig. \ref{fig:SU2+2F} (left) we need to fix K\"ahler parameters $Q_{m_1}=t\sqtq$ and $Q_{m_2}=\sqtq$. We then find
\begin{align}\label{openclosed1}
\begin{split}
&Z_{\tbrane}^{\text{open-closed}}(\mathbb{F}_0^2)=\sum\limits_{\v_1,\v_2,\v_3} (-1)^{|\v_3|} q^{\frac{||\v_3||^T }{2 }+ ||\v_2^T||^2   }  t^{ \frac{||\v_3||^2 }{2} + ||\v_1^T||^2  }  \times \\
&\qquad \times Q_3^{|\v_3|} Q_B^{|\v_1|+|\v_2|}  ||Z_{\v_1}(q,t  ) ||^2 ||Z_{\v_2}(t,q  ) ||^2 ||Z_{\v_3}(t,q  ) ||^2 \times \\
&\qquad \times
\frac{ N^{\half,-}_{\v_3^T}\left(  \frac{1}{q}\sqtq  \r)  
	N^{\half,+}_{\v_1}\left(   Q_2 Q_F  \r)    
	N^{\half,+}_{\v_2^T}\left(  Q_2  \r)  
	N_{\v_1,\v_3^T}\left(  Q_1\sqtq  \r)  
	N_{\v_2^T,\v_3^T}\left(  Q_1Q_F \sqtq  \r)  
}
{  N^{\half,+}_{\v_3^T} \left( Q_1Q_2Q_F \sqtq \r) 
	N_{\v_2^T,\v_1}\left(  Q_F  \r)  
	N_{\v_2^T,\v_1}\left(  Q_F \frac{t}{q} \r)  
}\,,
\end{split}
\end{align}
where $N_{\u,\v}(Q)$ is a shorthand notation for $N_{\u,\v}(Q, t^{-1},q^{-1})$, and $Q_3$ represents the open K\"ahler parameter. Note that the partition function \eqref{openclosed1} contains also closed string contributions
\begin{align}\label{su22f}
\begin{split}
Z^{\text{closed}}(\mathbb{F}_0^2)=&\sum\limits_{\v_1,\v_2} q^{||\v_2^T||^2} t^{ ||\v_1^T||^2} Q_B^{|\v_1|+|\v_2|} ||Z_{\v_1}(q,t) ||^2 ||Z_{\v_2}(t,q ) ||^2 \times \\
& \times \frac{
	N^{\half,-}_{\v_1}\left(   Q_1  \r)    
	N^{\half,-}_{\v_2^T}\left(  Q_1 Q_F  \r)  
	N^{\half,+}_{\v_1}\left(  Q_2 Q_F  \r)  
	N^{\half,+}_{\v_2^T}\left(  Q_2  \r)  
}
{ 
	N_{\v_2^T,\v_1}\left(  Q_F  \r)  
	N_{\v_2^T,\v_1}\left(  Q_F \frac{t}{q} \r)}.
\end{split}
\end{align}
The $t$-brane open partition function of our interest therefore takes form
\begin{align}
Z_\tbrane(\mathbb{F}_0^2) =\frac{ Z_{\tbrane}^{\text{open-closed}} (\mathbb{F}_0^2) }{Z^{\text{closed}}(\mathbb{F}_0^2)  }\,,
\end{align}
and associated open refined BPS invariants are listed in table \ref{tb:SU2+2F}. They are indeed non-negative integers, as expected. 

\begin{table}[h!]
	{\small
		\begin{tabular}{|l|c|c |}\hline
			$(d_1,d_3,d_B,d_F)$ & $2 r\setminus2s $&0  \\\hline
			$(0,1,0,0)$&0&1
			\\	\hline 
		\end{tabular}
		\begin{tabular}{|l|c|c|}\hline
			$(d_1,d_3,d_B,d_F)$ & $2 r\setminus2s $&1  \\\hline
			$(1,1,0,0)$&1&1
			\\	\hline 
		\end{tabular}
		\begin{tabular}{|l|c| c|}\hline
			$(d_1,d_3,d_B,d_F)$ & $2 r\setminus2s $&1  \\\hline
			$(1,1,1,0),(1,1,0,1)$&1&1
			\\	\hline 
		\end{tabular}
		\begin{tabular}{|l|c| c  c c|}\hline
			$(d_1,d_2,d_3,d_B,d_F)$ & $2 r\setminus2s$&-1&1&3 \\\hline
			$(2,0,1,0,1)$&1&&1&\\ \hline
			$(1,0,1,1,1)$&-1&1&&\\
			&1&&1&\\
			&3&&&1
			\\	\hline 
		\end{tabular}
		\begin{tabular}{|l|c|cccccc|}\hline
			$(d_1,d_2,d_3,d_B,d_F)$ & $2 r\setminus2s$& -3&-1&0&1&3&5   \\\hline
			$(2,0,1,1,1)$&0&&&1&&&\\
			&1&&&&1&&\\ \hline
			$(2,1,1,0,1),(2,0,2,1,0)$&3&&&&&1&
			\\\hline
			$(1,1,1,1,1)$&0&&&1&&&   \\
			&1&&&&1&&    \\\hline
			$(1,0,1,1,2),(1,0,1,2,1)$&-3&1&&&&&\\
			&-1&&1&&&&\\
			&1&&&&1&& \\
			&3&&&&&1&  \\
			&5&&&&&&1
			\\	\hline 
		\end{tabular}
		\begin{tabular}{|l|c|ccccccccccccccccc|}\hline
			$(d_1,d_2,d_3,d_B,d_F)$ & $2 r\setminus2s$& -7&-6&-5&-4&-3&-2&-1&0&1&2&3&4&5&6&7&8&9  
			\\\hline
			$(2,1,1,1,1)$&1&&&&&&&&&1&&&&&&&&		\\
			$(2,0,2,1,1)$ &3&&&&&&&&&&&1&&&&&&\\\hline
			$(2,1,1,0,2)$&3&&&&&&&&&&&1&&&&&&\\ \hline
			$(3,0,2,0,1)$&4&&&&&&&&&&&&1&&&&&\\
			$(2,1,2,0,1)$&&&&&&&&&&&&&&&&&&\\\hline
			$(2,0,1,1,2)$&-2&&&&&&1&&&&&&&&&&&\\
			$(2,0,1,2,1)$	&0&&&&&&&&1&&&&&&&&&\\
			$(1,1,1,1,2 )$	&2&&&&&&&&&&1&&&&&&&\\
			$(1,1,1,2,1 )$	&4&&&&&&&&&&&&1&&&&&\\\hline
			$(1,0,1,1,3)$	&-5&&&1&&&&&&&&&&&&&&\\
			$(1,0,1,3,1 )$&-3&&&&&1&&&&&&&&&&&&\\
			&-1&&&&&&&1&&&&&&&&&&\\
			&1&&&&&&&&&1&&&&&&&&\\
			&3&&&&&&&&&&&1&&&&&&\\
			&5&&&&&&&&&&&&&1&&&&\\
			&7&&&&&&&&&&&&&&&1&&\\\hline
			$(1,0,1,2,2)$	&-7&&&1&&&&&&&&&&&&&&\\
			&-5&1&&2&&1&&&&&&&&&&&&\\
			&-3&&&1&&3&&1&&&&&&&&&&\\
			&-1&&&&&1&&3&&1&&&&&&&&\\
			&1 &&&&&&&1&&3&&1&&&&&&\\
			&3 &&&&&&&&&1&&3&&1&&&&\\
			&5 &&&&&&&&&&&1&&3&&1&&\\
			&7 &&&&&&&&&&&&&1&&2&&1\\
			&9 &&&&&&&&&&&&&&&1&&
			\\	\hline 
		\end{tabular}
	} 
	\caption{Refined open BPS invariants for Hirzebruch surface $\mathbb{F}_0^2$  with a lagrangian brane; the index $d_3$ is the degree for open K\"{a}hler parameter $Q_3$. }\label{tb:SU2+2F}
\end{table}

\begin{figure}
	\centering
	\includegraphics[width=4.5in]{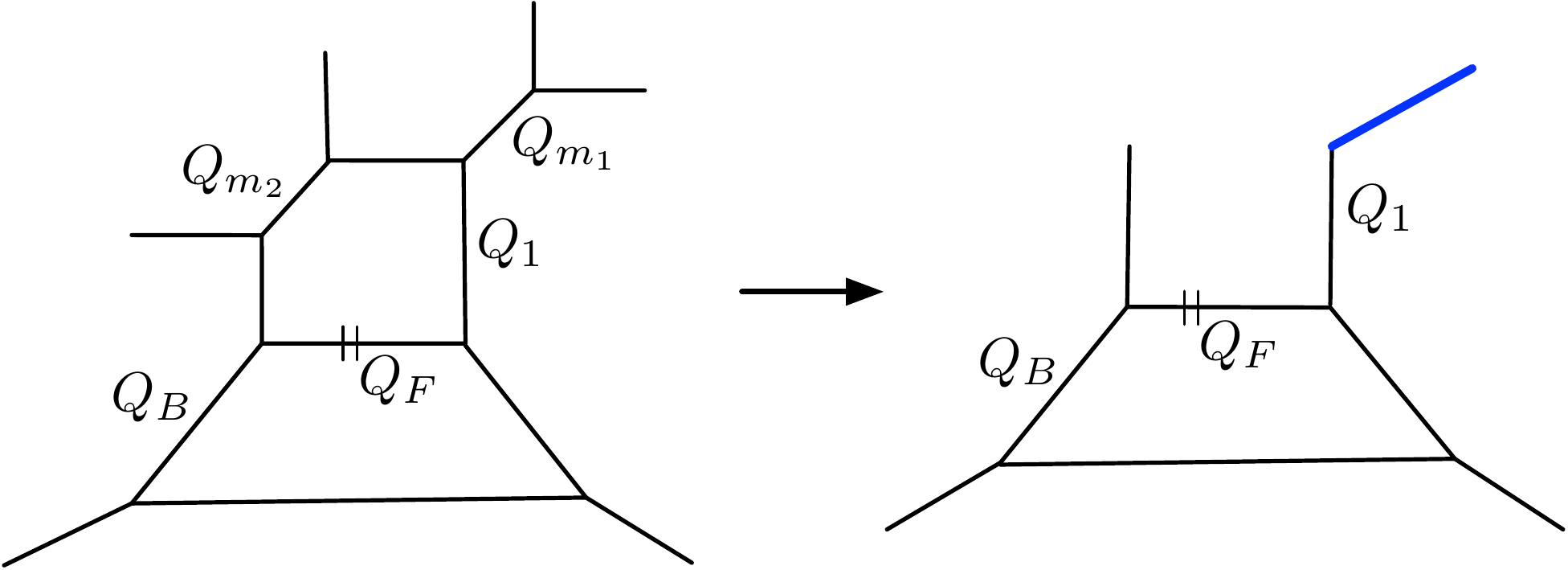}
	\caption{Geometric transition on $Q_{m_1}$ and $Q_{m_2}$ (in the left diagram) produces the Hirzebruch surface $\mathbb{F}_2$ with a lagrangian brane (right).}
	\label{fig:F2}
\end{figure}

As the second example we consider  the Hirzebruch surface $\mathbb{F}_2$ with a lagrangian brane, shown in fig. \ref{fig:F2} (right). It can be obtained through the geometric transition from the manifold whose diagram is shown in the left, upon fixing K\"ahler parameters $Q_{m_1}=t\sqtq$ and $Q_{m_2}=\sqtq$. This way we obtain the open-closed partition function 
\begin{align}
\begin{split}
Z_{\tbrane}^{\text{open-closed}}(\mathbb{F}_2)=&\sum\limits_{\v_1,\v_2,\v_3} q^{ -||\v_3||^2} t^{ ||\v_2||^2+ 2 ||\v_3^T||^2 }  Q_B^{2|\v_3| } Q_F^{ |\v_2|+|\v_3| }
||Z_{\v_2}(t,q)||^2 ||Z_{\v_3}(q,t) ||^2  \times\\
&\times  \frac{  N_{\v_2^T}^{\half,+} \left( Q_1 t \sqtq \r)  
	N_{\v_3}^{\half,+} \left( Q_1 Q_B t \sqtq \r)  
}
{  N_{\v_3}^{\half,+} \left( Q_1 Q_B  \sqtq \r)  
	N_{\v_2^T}^{\half,+} \left( Q_1 \sqtq \r)  
	N_{\v_2^T,\v_3} \left( Q_B t \r)  
	N_{\v_2^T,\v_3} \left( Q_B \frac{t}{q} \r)  
},
\end{split}
\end{align}
which also contains closed string contributions
\begin{align}
Z^{\text{closed}}(\mathbb{F}_2)&= \sum\limits_{\v_2,\v_3} 
\frac{q^{ -||\v_3||^2} t^{||\v_3||^2+2 ||\v_3^T||^2 } Q_B^{ 2 |\v_3|} ||Z_{\v_2}(t,q) ||^2 ||Z_{\v_3}(q,t)||^2   }
{  N_{\v_2^T,\v_3} \left( Q_B t \r)  
	N_{\v_2^T,\v_3} \left( Q_B \frac{t}{q} \r)    
}\,.
\end{align}
Since in this case $Q_1$ is the open K\"{a}hler parameter, it also appears in the prefactor (\ref{Z^M-factor}) 
\begin{align}
Z^{M}(\mathbb{F}_2)=\frac{ M(Q_1 t,q,t) M( Q_1 Q_B t,q,t  )}{ M( Q_1,q,t  ) M(Q_1 Q_B,q,t )  }.
\end{align}
Therefore, the $t$-brane partition function of our interest takes form
\begin{align}
Z_\tbrane (\mathbb{F}_2)=\frac{ Z^{M}  (\mathbb{F}_2) \cdot Z_{\tbrane}^{\text{open-closed}}   (\mathbb{F}_2) }{Z^{\text{closed}}  (\mathbb{F}_2) }\,.
\end{align}
The corresponding refined open BPS invariants (wrapping the compact divisor) are given in table \ref{tb:F2}. They are non-negative integers, as expected. 

\begin{table}[!h]
	\centering
	{\small
		\begin{tabular}{|c|c|ccccccccc|}\hline
			$(d_1,d_B,d_F)$&$2r \backslash 2s$& -7 & -5 &-3 & -1 & 1 & 3 & 5 & 7 & 9 \\
			\hline
			$(1,4,2)$&	-7 &   & 1 &   &   &   &   &   &     & \\
			&	-5 & 1 & 2 & 1 &   &   &   &   &   &    \\
			&	-3 &   & 1 & 4 & 1 &   &   &   &   &  \\
			&	-1 &   &   & 1 & 4 & 1 &   &   &   &   \\
			&	1 &   &   &   & 1 & 4 & 1 &   &   &  \\
			&	3 &   &   &   &   & 1 & 4 & 1 &   &  \\
			&	5 &   &   &   &   &   & 1 & 4 & 1 &   \\
			&	7 &   &   &   &   &   &   & 1 & 3 & 1 \\
			&	9 &   &   &   &   &   &   &   & 1 & 1 \\
			\hline
		\end{tabular}
		\begin{tabular}{|c|c|ccccccccc|}\hline
			$(d_1,d_B,d_F)$&$2r \backslash 2s$& -5 & -3 & -1 & 1 & 3 & 5 & 7 & 9 & 11 \\
			\hline
			$(2,4,2)$&-5 &   & 1 &   &   &   &   &   &   &   \\
			&	-3 & 1 & 2 & 1 &   &   &   &   &   &   \\
			&	-1 &   &   & 5 & 2 &   &   &   &   &   \\
			&	1 &   &   &   & 6 & 2 &   &   &   &   \\
			&	3 &   &   &   &   & 6 & 2 &   &   &   \\
			&	5 &   &   &   &   &   & 6 & 2 &   &   \\
			&	7 &   &   &   &   &   &   & 5 & 2 &   \\
			&	9 &   &   &   &   &   &   &   & 3 & 1 \\
			&	11 &   &   &   &   &   &   &   &   & 1 \\ \hline
		\end{tabular}
		\begin{tabular}{|c|c|ccccccccc|}\hline
			$(d_1,d_B,d_F)$&$2r \backslash 2s$ & -5 & -3 &-1 & 1 & 3 & 5 & 7 & 9 & 11 \\
			\hline
			$(1,4,3)$&	-5 & 1 & 1 &   &   &   &   &   &   &   \\
			&	-3 & 1 & 3 & 1 &   &   &   &   &   &   \\
			&	-1 &   & 1 & 5 & 1 &   &   &   &   &   \\
			&	1 &   &   & 1 & 5 & 1 &   &   &   &   \\
			&	3 &   &   &   & 1 & 5 & 1 &   &   &   \\
			&	5 &   &   &   &   & 1 & 5 & 1 &   &   \\
			&	7 &   &   &   &   &   & 1 & 5 & 1 &   \\
			&	9 &   &   &   &   &   &   & 1 & 3 & 1 \\
			&	11 &   &   &   &   &   &   &   & 1 & 1 \\ \hline
		\end{tabular}
		\begin{tabular}{|c|c|ccccccccccccc|}\hline
			$(d_1,d_B,d_F)$&$2r \backslash 2s$& -11 & -9 & -7 & -5 & -3 & -1 & 1 & 3 & 5 & 7 & 9 & 11 & 13 \\ 
			\hline
			$(1,5,2)$&	-11 &   &   & 1 &   &   &   &   &   &   &   &   &   &   \\
			&	-9 &   & 1 & 2 & 1 &   &   &   &   &   &   &   &   &   \\
			&	-7 & 1 & 2 & 4 & 3 & 1 &   &   &   &   &   &   &   &   \\
			&	-5 &   & 1 & 3 & 7 & 3 & 1 &   &   &   &   &   &   &   \\
			&	-3 &   &   & 1 & 3 & 8 & 3 & 1 &   &   &   &   &   &   \\
			&	-1 &   &   &   & 1 & 3 & 8 & 3 & 1 &   &   &   &   &   \\
			&	1 &   &   &   &   & 1 & 3 & 8 & 3 & 1 &   &   &   &   \\
			&	3 &   &   &   &   &   & 1 & 3 & 8 & 3 & 1 &   &   &   \\
			&	5 &   &   &   &   &   &   & 1 & 3 & 8 & 3 & 1 &   &   \\
			&	7 &   &   &   &   &   &   &   & 1 & 3 & 7 & 3 & 1 &   \\
			&	9 &   &   &   &   &   &   &   &   & 1 & 3 & 5 & 2 & 1 \\
			&	11 &   &   &   &   &   &   &   &   &   & 1 & 2 & 2 &   \\
			&	13 &   &   &   &   &   &   &   &   &   &   & 1 &   &   \\ \hline
		\end{tabular}
	}
	\caption{Open BPS invariants for Hirzebruch surface $\mathbb{F}_2$, with $Q_1$ playing the role of open K\"{a}hler parameter. The open strings around the compact divisor, whose length is $Q_1Q_B^4Q_F^2$, have degrees at least $(1,4,2)$. }\label{tb:F2}
\end{table}

\newpage

\clearpage

\acknowledgments
We thank Andrea Brini, Sung-Soo Kim, H\'{e}lder Larragu\'{i}vel, Mi{\l}osz Panfil, and Xin Wang for valuable discussions and comments on the manuscript. S.C. thanks DESY and ICTP for hospitality at different stages of this work. This work has been supported by the TEAM programme of the Foundation for Polish Science co-financed by the European Union under the European Regional Development Fund (POIR.04.04.00-00-5C55/17-00).

\appendix 

\appendix


\section{Various identities}\label{notation}

Identities involving Schur functions and Nekrasov factors:

\begin{align}
	&s_\u(q^\p)=q^{ \frac{||\u||^2-||\u^T||^2 }{2} } s_{\u^T}(q^\p)   \\
	&s_{\v/\la}(z)=\begin{cases}
		z^{|\v|-|\la|}   & \v \> \la\\
		0   & \text{others}\,.
	\end{cases} \\
	& \tZ_{\v}(t,q):=\prod\limits_{(i,j)\in \v   } 
	\left(     1-q^{  \v_i-j} t^{\v_j^T -i+1}       \r)^{-1}  \,, \\
	&	||\tZ_{\u}(t,q)||^2:= \tZ_{\u^T}(t,q) \tZ_{\u}(q,t)\,,\\
	&	||\tZ_{\u}(t,q)||^2= 
	||\tZ_{\u^T}(q,t)||^2\,,\\
	&N_{\u \v}(Q; t, q) :=\prod\limits_{i, j=1}^\inf \frac{1- Q~ q^{\v_i -j}~ t^{\u_j^T-i+1}	}{	1- Q~ q^{-j} ~t^{-i+1}	} \\
	&N_{\u \v}(Q; t, q) = 
	\prod\limits_{(i,j)\in \v}\left(1-Q~q^{\v_i - j }~t^{\u_j^T -i+1}	\r)  
	\prod\limits_{(i,j) \in \u} \left(	1-Q~q^{-\u_i+j-1}~t^{-\v_j^T+i} 	\right)\, ,	\\
	&	N_{\v}^{\rm half,+}(Q; t,q):= N_{\0 \v}(Q\sqrt{ \frac{q}{t}}, t,q)  \,, \\
	&N_{\v}^{\rm half,-}(Q; t,q):= N_{\v \0}(Q\sqrt{ \frac{q}{t}}, t,q) \,,\\  
	&N_{\v}^{\rm half,+}  \left(Q; t^{-1},q^{-1}  \r) =N_{\v^{T}}^{\rm half,-}\left(Q; q^{-1},t^{-1} \r)\,,\\
	& N_{\v}^{\rm half,+}  \left(Q; t^{-1},q^{-1}  \r) = N_{\0 \v}  \left(Q\sqrt{ \frac{t}{q}}, t^{-1},q^{-1}  \r), \\
	&N_{\v}^{\rm half,-}\left(Q; t^{-1},q^{-1} \r)=
	N_{\v \0}\left(Q\sqrt{ \frac{t}{q}}, t^{-1},q^{-1} \r) \,,\\
	& N_\v^{\half,+}(Q,t^{-1},q^{-1}  ) = (-Q)^{|\v|} t^{\frac{||\v^T||^2}{2}} q^{\frac{-||\v||^2}{2}  } N_{\v}^{\half, -}(Q^{-1}, t^{-1},q^{-1}) \,.
\end{align}

Identities involving $q$-Pochhammer symbols:
\begin{align} 
	& (q,q)_n=\prod_{d=1}^{n}(1-q^d)\,,~~(q,q)_0 :=1\,,~~(q,q)_1=(1-q) \,,   \\
	&(\a; q)_\inf=\prod\limits_{k=0}^{\inf}(1- \a q^k)=\prod\limits_{k=1}^{\inf} \left(1- \frac{\a}{q} q^k \r)  ,\\
	&   ( \a,q )_\inf=\sum_{d_i=0}^{\inf}  \frac{ (-1)^{d_i} q^{d_i^2 /2 }   } { (q,q)_{d_i}  }
	\left(\frac{\a}{\sqrt{q}} \r)^{d_i}
	= \sum_{d_i=0}^{\inf}   
	\left( -\sqrt{q}  \r)^{d_i^2} \frac{  \left(\frac{\a}{\sqrt{q}} \r)^{d_i}    } { (q,q)_{d_i}  }
	,\\
	&
	\frac{1}{( \a,q )_\inf}=\sum_{d_i=0}^{\inf} \frac{\a^{d_i} } { (q,q)_{d_i}  } \,,\\
	& (\a;q)_n=\frac{ (\a,q)_\inf  }{(\a q^n,q)_\inf   }  =\sum_{d_i,d_j=0}^{\inf}
	\left( 
	- \sqrt{q} \r)^{d_i^2+2 d_j n  } 
	\frac{ 
		\left(\frac{\a}{\sqrt{q}} \r)^{d_i}  
		\a^{d_j}  }{  (q;q)_{d_i}  (q;q)_{d_j}    }  \,,\\
	&\frac{1}{ (\b;q)}_n   = \frac{ (\b q^n,q)_\inf   }{ (\b,q)_\inf  } =\sum_{d_i,d_j=0}^{\inf}
	\left( - \sqrt{q} \r)^{d_i^2+2 d_i n  }  
	\frac{ \left(    \frac{\b}{\sqrt{q}} \r)^{d_i} \b^{d_j}    }{  (q;q)_{d_i}  (q;q)_{d_j}    }  \,,\\
	& \left( \frac{1}{t},  \frac{1}{q}  \r)_n=(-1)^n t^{-n} q^{\frac{n-n^2}{2}} (t;q)_n  ~, \\
	&\frac{1} {\left( \frac{1}{q},  \frac{1}{q}  \r)_n}=
	\frac{(-1)^n(\sqrt{q})^{n^2+n} }{ (q,q)_n} \,
	  , \\
	& \left( \frac{Q}{t}, \frac{1}{t} \r)_\inf  \cdot (Q,t)_\inf=1\,,\\
	&   \left(Q;   q^{-1}  \r)_n=(-Q)^n q^{\frac{n-n^2}{2}} (Q^{-1};q)_n  =      (-\sqrt{q})^{-n^2}   (\sqrt{q}Q)^n   (Q^{-1};q)_n          \,,\\
	& \frac{ M(Q;t,q)  }{M(t Q; t,q)   } =\prod\limits_{i=0}^\inf(1- Qq~ q^i)=(Q q; q)_\inf\,,\\
	&  \frac{ M(Q;t,q)  }{M(q Q; t,q)   } = \prod\limits_{i=0}^\inf(1- Q q~ t^i)   =(Q q; t)_\inf\,.
\end{align}


\bibliographystyle{JHEP}

\newpage
\bibliography{abmodel}

\end{document}